\documentclass[12pt]{article}

\pdfoutput=1
\usepackage{verbatim}
\usepackage[top=80pt,bottom=85pt,left=57pt,right=57pt]{geometry}
\usepackage{amssymb,amsmath,xypic}
\usepackage{slashed}
\usepackage{graphicx,color,subfigure}
\usepackage{float}
\usepackage{sectsty}
\usepackage{cite}
\usepackage[debug,pageanchor=false]{hyperref}
\definecolor{link}{rgb}{.8,.15,.1}
\hypersetup{colorlinks=true,linkcolor=link,citecolor=link,urlcolor=link,linktocpage}

\makeatletter
\@addtoreset{equation}{section}
\makeatother

\newcommand{\beq}{\begin{equation}}
\newcommand{\eeq}{\end{equation}}
\newcommand{\bea}{\begin{eqnarray}}
\newcommand{\eea}{\end{eqnarray}}
\newcommand{\nn}{\nonumber}

\begin{document}

\begin{titlepage}

\begin{center}

\vskip .5in 
\noindent

{\Large \bf{All $\mathcal{N}=(8,0)$ AdS$_3$ solutions in 10 and 11 dimensions}}

\bigskip\medskip

Andrea Legramandi$^{a,b,}$\footnote{andrea.legramandi@swansea.ac.uk}, Gabriele Lo Monaco$^{c,d,}$\footnote{gabriele.lomonaco@ipht.fr}, Niall T. Macpherson$^{e,f}$\footnote{ntmacpher@gmail.com}\\

\bigskip\medskip
{\small

$a$: Dipartimento di Fisica, Universit\`a di Milano--Bicocca, \\ Piazza della Scienza 3, I-20126 Milano, Italy \\ and \\ INFN, sezione di Milano--Bicocca\\
\vskip 3mm
$b$: Department of Physics, Swansea University, Swansea SA2 8PP, United Kingdom\\
\vskip 3mm
$c$: Department of Physics, Stockholm University, AlbaNova, 10691 Stockholm, Sweden\\
\vskip 3mm
$d$: Institut de Physique Th\'eorique, Universit\'e Paris Saclay, CEA, CNRS, \\ Orme des Merisiers, 91191 Gif-sur-Yvette CEDEX, France}\\
\vskip 3mm
$e$: International Institute of Physics, Universidade Federal do Rio Grande do Norte,
Campus Universitario - Lagoa Nova, Natal, RN, 59078-970, Brazil\\
\vskip 3mm
$f$: Department of Physics, University of Oviedo, Avda. Federico Garcia Lorca s/n, 33007 Oviedo,
Spain\\
\vskip 3mm

\vskip .9cm 
     	{\bf Abstract }\\
			~\\
			\end{center}
We classify AdS$_3$ solutions preserving $\mathcal{N}=(8,0)$ supersymmetry in ten and eleven dimensions and find the local form of each of them. These include  the  AdS$_3\times$S$^6$ solution of \cite{Dibitetto:2018ftj} and the embeddings of AdS$_3$ into AdS$_4\times$S$^7$, AdS$_5\times$S$^5$, AdS$_7/\mathbb{Z}_k\times$S$^4$ and its IIA reduction within AdS$_7$. More interestingly we find solutions preserving the superconformal algebras $\mathfrak{f}_4$, $\mathfrak{su}(1,1|4)$, $\mathfrak{osp}(4^*|4)$ on certain squashings of the 7-sphere. These solutions asymptote to AdS$_4\times$S$^7$  and are promising candidates for holographic duals to defects in Chern-Simons matter theories.

\vskip .1in

\noindent

\noindent

\vfill
\eject

\end{titlepage}

\tableofcontents

\section{Introduction and summary}

In this work we will concern ourselves with the construction  of supergravity solutions in ten and eleven dimensions that give a dual description of 2-dimensional supersymmetric conformal field theories (SCFTs) in terms of the AdS$_3$/CFT$_2$ correspondence. Two dimensional SCFTs are consistent with a far wider array of superconformal algebras than their higher dimensional counter-parts. This is intimately related to the symmetry algebra of CFT$_2$s, the Virasoro algebra, which being infinite dimensional leads to many possibilities for supersymetric extensions - see \cite{Fradkin:1992bz} for a classification of these.  Given this wealth of options, the classification and construction of AdS$_3$ solutions realising them should be rich and varied, but it is at this time mostly unknown. In \cite{Beck:2017wpm} all superconformal algebras that may be embedded into AdS$_3$ solutions of $d=10,11$ supergravity are found, many cases have no known example, many more only a small number - we aim to expand upon these.

The best understood avatars of the AdS/CFT correspondence provide a map between geometries and CFTs preserving maximal supersymmetry, which for AdS$_3$ solutions in ten or eleven dimensions is 16 real supercharges \cite{Haupt:2018gap}. These maximal cases are thus a well motivated place to focus a classification effort. Since $d=2$ superconformal algebras are chiral and a solution can support two distinct algebras of opposing chirality, there are  multiple distinct ways to construct  maximally supersymmetric AdS$_3$ solutions. The canonical examples are the D1-D5 and D1-D5-D1-D5  near horizons of \cite{Maldacena:1997re,deBoer:1999gea}, which preserve $\mathcal{N}=(4,4)$ symmetry with small and large superconformal symmetries respectively. (A classification of AdS$_3$ solutions with large $\mathcal{N}=(4,4)$ in M-theory is given across \cite{DHoker:2008lup,Estes:2012vm,Bachas:2013vza} and in type II in \cite{Macpherson:2018mif}). Generically, any solution that is  $\mathcal{N}=(n,8-n)$ supersymmetric, for $n=0,...8$, is a maximal case - they are all deserving of study, but we will not attempt to do justice to them all here. Instead we shall set ourselves the more modest goal of classifying all solutions in ten and eleven dimensions preserving $\mathcal{N}=(8,0)$ supersymmetry\footnote{Note that the difference between this and $\mathcal{N}=(0,8)$ is essentially just a matter of convention. Additionally we need only consider solutions in type II and M-theory explicitly, as while AdS$_3$ solutions exist in Heterotic supergravity, they are incompatible with $\mathcal{N}=(8,0)$ \cite{Beck:2017wpm}.} - we will in fact find the local form of each of them.  For related work on AdS$_3$ solutions preserving various supersymmetries see for instance \cite{Martelli:2003ki,Tsimpis:2005kj,Kim:2005ez,Kim:2007hv,Figueras:2007cn,Donos:2008hd,Colgain:2010wb,Jeong:2014iva,Lozano:2015bra,Kelekci:2016uqv,Couzens:2017way,Couzens:2017nnr,Eberhardt:2017uup,Dibitetto:2018iar,Macpherson:2018mif,Legramandi:2019xqd,Lozano:2019emq,Lozano:2019jza,Lozano:2019zvg,Lozano:2019ywa,Couzens:2019mkh,Couzens:2019iog,Passias:2019rga,Filippas:2019ihy,Speziali:2019uzn,Lozano:2020bxo,Farakos:2020phe,Couzens:2020aat,Rigatos:2020igd,Faedo:2020nol,Dibitetto:2020bsh,Filippas:2020qku,Passias:2020ubv,Faedo:2020lyw,Eloy:2020uix}.

To our knowledge, the  first\footnote{Strictly speaking, this is the first example that is not merely a reparameterisation of a higher dimensional AdS space.} example of an AdS$_3$ solution preserving $\mathcal{N}=(8,0)$ was found in \cite{Dibitetto:2018ftj}. This solution is in massive IIA and realises maximal supersymmetry in terms of the superconformal algebra $\mathfrak{f}$(4). Beyond this, a systematic classification  effort focused on $\mathcal{N}=(8,0)$ AdS vacua of 3-dimensional gauged supergravity was recently performed in \cite{Deger:2019tem}, finding many such examples exhibiting all consistent maximal superconformal algebras - namely  $\mathfrak{osp}$(8$|$2), $\mathfrak{f}$(4), $\mathfrak{su}$(1,1$|$4) and $\mathfrak{osp}$(4$^*|$4) \cite{Fradkin:1992bz}. Interestingly, the majority of these solutions come with symmetry groups too  large to be lifted to ten or eleven dimensions. However this need not  discourage our efforts here as a single gauged supergravity vacuum can be lifted to many distinct solutions in higher dimensions.\footnote{As an example, there is a single AdS$_6$ vacuum of Romans F(4) gauged supergravity that lifts \cite{Hong:2018amk} to the unique \cite{Passias:2012vp} AdS$_6$ local solution in type IIA \cite{Brandhuber:1999np} and infinitely many local solutions in IIB \cite{DHoker:2016ujz}. For alternative classifications of AdS$_6$ see also \cite{Apruzzi:2014qva,Kim:2015hya,Apruzzi:2018cvq,Legramandi:2021uds}}\\
~\\
The lay out of this paper is as follows, a summary of our results is also give in table \ref{table:1}\\
~~\\
In section \ref{sec:AllSolutions} we review which superconformal algebras may be  realised by AdS$_3$ solutions with ${\cal N}=(8,0)$ supersymmetry. We use group theoretical arguments to reduce the problem of classifying them down to a few distinct ansatze that need to be studied explicitly. We then proceed to find the local form of each solution residing within these in the rest of the paper. We study solutions with AdS$_3\times$S$^5$ factors in type II in section \ref{sec:ads3s5typeII} where we find the only possibilities are the AdS$_3\times$S$^6$ solution of \cite{Dibitetto:2018ftj} and AdS$_5\times$ S$^5$ realising $\mathfrak{f}(4)$ and  $\mathfrak{su}$(1,1$|$4)  respectively. In section \ref{sec:ads3s4s2typeII} we look for type II solutions realising  $\mathfrak{osp}$(4$^*|$4) in terms of AdS$_3\times$S$^4\times$S$^2$ factors, and find there is only the IIA  reduction of a $\mathbb{Z}_k$ orbifold\footnote{Both the orbifolding and reduction are performed on the hopf fiber of a 3-sphere factor within AdS$_7$} of AdS$_7\times$S$^4$. Having exhausted the possibilities in 10 dimensions, we next study M-theory solutions with AdS$_3\times$S$^4$ factors in section \ref{sec:M-theory} finding just AdS$_4\times$S$^7$,  realising $\mathfrak{osp}$(8$|$2), and the $\mathbb{Z}_k$ orbifolded AdS$_7\times$S$^4$ exist. Finally in section \ref{sec:squshed7spheres} we consider the more interesting, and technically challenging, possibility of realising the algebras  $\mathfrak{f}$(4), $\mathfrak{su}$(1,1$|$4) and $\mathfrak{osp}$(4$^*|$4) on 3 distinct squashings of the 7-sphere, finding a new solution for each. These solutions are all foliations over semi-infinite intervals which tend to AdS$_4\times$S$^7$  at infinity. They are thus good candidates for holographic duals to defects in Chern-Simons matter theories. The main text is supplemented by extensive technical appendices.

\begin{table}[h!]
\centering
\begin{tabular}{||c | c | c | c | c||} 
 \hline
 Geometry & Algebra  & Supergravity & Location & Comment\\ [0.5ex] 
 \hline\hline
 AdS$_4\times$S$^7$ & $\mathfrak{osp}(8|2)$ &    M theory &  \eqref{eq:mtheorysol1} & locally higher dim AdS\\ 
 AdS$_3\times\widehat{\text{S}}^7\times I$ & $\mathfrak{f}(4)$ & M theory & \eqref{eq:squshed2} & conformal defect \\
 AdS$_3\times$S$^6\times I$ & $\mathfrak{f}(4)$ & IIA & \eqref{eq:AdS3XS6}& \cite{Dibitetto:2018ftj} \\
 AdS$_3\times\widehat{\text{S}}^7\times I$ & $\mathfrak{su}(1,1|4)$ & M theory &\eqref{eq:squshed1}& conformal defect\\
 AdS$_5\times $S$^5$& $\mathfrak{su}(1,1|4)$ & IIB & \eqref{eq:AdS5xS5} & locally higher dim AdS\\
 AdS$_3\times\widehat{\text{S}}^7\times I$ & $\mathfrak{osp}(4^*|4)$ &  M-theory & \eqref{eq:squshed3}& conformal defect \\ 
 AdS$_7/\mathbb{Z}_k\times$S$^4$ &  $\mathfrak{osp}(4^*|4)$&  M theory &  \eqref{eq:mtheorysol2} & locally higher dim AdS\\ 
 AdS$_3\times$S$^4\times$S$^2\times I$ &  $\mathfrak{osp}(4^*|4)$ &  IIA & \eqref{eq:S2S4sol}  & reduction of former \\ 
 \hline
\end{tabular}
\caption{Summary of $\mathcal{N}=(8,0)$ solutions, the maximal $d=1+1$ superconformal algebras they preserve and where they are presented in the paper. Here $\widehat{\text{S}}^7$ represents a squashing of the 7-sphere and when $I$ appears it is an interval over which the rest of the geometry is foliated.  The higher-dimensional AdS spaces should be understood as  foliations containing an AdS$_3$ factor}
\label{table:1}
\end{table}

\section{ Realising  $\mathcal{N}=(8,0)$ supersymmetry for AdS$_3$ solutions}
\label{sec:AllSolutions}
In this section we review the possible superconformal algebras that $\mathcal{N}=(8,0)$ solutions should realise on their internal spaces. We then narrow down the types of geometry one needs to classify in order to find all such solutions. We begin by giving a generic recipe for realising extended supersymmetry in terms of $\mathcal{N}=1$ conditions and spinors that transform under some representation of an R-symmetry - which we will make much use of later in the paper.

\subsection{A recipe for AdS$_3$ with extended supersymmetry}\label{sec:arecipe}
Superconformal algebras in $1+1$ dimensions are chiral, and may be classified in terms of their bosonic sub-algebra\footnote{Strictly speaking this decomposition only applies to simple algebras \cite{Fradkin:1992bz}, but all algebras that may be embedded within a supergravity solution with an AdS$_3$ factor are simple.}
\beq\label{eq:bosinicsubalgebra}
\mathfrak{sl}(2)\oplus \mathfrak{g}_R
\eeq 
and a number of fermionic generators transforming in the $(\mathbf{2},\mathbf{\rho})$  representation of this algebra, with $\mathbf{\rho}$ a unique dimension $n$ representation of $\mathfrak{g}_R$, the Lie algebra of the R-symmetry group. Here $\mathfrak{sl}(2)$ is one factor of the conformal algebra $\mathfrak{so}(2,2)= \mathfrak{sl}(2)_{+}\oplus \mathfrak{sl}(2)_{-}$, with the specific factor determining the chirality of the superconformal algebra. A SCFT may realise two such algebras, provided they have opposing chiralities, thus a completely generic  SCFT may preserve $\mathcal{N}=(n_+,n_-)$ supersymmetry in terms of such left and right algebras.

In $d=10,11$ supergravity, one realises an SO(2,2) isometry group with an AdS$_3$ factor in the metric so that it decomposes as a warped product  AdS$_3\times$M$_{7/8}$ with M$_{7/8}$ an internal space to be determined. The Killing spinors on (unit) AdS$_3$ $\zeta$ transform in the \textbf{2} of an SL(2)$_{\pm}$ subgroup of this isometry when they obey
\beq\label{eq:AdS3KSE}
\nabla_{\mu}\zeta= \pm\frac{1}{2}\gamma_{\mu} \zeta
\eeq
and so realise the first factor in \eqref{eq:bosinicsubalgebra}. The R-symmetry isometry is realised by an additional factor in the internal space M$(\mathfrak{g}_R)$, so we should further decompose M$_{7/8}=$ M$(\mathfrak{g}_R)\times$M$_{\text{co-dim}}$. The manifold M$(\mathfrak{g}_R)$ could be a group manifold, coset space, or some fiber bundle and/or product involving these. Further, the AdS$_3$ and M$(\mathfrak{g}_R)$ factors may both be foliated over M$_{\text{co-dim}}$ if this does not break the required isometry. To fully realise the R-symmetry, all the bosonic supergravity fields need to be $\mathfrak{g}_R$ singlets, while there should be dim$(\mathbf{\rho})$ independent spinors on the internal space\footnote{Here we are being schematic. As we shall explain later, in M-theory it is sufficient to consider a single multiplet of 8d spinors on the internal space, but in type II one should have two independent multiplets - this changes nothing substantive in what follows though.} $\chi^I$ for $I=1,...$dim$(\mathbf{\rho})$, transforming in the $\mathbf{\rho}$ representation of $\mathfrak{g}_R$. This last point implies that under the spinoral Lie derivative the internal spinors obey
\beq\label{eq:SLD}
{\cal L}_{K_i}\chi^I = (\mathfrak{d}(\mathbf{\rho})_i)^I_{~J} \chi^J
\eeq 
with $K_i$ a set of Killing vectors on M$(\mathfrak{g}_R)$, and $\mathfrak{d}(\mathbf{\rho})_a$ some dim$(\mathbf{\rho})$ basis of $\mathfrak{g}_R$. As \eqref{eq:SLD} provides a map between each of the $n=$ dim$(\mathfrak{g}_R)$ supercharges, and we demand that the bosonic supergravity fields are $\mathfrak{g}_R$ singlets, it follows that if a solution is supersymmetric with respect to a single component of $\chi^I$, parameterising an ${\cal N}=1$ sub-sector, it is supersymmetric with respect to the whole of $\chi^I$. This can be seen as follows: We impose an AdS$_3$ factor on the solutions we seek, so the spinoral supersymmetry conditions in 10 or 11 dimensions will be implied by a reduced set of spinoral conditions on the 7/8 dimensional internal space. Schematically these reduced conditions will take one of two forms 
\beq\label{eq:scematicconds}
\Delta_1 \chi^I=0,~~~~ \nabla_a\chi^I= (\Delta_2 \gamma_a+\gamma_a \Delta_3)\chi^I
\eeq 
where $\Delta_{1,2,3}$ are operators formed of contractions of the physical supergravity fields and their derivatives with the internal gamma matrices -  as such they are $\mathfrak{g}_R$ singlets. If $\chi$ solves the first of \eqref{eq:scematicconds}, clearly then so does ${\cal L}_{K_i}\chi$. Likewise one can show that
\beq
\nabla_a({\cal L}_{K_i}\chi)= (\Delta_2 \gamma_a+\gamma_a \Delta_3)({\cal L}_{K_i}\chi),
\eeq 
when $\chi$ solves the second of \eqref{eq:scematicconds} (the identities  required  to prove this equality can be found in appendix \ref{sec:andreaextras}). As such if an ${\cal N}=1$ sub-sector of $\chi^I$ solves \eqref{eq:scematicconds}, \eqref{eq:SLD} then implies that the remaining components of $\chi^I$ also do - this allows us to exploit classifications of $\mathcal{N}=1$ AdS$_3$ solutions to find solutions with extended supersymmetry throughout the paper. 
 
\subsection{Realising $\mathcal{N}=(8,0)$ superconformal algebras}
In this work we are interested in $\mathcal{N}=(8,0)$ solutions, we will thus take the $+$ sign in \eqref{eq:AdS3KSE} - let us stress that this  is just a matter of conventions. The possible superconformal algebras consistent with $\mathcal{N}=(8,0)$ are the following
\beq\label{eq:Neq8alebras}
\begin{array}{c|c|c}
\text{Superconformal algebra} & \mathfrak{g}_R & \mathbf{\rho}~\\
\hline
\hline
\mathfrak{osp}(8|2)& \mathfrak{so}(8)&\textbf{8}\\
\mathfrak{f}(4)&\mathfrak{spin}(7) & \textbf{8}_s\\
\mathfrak{su}(1,1|4)& \mathfrak{su}(4)\oplus \mathfrak{u}(1)& \mathbf{4_{-1}}\oplus\overline{\mathbf{4_1}}\\
\mathfrak{osp}(4^*|4)& \mathfrak{sp}(2)\oplus \mathfrak{sp}(1)&(\textbf{4},\textbf{2})
\end{array}
\eeq 
where we include the R-symmetry algebra and the representation that needs to be realised by the internal space and spinors. Our first job is to identify which manifolds can realise the R-symmetry. Typically this is an expansive problem, with many distinct ways to realise $\mathfrak{g}_R$. However all the R-symmetry algebras in  \eqref{eq:Neq8alebras} are rather large, and we ultimately need to embed them into a 7 or 8 dimensional internal space, this leaves limited data to be determined by non group theoretical constraints -  we will thus be able to give the local form of all such solutions.

First we note that there is no group space of dim$\leq 8$ realising the entire of any of the R-symmetries\footnote{Of course one can realise $\mathfrak{sp}(1)=\mathfrak{su}(2)$ and $\mathfrak{u(1)}$ with S$^3$ and S$^1$, but these are only part of two of the R-symmetry algebras, the other factors cannot be realised as group spaces of small enough dimension.} in \eqref{eq:Neq8alebras}, so we will need to consider coset spaces. Putting aside the possibility of fibrations and deformations for the moment, we can  realise the R-symmetry groups in a solution by assuming that it contains one of the following factors
\beq
\label{tab:cosets}
\begin{array}{c|c}
\text{R-symmetry} & \text{M}(\mathfrak{g}_R) \\
\hline
\hline
\text{SO}(8)& \text{S}^7\\
\text{Spin}(7) & \text{S}^6\\
 \text{U}(4)& \text{S}^5\times \text{S}^1,~~~ \mathbb{CP}^3\times \text{S}^1,\\
\text{Sp}(2)\times\text{Sp}(1)&  \text{S}^4\times \text{S}^2,
\end{array}
\eeq
which list the minimal ways to achieve this, {\it i.e.} the 7-sphere actually contains all of these R-symmetry groups, but it can be decomposed in terms of each of these factors; we will comment more on this particular case in section \ref{sec:squashS7}. Observe that vacua containing round $\text{AdS}_3\times\text{S}^{5,6,7}$  factors can be thought as particular cases of the more general warped $\text{AdS}_3\times \text{S}^4$ ansatz: in fact $\frak{sp}(2)$ is a sub-algebra of each of $\frak{so}(8)$, $\frak{spin}(7)$ and $\frak{u}(4)$ but importantly it is not a viable R-symmetry on its own, so will always experience an enhancement\footnote{One can view this as a consistency constraint.}.  So one can explore the possibility of solutions preserving all but one  entry in \eqref{tab:cosets} in a unified way by classifying  AdS$_3\times$S$^4$ in 10 and 11 dimensions. For M-theory we shall do just that in section \ref{sec:M-theory}, however for mundane technical reasons, we find it easier to split type II into solutions preserving S$^5$ and $\text{S}^4\times \text{S}^2$ in sections \ref{sec:ads3s5typeII} and \ref{sec:ads3s4s2typeII} respectively.  The final possibility in \eqref{tab:cosets}, $\mathbb{CP}^3\times \text{S}^1$,  can be ruled out as follows:  Spinors on $\mathbb{CP}^3$ can be studied by considering their embedding in $\text{S}^7$ \cite{Nilsson:1984bj}; depending on the choice of embedding of U(4) in SO(8), the Killing spinors on $\text{S}^7$ transform either in the $\mathbf{6}_0\oplus\mathbf{1}_{2}\oplus\mathbf{1}_{-2}$ representation or in the $\mathbf{4}_{1}\oplus\overline{\mathbf{4}}_{-1}$, where the subscript denotes the charge under the Hopf-circle isometry.\footnote{The two choices correspond to different possible squashings of the seven-sphere.} A non-trivial charge signals the obstruction of a spinor to be a well-defined section of the spin-bundle on $\mathbb{CP}^3$. For this reason, the unique well-defined spinor multiplet on round $\mathbb{CP}^3$ transforms in the $\mathbf{6}$ representation of SU(4) and cannot serve as supersymmetry parameter in the case at hand. On the other hand, if one considers a U(1) fibration over $\mathbb{CP}^3$, this obstruction can be evaded. There are also other ways to realise the R-symmetries if we also consider squashing the seven-sphere and fibre bundles, to deal with the latter we will need to introduce some technology reviewed in appendix \ref{sec:lifting}.

\subsubsection{Realising the algebras using fibrations}\label{sec:fibrations}
Assuming the internal space decomposes in terms of a round sphere or spheres is the easiest way to realise the R-symmetries of the four distinct $(8,0)$ superconformal algebras, but this is also possible using fiber bundles. This complicates the problem, but one is still able to approach it systematically as we explain here. In order to work out which fibrations are possible one must establish under which condition an isometry of a base manifold may be lifted to the entire space.\\
~\\
As explained at greater length in appendix  \ref{sec:lifting} one can express a generic fiber bundle metric in the form 
\beq
d s^2 = (g_B)(x)_{mn} d x^m d x^n   + (g_F)(y)_{ij} D y^i D y^j ,~~~ D y^i = d y^i + {\cal A}_a K_F^{a i} \,
\eeq
where $g_B$ and $g_F$ are metrics on the base and fiber respectively and $K^{ai}_F\partial_{y^i}$ are a set of Killing vectors on the fiber. It is possible to show that an isometry of the base, parameterised by a Killing vector $K_B$, is lifted to an isometry of $ds^2$ if and only if ${\cal A}_a$ transforms as a gauge field with respect to the base isometry, {\it i.e.}
\beq\label{eq:liftingcondtion}
{\cal L}_{K_B} {\cal A}_a= d\lambda_a+ f^{bc}_{~a}\lambda_b {\cal A}_c
\eeq
where $f^{bc}_{~a}$ are the structure constants of the fiber isometry, and $\lambda_a$ are functions which determine the lift of the Killing vector in the full space
\beq
K = K_B- \lambda_a(x)K^{a i}_F\partial_{y_i}.
\eeq
The condition \eqref{eq:liftingcondtion} is a powerful way to  constrain the possible fiber bundles realising the R-symmetries in  \eqref{eq:Neq8alebras} - this is principally because they necessitate large base manifolds.  
~~\\
The most simple fiber bundle is a U(1) fibration. An immediate consequence of \eqref{eq:liftingcondtion} is that  the isometries of the base only get lifted when $d{\cal A}$ is a singlet with respect to them - this means for the case at hand that only U(1) fibrations over $\mathbb{CP}^3$ and S$^2$ are relevant for us,\footnote{As only these come equipped with the required invariant 2-forms whilst also being small enough to fit in at most 8 dimensions.} {\it i.e.} U(4) and Sp(1) preserving squashed 7 and 3-spheres. As we will see the first option leads to a new solution, while the second does not. Indeed it is relatively simple to establish that if a Sp(2)$\times$Sp(1) solution with a U(1) fibration over S$^2$ were to exist in 10d - it must necessarily be related by duality to a solution in 11d.\footnote{In such a fibration the U(1) must be a flavour symmetry so just Sp(1) is realised, thus T-duality on this direction breaks no supersymmetry. If such a solution had no NS flux component along the U(1), T-duality would map it to the symmetric space solution AdS$_3\times$S$^4\times$S$^2\times$S$^1$ - but this does not exist  \cite{Wulff:2017zbl}. If there is a leg in the U(1), then T-duality maps between U(1) fibrations so it is sufficient to consider IIA where the Bianchi identity of the RR 2-form would then force $F_0=0$, allowing a lift to 11d.} This is covered by the classification of AdS$_3\times$S$^4$ solutions in section \ref{sec:M-theory}, the only example is AdS$_7/Z_k\times$S$^4$.  

Although it is possible to realise the requisite R-symmetry groups on may other fiber bundles, the need to embed these into a 7 or 8 dimensions means that almost all of them are not possible. Indeed one soon realises that the only other viable option is to fiber a S$^3$ over S$^4$, which gives a Sp(2)$\times$Sp(1) preserving squashed 7-sphere, which appeared first in the AdS$_4$ solution of \cite{Awada:1982pk}.\\
~~
We have established that the only fiber bundles we need to explicitly consider are certain squashings of the 7-sphere, we shall discuss them and the spinors they preserve in the next section.\\

\subsubsection{Squashing the seven-sphere}
\label{sec:squashS7}
We established in the previous section that \eqref{tab:cosets} is not actually complete. In fact, in deriving this table we assumed the usual realisation of the round seven-sphere as the coset $\frac{\text{Spin}(8)}{\text{Spin}(7)}$ but it actually admits other coset realizations; in such cases, the metric is squashed or the isometries are broken by particular structures. We collect in the following table all possibilities:
\beq
\label{tab:squashS7}
\begin{array}{c|c|c}
	\text{Coset }&\text{  R-sym. algebra} & \text{Comments} \\
	\hline
	\hline
	\text{Spin}(8)/\text{Spin}(7)& \mathfrak{spin}(8)& \text{round S$^7$}\\
	\text{Spin}(7)/G_2 & \mathfrak{spin}(7) & \text{Spin(7) weak G$_2$ holonomy}\\
	\text{Sp}(2)/\text{Sp}(1) & \mathfrak{sp}(2)\oplus \mathfrak{sp}(1) & \text{$\text{S}^3$ fibration over $\text{S}^4$}\\
	\text{U}(4)/\text{U}(3) & \mathfrak{su}(4)\oplus \mathfrak{u}(1) & \text{U(1) fibration over $\mathbb{CP}^3$}
\end{array}
\eeq
The last two rows in \eqref{tab:squashS7} correspond to actual squashing of the round seven-sphere metric, those discussed in the previous section, while in the case of the coset $\text{Spin}(7)/G_2$, the metric is the same as in the round case, but isometries are broken by a Spin$(7)$ invariant  G$_2$ structure that may enter the flux; nevertheless, we will refer to this last case improperly as a squashed $\text{S}^7$.  Given the dimension of such manifolds, they fit into ansatze for M-theory that are foliations  of $\text{AdS}_3 \times \text{S}^7_{\text{Squash}}$ over an interval $I$, with appropriate invariant forms appearing in the internal flux. In type II supergravity there is no such interval so all warping are constant and, through explicit computation, it is quick to establish that they cannot solve the equations of motion, so there is no need to check whether supersymmetry is preserved. 

One can wonder if the cosets \eqref{tab:squashS7} actually host spinors transforming appropriately under the residual isometries, in order to preserve the right amount of supersymmetry \eqref{eq:Neq8alebras}. This problem can be easily addressed in the following way \cite{Duff:1986hr}. Any spinor on the coset spaces under consideration can be decomposed on a basis of eigenfunctions of the Dirac operator $\slashed D$. Since $\slashed D$ is an invariant operator, a complete basis decomposes into representations of the isometry group. 
Since the Killing vectors on the squashed S$^7$ are a subgroup of the S$^7$ ones, as showed explicitly for the Sp(2)$\times$Sp(1) case in appendix \ref{Appendix C}, we can use the Dirac operator of the round S$^7$ to classify $\text{Sp}(2) \times \text{Sp}(1)$ eigenspaces. This spinor basis is independent of the manifold we used to define the Dirac operator, which can be therefore chosen in a convenient way. 
The round seven-sphere is quite well-understood: in this case the spectrum can be decomposed into multiplets transforming in the representation $[n,0,0,1]$ or $[n,0,1,0]$ of the $\text{Spin}(8)$ isometry group; in particular there is exactly one multiplet for each $n$; the Killing spinors belong to the multiplets with $n=0$, {\it i.e.} the spinorial representations $\mathbf{8_s}$ and $\mathbf{8_c}$ respectively.
Supersymmetry usually selects one of the two representations or equivalently an orientation of the seven-sphere.

Since $\text{Sp}(2) \times \text{Sp}(1) \subset \text{Spin}(8)$, the eigenmodes which transforms with respect to $\text{Sp}(2) \times \text{Sp}(1)$ on the squashed S$^7$ can be obtained by simply branching the multiplets of the round sphere. In the $\text{Sp}(2)/\text{Sp}(1)$ case we have that:
\beq
\label{eq:branchSpin8-Sp2}
\begin{split}
	[n,0,0,1]\rightarrow \sum_{j=0}^{\lfloor n/2\rfloor}&\left\{(n+1-2j,j|n+1-2j)\oplus(n+1-2j,j|n-1-2j)\oplus\right.\\
	&\left.\oplus(n-1-2j, j+1|n+1-2j)\oplus(n-1-2j,j|n-1-2j)\right\}\,,\\
	[n,0,1,0]\rightarrow \sum_{j=0}^{\lfloor n/2\rfloor}&\left\{(n-2j,j+1|n-2j)\oplus(n-2j,j|n-2j)\oplus\right.\\
	&\left.\oplus(n-2j, j|n+2-2j)\oplus(n-2j,j|n-2-2j)\right\}\,,\\
\end{split}
\eeq
where $(n,m|k)$ denotes the representation with Dynkin coefficients $[n,m]$ with respect to $\text{Sp}(2)$ and Dynkin coefficient $[k]$ with respect to $\text{Sp}(1)$; in this notation, the representation $(\mathbf{4,2})$ of $\text{Sp}(2)\times \text{Sp}(1)$ is $(1,0|1)$. Using \eqref{eq:branchSpin8-Sp2}, it is evident that we have exactly three possible multiplets transforming in the $(\mathbf{4,2})$ representation, coming respectively from the branching of the $[0,0,0,1]$, $[1,0,1,0]$ and $[2,0,0,1]$ multiplets of $\text{Spin}(8)$. This immediately tells us that half of the Killing spinors for the round S$^7$ will still form a multiplet for the squashed one, as we will explicitly check in appendix \ref{Appendix C}.

In the same way, we can study the spectrum of the $\text{Spin(7)}/G_2$ coset looking at the branching rules of the spinor representations of $\text{Spin}(8)$ under $\text{Spin}(7)$. In this case we have:
\beq
\label{eq:branchSpin8-7}
\begin{split}
	[n,0,0,1]\,\,\rightarrow\,\,& [0,0,n+1]\oplus [0,1,n-1]\,,\\
	[n,0,1,0]\,\,\rightarrow\,\,& [1,0,n]\oplus [0,0,n]\,,\\
\end{split}
\eeq
where, in terms of Dynkin coeffients, the spinorial representation $\mathbf{8}$ of $\text{Spin}(7)$ is $[0,0,1]$; thus there exist exactly two multiplets transforming in this representation, coming respectively from the branching of the $[0,0,0,1]$ and $[1,0,1,0]$ multiplets of $\text{Spin}(8)$. 

Finally, in the $\text{U}(4)/\text{U}(3)$ case, the branching rules are:
\begin{align}
	[n,0,0,1]\rightarrow &\sum_{j=0}^{\lfloor \frac{n+1}{2}\rfloor}\left\{[n-j,0,j]_{n-2-2j}\oplus[j,0,n-j]_{-n+2+2j}\right\}\oplus\\
	&\oplus\sum_{j=1}^{\lfloor \frac{n}{2}\rfloor}\left\{[n+1-j,0,j]_{n+3-2j}\oplus[j,0,n+1-j]_{-n-3+2j}\right\}\oplus \sum_{j=1}^n [n-j,1,j-1]_{n+1-2j}\,,\nn\\
	[n,0,1,0]\rightarrow &\sum_{j=0}^{\lfloor \frac{n+1}{2}\rfloor}\left\{[n+1-j,0,j]_{n-1-2j}\oplus[j,0,n+1-j]_{-n+1+2j}\right\}\oplus\nn\\
	&\oplus\sum_{j=1}^{\lfloor \frac{n}{2}\rfloor}\left\{[n+1-j,0,j-1]_{n+4-2j}\oplus[j-1,0,n+1-j]_{-n-4+2j}\right\}\oplus \sum_{j=0}^n [n-j,1,j]_{n-2j}\,,\nn
\end{align}
where the subscripts denote the U(1) charge. In this case we look for the representation $\mathbf{4_{-1}}\oplus\mathbf{\overline{4}_1}$, corresponding in the previous notation to $[1,0,0]_{-1}\oplus[0,0,1]_{1}$. We can easily recognise that we have two $\mathbf{4_{-1}}$ multiplets and and two $\mathbf{\overline{4}_{1}}$ multiplets  coming from the branching of the $[0,0,0,1]$ and $[1,0,1,0]$ representations; as we will show, using these four multiplets it is possible to construct four sets of spinors transforming in the representation $\mathbf{4_{-1}}\oplus \mathbf{\overline{4}_{1}}$ appropriate for the $\mathfrak{su}(1,1|4)$ superconformal algebra.

We will provide additional details about the form of the spinors, the metric and the invariant forms on each coset in the dedicated sections, \ref{sec:u4cases}, \ref{sec:spin7cases} and \ref{sec:sp2sp1cases} respectively. We shall now proceed to the classification of solution in type II.

\section{AdS$_3\times$S$^5$ in type II}\label{sec:ads3s5typeII}
In this section we classify AdS$_3\times$S$^5$ solutions of type II supergravity dealing with IIB in section \ref{sec:S5IIB} and section \ref{sec:S5IIA} . To do this we make use of an existing ${\cal N}=1$ AdS$_3$ classification that we review in the next section. We shall also make the simplifying assumption that the electric component of NS 3-form is not turned on. In IIB one can show this is simply a choice of SL$(2,\mathbb{R})$ duality frame\footnote{As we find no solution with non trivial RR 3-form, such a solution can be ruled out.} while in IIA one can show that electric NS flux implies $F_0=0$, so if such a solution did exist,\footnote{Spoiler: They do not.} it could be lifted to a solution of the type we deal with in section \ref{sec:M-theory} we will come back to this point in the next section (we have a more detailed discussion around 3.9).

\subsection{Generalized $G$-structures and supersymmetry conditions}\label{eq:7dbispinors}
It is well known that the existence of a pair of ten-dimensional nowhere-vanishing chiral spinors satisfying the supersymmetry equations can be translated in the language of generalized G-structures, namely G-structures on the generalized tangent bundle $T\oplus T^*$ \cite{Tomasiello:2011eb}. Such structures can be described in terms of polyforms and supersymmetry equations translate into differential constraints for the bispinors defined by the supercharges. Let us give some more details.

We are interested in finding solutions preserving an SO(2,2) symmetry factor, {\it i.e.} solutions whose metrics can be written as a warped product of $\text{AdS}_3$ and some unknown seven-dimensional compact spin manifold:
\begin{equation}
\text{d}s^2\,=\,e^{2A}\,\text{d}s^2(\text{AdS}_3)\,+\,\text{d}s^2(\text{M}_7)\,.
\end{equation}
Since we want to preserve the full $\text{AdS}_3$ isometry group, the RR fluxes must have the following form:
\begin{equation}
F_{\pm}\,=\,f_{\pm}\,+\,e^{3A}\text{vol}(\text{AdS}_3)\wedge \star_7 \lambda f_{\pm}\,,
\end{equation}
where $f_{\pm}$ has legs on the internal manifold only ({\it i.e.} it is magnetic); the operator $\lambda$ acts on each form as $(-1)^{\lfloor \text{deg} \rfloor}$. If such background preserves at least $\mathcal{N}=1$ supersymmetry for some choice of internal manifold M$_7$, the transformations preserving it have parameters consisting of a pair of chiral ten-dimensional spinors of the form
\begin{equation}
\epsilon_{1}\,=\, \zeta \otimes \chi_{1}\otimes |+\rangle\,, \quad \epsilon_{2}\,=\, \zeta \otimes \chi_{2}\otimes |\mp\rangle
\end{equation}
where $\zeta$ is an $\text{AdS}_3$ Killing spinor:
\begin{equation}
\nabla_\nu\,\zeta\,=\,\frac{\mu}{2}\,\gamma_\nu\,\zeta\,,
\end{equation}
$\chi_{1,2}$ are Majorana spinors on M$_7$ and $|\pm\rangle$ are two-dimensional auxiliary spinors selecting the chirality. In type IIA, the Killing spinor $\epsilon_i$ have opposite chirality while in type IIB they have the same chirality. In the case at hand, the supercharges define a $G_2\times G_2$-structure \cite{Witt:2004vr} that can be described in term of poly-forms on M$_7$ defined by $\chi_i$ through the Clifford map:
\begin{equation}
\label{eq:7d_bispinors}
\chi_1\otimes \chi_2^\dagger\,= \frac{1}{8}\sum_{n=1}^7\frac{1}{n!}\,\chi_2^{\dagger}\,\gamma^{(7)}_{a_n...a_1}\,\chi_1\,e^{a_1}\wedge...\wedge e^{a_n}=\,\Psi_+\,+\,i\,\Psi_-\,,
\end{equation}
where $\gamma^{(7)}_a$ are a basis of 7-dimensional flat space gamma matrices, $e^{a}$ a vielbein on M$_7$ and 
where $\Psi_+$ is even and $\Psi_-$ is odd; they are related by Hodge duality:
\begin{equation}
\Psi_-\,=\,-\star_7\,\lambda \Psi_+\,,
\end{equation}
The constraints imposed by supersymmetry are reduced to a set of conditions to be satisfied by the $G_2\times G_2$ structure \cite{Dibitetto:2018ftj}:
\begin{eqnarray}
\label{eq:GeneralDiff}
&\text{d}_{H}\left(e^{A-\phi}\Psi_\mp \right)\,=\,0\,,\quad &\text{d}_H\left( e^{2A-\phi} \Psi_\pm\right)-2\mu\,e^{A-\phi}\Psi_\mp\,=\,\frac{1}{8}e^{3A}\star \lambda f_{\pm}\,,\\
\label{eq:GeneralPair}
&\left( \Psi_\mp, f_\pm\right)\,=\,\frac{\mu}{2} e^{-\phi}\,, &\left( \Psi_+,\Psi_- \right)\,=\,\frac{1}{8} e^{2A}\,,
\end{eqnarray}
where the upper sign holds in type IIA  case while the lower case must be picked in the IIB context, $\phi$ is the dilaton, $\text{d}_{H} \equiv \text{d}-H\wedge$ and we introduced the Mukai pairing $(\Psi, \Phi)\,=\,\frac{(\Psi\wedge \lambda \Phi)_7}{\text{vol}_{\text{M}_7}}$. Moreover, in this and the next sections, we will take $\mu=1$. Observe the system \eqref{eq:GeneralDiff}-\eqref{eq:GeneralPair} assumes $H$ to be purely magnetic. However, it must be stressed that the NS flux can also contain an electric component of the form $h\,\text{vol}_{\text{AdS}_3}$, with $h$ a constant; in fact, such term is closed and still preserves the full $SO(2,2)$ isometry group. However, we can assume $h=0$ without loss of generality for the following reason. In the type IIB it is always possible to find an SL$(2, \mathbb{R})$ duality frame where the electric component vanishes. In the type IIA case, instead, let us assume that $h\neq 0$; as a consequence, the first equation in \eqref{eq:GeneralDiff} gets a correction:
\begin{equation}
\text{d}_{H_0}\left( e^{A-\phi}\Psi_{-}\right)\,\propto\, h f_+\,.
\end{equation}
where $H_0$ is the magnetic component of $H$. We recognize that, if $h\neq 0$, this immediately implies the Romans mass must vanish, $f_0=0$, and any solution  admits a lift to M-theory, whose analysis is performed in section \ref{sec:M-theory}. In this sense, we can assume $h=0$.

As discussed in section \ref{sec:AllSolutions}, backgrounds realizing the super-algebras $\frak{osp}(8|2)\,,\frak{f}(4)$ or $\frak{su}(1,1|4)$ in terms of round spheres must contain an (warped) $\text{AdS}_3\times \text{S}^5$ factor:
\begin{equation}
\text{d}s^2\,=\,e^{2A}\,\text{d}s^2(\text{AdS}_3)+e^{2Q}\,\text{d}s^2(\text{S}^5)\,+\,\text{d}s^2(\text{M}_2)\,,
\end{equation}
where $M_2$ is a two-dimensional manifold to be determined; the warpings $A\,,Q$ only depend on the M$_2$ coordinates. The SU(4) sub-group of the R-symmetry is realised as isometry group of a round five-sphere. The Killing spinors on S$^5$ transform in the $\mathbf{4}$ of this symmetry, which is indeed a sub-representation of the first three entries in \eqref{eq:Neq8alebras}. As a consequence, the seven-dimensional Majorana spinors defining the $G_2\times G_2$ structure \eqref{eq:7d_bispinors} can be written as follows:
\begin{equation}
\chi_i\,=\,e^{\frac{A}{2}}\,\Big(\xi^{\text{S}^5}\otimes  \eta_i\,+\big(\xi^{\text{S}^5 }\otimes  \eta_i)^c\Big)\,,
\end{equation}
where $\xi^{\text{S}^5}$ is a Killing spinor on the five sphere, $\eta_i$ are unknown spinors on M$_2$ and the $c$ superscript denotes Majorana conjugation. In order to define the $G_2\times G_2$ structure we need to introduce a suitable splitting of the $\gamma$-matrices:
\begin{equation}
\label{eq:GammaS5xM2}
\gamma^{(7)}_{\alpha}\,=\,e^{Q}\gamma^{(5)}_{\alpha}\otimes \sigma_3\,,\quad \gamma^{(7)}_{i}\,=\,\mathbb{I}\otimes \sigma_i\,,
\end{equation}
where $\sigma_i$ denotes the Pauli matrices. The seven-dimensional conjugation matrix can be taken to be $B_{(7)}=B_{(5)}\otimes\,i\sigma_2$ where $B_{(5)}$ is real, antisymmetric and $B_{(5)}^2=-\mathbb{I}$. Let us define the following forms:
\begin{equation}
\begin{split}
&v\,=\, \sum_i^2 \eta_2^\dagger\,\sigma_i\,\eta_1\,e_i\,, \quad \, j\,=\,\sum_i \eta_2^\dagger\,(1-i\text{vol}(\text{M}_2)\,\sigma_3)\,\eta_1\,,\\
&w\,=\,\sum_i^2\overline \eta_2\,\sigma_i\,\eta_1\,e_i\,,\quad\, \omega\,=\, \overline\eta_2\,(1-i \text{vol}(\text{M}_2)\,\sigma_3)\,\eta_1\,,
\end{split}
\end{equation}
where $e_i$ are vielbein on M$_2$, $\text{vol}(\text{M}_{2})=e_1\wedge e_2$ and $\overline{\eta}_2=-i \,\eta^t\sigma_2$. Using the Clifford map and the decomposition \eqref{eq:GammaS5xM2} it is straightforward to show that the $G_2\times G_2$ structure can be written as follows:
\begin{equation}
\label{eq:BispS5}
\begin{split}
&\Psi_+\,=\,e^A\,\text{Re}\left( \left(j-i\,e^Q\,\star_2\!v\wedge V\right)\wedge e^{-i\,e^{2Q}J}+\left(\omega-i\,e^Q\,\star_2\!w\wedge V\right)\wedge e^{3Q}\,\Omega\right)\,,\\
&\Psi_-\,=\,e^A\,\text{Im}\left( \left(v+i\,e^Q\,\lambda\!\star_2\!j\wedge V\right)\wedge e^{-i\,e^{2Q}J}\,+\,\left(w+i\,e^Q\,\lambda\!\star_2\!\omega\wedge V\right)\wedge e^{3Q}\Omega \right)\,,
\end{split}
\end{equation}
where $(V,J, \Omega)$ is one of the Sasakian structures on $\text{S}^5$,  $(V_{\pm}, J_{\pm}, \Omega_{\pm})$ as reviewed in the appendix \ref{sec:BispSphere}.

\subsection{Reducing the supersymmetry equations in type IIB}\label{sec:S5IIB}

Let us focus on the choice $(V_+, J_+, \Omega_+)$ and $\mu=1$ for sake of clarity; the analysis goes along the same lines for other choices. The unique invariant form on S$^5$ is the volume form $\frac{1}{2}V_+\wedge J_+^2$. In following, we will omit the subscript denoting the choice of structure.  As there are no SU(4) invariant forms on the 5-sphere but the volume form, we must set the NS flux to zero, $H=0$, and the unique consistent choice of RR fluxes is the following:
\begin{equation}
f_-\,=\,\text{d}q_1+\,p\,\text{vol}(\text{S}^5)\,+\,q_7\,\text{vol}(\text{M}_2)\wedge \text{vol}(\text{S}^5)\,,
\end{equation}
where $q_1$ and $q_7$ are functions on M$_2$ while $p$ is a constant; with this choice, the Bianchi identity $\text{d} F=0$ is automatically satisfied. Let us observe that the bispinors \eqref{eq:BispS5} can be divided into two different pieces, one depending on $J$ and one depending on $\Omega$. Such sectors are not mixed under the action of the exterior derivative and the supersymmetry equation \eqref{eq:GeneralDiff} can be written as:
\begin{align}
\label{eq:S5one}
&\text{d}\left(e^{2A-\phi+3Q}\text{Re}(\omega-e^Q\,\star_2\!w\wedge V)\wedge \Omega\right)\,=\,0,\\
\label{eq:S5two}
&\text{d}\left(e^{2A-\phi}\text{Re}(j-e^Q\,\star_2\!v\wedge V)\wedge e^{-i\,e^{2Q}J} \right)\,=\,0\,,\\ 
\label{eq:S5three}
&\text{d}\left(e^{3A-\phi+3Q}\text{Im}(w+e^Q\,\lambda\!\star_2\!\omega\wedge V)\wedge \Omega\right)-2\,e^{2A-\phi+3Q}\,\text{Re}(\omega-e^{Q}\star_2\!w\wedge V)\wedge \Omega )\,=\,0,\\
\label{eq:S5four}
&\text{d}\left(e^{3A-\phi}\text{Im}(v+e^Q\,\lambda\!\star_2\!j\wedge V)\wedge e^{-i\,e^{2Q}J}\right)-2\,e^{2A-\phi}\,\text{Re}(j-e^Q\,\star_2\!v\wedge V)\wedge e^{-i\,e^{2Q}J}\,=\,\frac{1}{8}e^{3A}\star\lambda f_-\,.
\end{align}
The three-form part of the equation \eqref{eq:S5three} immediately implies that the zero-form component of $\omega$ must vanish, $\omega_0=0$. This allows us to parametrise the spinors $\eta_1$ and $\eta_2$ as follows:
\begin{equation}
\label{eq:spinorS5IIB}
\eta_1\,=\,e^{i(\alpha-\beta)/2}\,\binom{\cos(\frac{\rho}{2})}{\sin(\frac{\rho}{2})}\,,\quad \eta_2\,=\,e^{i(\alpha+\beta)/2}\,\binom{\cos(\frac{\rho}{2})}{\sin(\frac{\rho}{2})}\,,
\end{equation}
where $\alpha\,,\beta$ and $\rho$ are functions on M$_2$. Given this parametrisation, the two-dimensional bispinors assume the following form:
\begin{eqnarray}
\label{eq:BispM2Expl1}
&v\,=\,e^{i\beta}\sin\rho\,e_1\,, \quad &j\,=\,e^{i\beta}\,\left(1-i\,\cos\rho\,\text{vol}(\text{M}_2)\right)\,,\\
\label{eq:BispM2Expl2}
&w\,=\,-i\,e^{i\alpha}(\cos\rho\,e_1\,+\,i\,e_2)\,,\quad &\omega\,=\,-i\,e^{i\alpha}\,\sin\rho\,\text{vol}(\text{M}_2)\,.
\end{eqnarray}
Let us now look to the equation \eqref{eq:S5four}, picking in particular the components proportional to $J\wedge J$; these implies:
\begin{equation}
\sin\beta\,(e^{Q}-\cos\rho\,e^A)\,=\,0\,,\quad \cos\beta\,(e^{Q}-2\cos\rho\,e^A)\,=\,0\,.
\end{equation}
It is evident the such equations can be simultaneously satisfied only if $\beta=0$ or $\beta=\frac{\pi}{2}$. 

\subsubsection*{Case 1: $\boldsymbol{\beta=\frac{\pi}{2}}$}
In this case the warping $Q$ can be expressed in terms of the warping $A$ and the angle $\rho$ as $e^{Q}=\cos\rho\,e^A$. Because the warpings cannot identically vanish by definition, the cases $\rho=\pm \frac{\pi}{2}$ are forbidden. We can now look at the equation \eqref{eq:S5three}, picking in particular the component $V\wedge \Omega$, implying that:
\begin{equation}
\label{eq:VielS5}
3i\,e^{3A-\phi+Q} w\,+\,i\,\text{d}(e^{3A-\phi+4Q}\star_2\!\omega_2)+2\,e^{2A-\phi+4Q}\star_2\! w=0
\end{equation}
We can use this equation together with \eqref{eq:BispM2Expl1}-\eqref{eq:BispM2Expl2} in order to express the vielbein on M$_2$ in terms of $\alpha,\rho,\phi$ and $A$:
\begin{equation}
e_1\,=\,-e^A\,\sin\rho\,\text{d}\alpha\,,\quad e_2\,=\,-\frac{e^A}{2}\,\frac{\left( 2-\cos2\rho \right)\text{d}\rho+\sin\rho\left( \text{d}\phi-6\,\text{d}A \right)}{2-\cos2\rho}\,.
\end{equation}
Notice that the angle $\alpha$ can now be taken to be a local coordinate on M$_2$. Given this parametrisation of the vielbeine, \eqref{eq:S5four} only provides constraints on the fluxes while \eqref{eq:S5one}-\eqref{eq:S5three} boil down to the simple conditions:
\begin{equation}
\text{d}\phi\,=\,\text{d}(\cos 2\rho)+4\cos\rho^2\,\text{d}A\,,\quad\text{d}A\wedge \text{d}\rho\,=\,0\,.
\end{equation}
The first of these serves to define the dilaton $\phi$ while the second implies that $A= A(\rho)$; as a consequence we can now take $(\alpha,\rho)$ to be local coordiantes spanning  M$_2$. As we mentioned before, the fluxes are completely fixed by \eqref{eq:S5four}:
\begin{equation}
\begin{split}
&q_7=0\,,\quad p=\,4e^{-\phi+4A}\,\frac{\cos\rho^5(\cos \rho+\sin\rho\,A')}{\cos2\rho\,+\,\sin2\rho\,A'}\\
&\star_2\text{d}q_1=-4\,\,e^{-\phi-A}\,\frac{(2-\cos2\rho)(\sin \rho-\cos\rho\,A')}{1-4 \sin z-\cos z\sin z(\phi'-6A')}\,e_1\,,
\end{split}
\end{equation}
where we recall that $A$ and $\phi$ are both functions of $\rho$ and $f'= \frac{\text{d}f}{\text{d}\rho}$. Once the fluxes have been expressed in terms of the vielbein, we can plug them in \eqref{eq:GeneralPair}: this is the last constraint we need to solve. We obtain
\begin{equation}
\sin \rho-\cos\rho\,A'\,=\,0\quad \Rightarrow\quad A\,=\,\log\left(\frac{L}{\cos \rho}\right)\,,
\end{equation}
where $L$ is an integration constant. Such a condition has strong implications. In fact, it sets $q_1=0$ while the dilaton turns out to be constant, $\phi=\phi_0$. Finally, the solution is:
\begin{equation}
\begin{split}
\label{eq:AdS5xS5}
&\text{d}s^2\,=\,L^2\left(\frac{\mu^2}{\cos\rho^2}\text{d}s^2(\text{AdS}_3)+\text{d}s^2(\text{S}^5)+\frac{1}{\cos\rho^2}\left( \text{d}\rho^2+\sin\rho^2\,\text{d}\alpha^2 \right)  \right)\\
&\phi\,=\,\phi_0\,,\quad f_-\,=\,4\,L^4\,e^{-\phi_0}\,\text{vol}(\text{S}^5).
\end{split}
\end{equation}
Although it might not be obvious in these coordinates, this is actually the well-known $\text{AdS}_5\times \text{S}^5$ solution. In fact, if one consider the change of variable $\cos\rho^{-1}\text{d}\rho=\text{d}r$ we obtain:
\begin{equation}
\frac{1}{\cos\rho^2}(\text{d}s^2(\text{AdS}_3)+\text{d}\rho^2+\sin\rho^2\text{d}\alpha^2)\,=\,\text{d}r^2+\cosh r^2\,\text{d}s^2(\text{AdS}_3)+\sinh r^2\,\text{d}\alpha^2\,=\,\text{d}s^2(\text{AdS}_5)\,.
\end{equation}
The other choices of Sasakian structure and $\mu$ lead again to $\text{AdS}_5\times \text{S}^5$ with only minor changes required: for instance the case $(V_-, J_-, \Omega_-)$, $\mu=1$, can be solved following the very same lines as before, with the unique difference that one needs to pick $\beta=\frac{3\pi}{2}$. This is consistent with a recent result of  \cite{Corbino:2020lzq}, that found that the only IIB solution with a warped AdS$_2\times$S$^5\times$S$^1$ factor is AdS$_5\times$S$^5$.


\subsubsection*{Case 2: $\boldsymbol{\beta=0}$}
In this case, the five-sphere warping is fixed to be $e^{Q}=2\cos\rho\,e^A$; the one-form component of \eqref{eq:S5two} is non-vanishing and implies $\phi=2A+\phi_0$, where $\phi_0$ is an integration constant. As before, local expressions for the vielbein can be found:
\begin{equation}
e_1\,=\,2\,e^A\,\sin\rho\,\text{d}\alpha\,,\quad e_2\,=\,2\,e^A\,(\cot \rho\,\text{d}A\,-\,\text{d}\rho)\,,
\end{equation}
while the remaining equations boils down to:
\begin{equation}
e^{A}=\text{const.}\equiv L\,,\quad p=q_1=0\,,\quad q_7\,=\,\frac{64\,L^2\,e^{-\phi_0}}{\mu^4}\cos\rho^5\,.
\end{equation}
Observe that the M$_2$ and M$_5$ combine forming a seven-sphere:
\begin{equation}
\label{eq:noSOLIIB}
\text{d}s^2\,=\,L^2\,\left( \text{d}s^2(\text{AdS}_3)+4\,\text{d}s^2(\text{S}^7) \right)\,,\quad F_7\,=\,2^8\,L^2e^{-\phi_0}\,\text{vol}(\text{S}^7)
\end{equation}
However, it is straightforward to show that such a background cannot satisfy the first pairing equation in \eqref{eq:GeneralPair}. Another way to see that \eqref{eq:noSOLIIB} cannot solve the supersymmetry equations nor type IIB EOM is noticing that, as the dilaton is constant and NS flux trivial, the Ricci scalar should vanish  (consider the dilaton EOM) which would require the ratio of AdS to 7-sphere radii squared to be 7, here we have 4. Thus there are no solutions within case 2.

\subsection{Reducing the supersymmetry equations in type IIA}\label{sec:S5IIA}
To solve the supersymmetry equations in type IIA we will follow the same steps we have seen for type IIB. Again preservation of S$^5$ isometries imposes that internal fluxes must contain just invariant forms, namely
\begin{equation}
\label{eq:IIA_fluxes}
H=0, \qquad f_+ = F_0 + p \, \text{vol}(\text{M}_2) + (q_1 e^1 + q_2 e^2) \wedge \text{vol}(\text{S}^5) ,
\end{equation}
where $F_0$ and $q$ must be constant away from sources according to the Bianchi identities $\text{d} f_+=0$, while $q_i$ are two functions on M$_2$ such that $(q_1 e^1 + q_2 e^2)$ is closed. 

The type IIA supersymmetry constraints are provided by using \eqref{eq:BispS5} inside \eqref{eq:GeneralDiff}. Again we can see that the equations  split depending on whether they contain $J$ or $\Omega$:
\begin{align}
\label{eq:S5one1}
&\text{d}\left(e^{2A-\phi+3Q}\text{Im}(w+e^Q\,\lambda\!\star_2\!\omega\wedge V)\wedge \Omega\right)=\,0, \\
\label{eq:S5two1}
&\text{d}\left(e^{2A-\phi}\text{Im}(v+e^Q\,\lambda\!\star_2\!j\wedge V)\wedge e^{-i\,e^{2Q}J}\right) =\,0, \\
\label{eq:S5three1}
&\text{d}\left(e^{3A-\phi+3Q}\text{Re}(\omega-e^Q\,\star_2\!w\wedge V)\wedge \Omega\right)-2\,e^{2A-\phi+3Q}\text{Im}(w+e^Q\,\lambda\!\star_2\!\omega\wedge V)\wedge \Omega=\,0,\\
\label{eq:S5four1}
&\text{d}\left(e^{3A-\phi}\text{Re}(j-e^Q\,\star_2\!v\wedge V)\wedge e^{-i\,e^{2Q}J} \right)-2 \, e^{2A-\phi}\text{Im}(v+e^Q\,\lambda\!\star_2\!j\wedge V)\wedge e^{-i\,e^{2Q}J}=\,\frac{1}{8}e^{3A}\star\lambda f_+\,.
\end{align}
This procedure allows us to immediately find two scalar equations by looking at the two- and four-form part of \eqref{eq:S5two1}, which together imply that the two-form part of $j$ is zero. Using this condition we can parameterise the two spinors $\eta^{1,2}$ on M$_2$ as
\begin{equation}
\label{eq:spinorS5IIA}
\eta_1\,=\,e^{i(\alpha-\beta)/2}\,\binom{\cos(\frac{\rho}{2})}{\sin(\frac{\rho}{2})}\,,\quad \eta_2\,=\,e^{i(\alpha+\beta)/2}\,\binom{\sin(\frac{\rho}{2})}{\cos(\frac{\rho}{2})}\,.
\end{equation}
Besides the definition of fluxes, \eqref{eq:S5one1}-\eqref{eq:S5four1} provide some constraints on the M$_2$ vielbein and the functions in the spinor definition \eqref{eq:spinorS5IIA}. A large number of equations arise, however after some manipulations they can be reduced to a small amount set of constraints implying them. These manipulations  require that $\rho$ and $\beta$ are non-vanishing, but as one can easily check $\beta = 0$ leads to a zero value for one vielbein while $\rho = 0$ is never met due to \eqref{eq:GeneralPair}, so one may assume $\rho,\beta \neq 0$ without losing generality. The final result of this operation is that we can restrict ourselves to consider a scalar constraint
\begin{equation}
\label{eq:scalar_IIAS5}
2  + 3 e^{A-Q} \cos \rho = 0 ,
\end{equation} 
two differential one-form conditions
\begin{equation}
\label{eq:1form_IIAS5}
\text{d} \alpha = 0 , \qquad e^{-5A+ \phi} \text{d} e^{5A- \phi} = 3 e^{-2A} \text{d}(e^{2A} \cos^2 \beta \sin^2 \rho),
\end{equation}
the definition of the vielbein on M$_2$,
\begin{equation}
\text{d} e^{2A+3Q-\phi} = 3 e^{2A+2Q-\phi}  \sin^2 \rho \, e^2 , \qquad \text{d} (e^{2A} \sin (2 \beta)\sin^2 \rho) = \frac{4}{3}  \, e^{A} \sin \rho \, e^1
\end{equation}
and a two form condition
\begin{equation}
\text{d}e^{A} \wedge \text{d} (\cos \beta \sin \rho)=0,
\end{equation}
which implies that $A = A(\cos \beta \sin \rho)$. Using these equations inside \eqref{eq:S5four1} and performing some manipulation we are able to define the functions appearing in the RR-fluxes \eqref{eq:IIA_fluxes}, in particular we get:
\begin{align}
& q_2 = -64 \mathcal{D} \,  \, e^{-A-\phi } \sin \beta   \left(A' \cos \beta \sin\rho+1\right), \qquad \qquad q_1 =  \cot \beta  \cos \rho \,  q_2 , \nonumber\\
&p\,=0 , \qquad F_0= -2 \mathcal{D} \,  \, e^{-A-\phi } (8 A' \left(-2 \cos ^2\beta \cos (2 \rho )+\cos (2 \beta )-3\right)+32 \cos\beta \sin \rho), \\
&\mathcal{D} = \left(3A' \left(4 \cos ^3\beta \sin (3 \rho )+7 \cos \beta  \sin \rho -3 \cos (3 \beta) \sin \rho\right)+24 \left(\cos ^2\beta  \cos (2 \rho )+\sin ^2\beta \right)\right)^{-1} \nonumber
\end{align}
Now we are left with just \eqref{eq:GeneralPair} to solve. Using the fluxes definition and \eqref{eq:scalar_IIAS5} this equation reduces to an ODE condition for $A$:
\begin{equation}
1- 6(\cos \beta \sin \rho)^2+6\cos \beta \sin \rho(1-(\cos \beta \sin \rho)^2)A'=0
\end{equation}
which is solved by 
\begin{equation}
\label{eq:A_def_S5IIA}
e^A = e^{A_0} (\cos \beta \sin \rho)^{-\frac{1}{6}}(1-\cos^2 \beta \sin^2 \rho)^{-\frac{5}{12}} .
\end{equation}
The last field we have to determine is the dilaton, which can be found by solving \eqref{eq:1form_IIAS5}:
\begin{equation}
\label{eq:phi_def_S5IIA}
e^{\phi} = e^{\phi_0} (\cos \beta \sin \rho)^{-\frac{5}{6}}(1-\cos^2 \beta \sin^2 \rho)^{\frac{5}{12}}.
\end{equation}
Finally the solution is made easier to interpret through the following change of coordinates
\begin{equation}
\cos \beta \sin \rho = \sin \theta, \qquad \frac{\cos \rho}{\sqrt{1-(\cos \beta \sin \rho)^2}} = \sin \delta.
\end{equation}
Using this parametrisation and \eqref{eq:A_def_S5IIA},\eqref{eq:phi_def_S5IIA} it is easy to see that the RR fluxes automatically satisfy their Bianchi identities.\\
~~\\
In summary the solution is given by
\begin{align}
d s^2 &= e^{2A} \left( \text{d}s^2(\text{AdS}_3)+ d \theta^2 +\frac{9}{4} \cos^2 \theta (\sin^2 \delta \text{d}s^2(\text{S}^5)+\text{d} \delta^2 ) \right) , \qquad e^A = e^{A_0} (\sin \theta \cos^5 \theta)^{-\frac{1}{6}},\nn\\[2mm] 
F_0 &= -\frac{2}{3}  e^{-A_0-\phi_0}, \qquad F_6 = - 5 e^{5A_0-\phi_0} \left(\frac{3 \sin \delta}{2 }\right)^5 \text{d} \delta \wedge \text{vol}_{\text{S}^5}\label{eq:AdS3XS6}
\end{align}
so that the vielbein on M$_2$ and the warping functions organize themselves so as to define a six-dimensional sphere in the internal space, as one can see comparing these expressions with \eqref{eq:metric_even} and \eqref{eq:volume_even}. Indeed one can check that this solution is equivalent to the one presented in \cite[Section 4]{Dibitetto:2018ftj}. So we have that in IIA the R-symmetry is enhanced from $\mathfrak{su}(4)$ to $\mathfrak{spin}(7)$.\\
~\\
This completes our classification of type II solutions realising the first 3 entries in \eqref{eq:Neq8alebras}, what is left for type II is to find the solutions with the superconformal algebra $\mathfrak{osp}(4^*|4)$ , we shall do so in the next section.

\section{AdS$_3\times$S$^4\times$S$^2$ in type II}\label{sec:ads3s4s2typeII}
In this section we will derive the unique type II solution realising the superconformal algebra $\mathfrak{osp}(4^*|4)$. Here we will focus on solutions with only magnetic NS 3-form turned on that also realise the $\mathfrak{sp}(2)\oplus \mathfrak{sp}(1)$ R-symmetry  with a round S$^2\times$S$^4$ factor. We argue in section \ref{sec:fibrations} that this is the only option to realise the R-symmetry that cannot be lifted to M-theory, which we cover in \ref{sec:M-theory} and section \ref{sec:squshed7spheres}.\\
~\\
The metric, NS flux $H$ and the magnetic components of the IIA/IIB RR polyform fluxes $f_{\pm}$ of solutions on AdS$_3\times$S$^2\times$S$^4$  decompose in the form
\begin{align}\label{S2S4ansatz}
ds^2&= e^{2A}ds^2(\text{AdS}_3)+ e^{2C_1}ds^2(\text{S}^2)+ e^{2C_2}ds^2(\text{S}^4)+ e^{2k}dr^2,~~~H= h dr\wedge \text{vol}(\text{S}^2),\nn\\[2mm]
f_+&= u_1+ u_2\text{vol}(\text{S}^2)+u_3\text{vol}(\text{S}^4)+u_4\text{vol}(\text{S}^2)\wedge\text{vol}(\text{S}^4),~~~  f_- =d\rho\wedge f_+,
\end{align}
with $(u_i,h,e^A,e^{C_1},e^{C_2})$ and the dilaton $e^{\phi}$ functions of $r$ only, which are constrained by the Bianchi identities of the fluxes ($d_H f_{\pm}=0$ away from the loci of potential sources), and by supersymmetry. We assume that the sphere and AdS factors have unit radius.

To find such supersymmetric solutions we again use the bispinor approach of \cite{Dibitetto:2018ftj}, summarised in section \ref{eq:7dbispinors}. We start by writing a basis for the gamma matrices consistent with these factors and appendix \ref{eq:evenspherebilinears}, namely
\beq
\gamma^{(7)}_{a_1} = e^{C_1}\sigma_{a_1}\otimes \mathbb{I}\otimes \mathbb{I},~~~\gamma^{(7)}_{a_2} = e^{C_2}\sigma_3\otimes \Gamma^{(4)}_{a_2},~~~~ \gamma^{(7)}_{r}= e^{k}\sigma_3\otimes \hat\Gamma^{(4)}.
\eeq
Here $\sigma_{a_2}$ ({\it i.e.} the Pauli matrices with $a_2=1,2$) are the gamma matrices on S$^2$ with chiality matrix $\sigma_3$, while $(\Gamma^{(4)}_{a_2},\hat\Gamma^{(4)})$ for $a_4=1,...4$ are defined on S$^4$ as in \eqref{eq:splitgamma}  -  the intertwiner defining Majorana conjugation (m.c.) is $B_{(7)}= \sigma_2\otimes\sigma_2\otimes \sigma_3$. Next we need to write down spinors transforming in the $(\mathbf{4},\mathbf{2})$ of $\mathfrak{sp}(2)\oplus \mathfrak{sp}(1)$, which in this case is clearly going to involve a product of the Killing spinors on S$^4$ and S$^2$\\
~\\
The most general way to decompose  a 7d Majorana spinor in terms of these factors is\footnote{One might also consider adding products of $\xi^{\text{S}^2}$ with the Majorana conjugate of $\xi^{\text{S}^4}$ and vice versa but,  given the ansatz \eqref{S2S4ansatz}, such terms do not mix with the terms we do include in $\chi$ below, within the gravitino or Dilatino conditions of type II supergravity. As such, any solution consistent with $\chi$ and these extra terms must be consistent with $\chi$ alone. The converse is not necessarily true, so by including extra terms we constrain rather than generalise the system. Finally if one were to only include additional terms of the form $\xi^{\text{S}^2 c}\otimes \xi^{\text{S}^4}$,... the result would be equivalent to $\chi$.}   
\beq\label{eq:genricspinorS2S4}
\chi= g_1 \xi^{\text{S}^2}\otimes \xi^{\text{S}^4}-g_2 \sigma_3\xi^{\text{S}^2}\otimes \hat\Gamma^{(4)}\xi^{\text{S}^4}-g_3 \xi^{\text{S}^2}\otimes \hat\Gamma^{(4)}\xi^{\text{S}^4}+g_4\sigma_3\xi^{\text{S}^2}\otimes \xi^{\text{S}^4}+\text{m.c},
\eeq
where $g_{1},..,g_{4}$ are arbitary complex functions of $r$ only, the signs on $g_{2,3}$ are set for convenience. A feature of even spheres, is that their bilinears contain charged zero forms (see \eqref{eq:evenbi}) that generically enter the AdS warp factor through the condition $e^{A}= |\chi_1|^2=|\chi_2|^2$, which imposes some additional constraints when one demands that the full solution does not break the isometry group of the spheres. We thus demand that our general spinor obey $|\chi|^2=e^{A}$, which is independent of the S$^2\times$S$^4$ directions - this imposes
\beq\label{eq:eAconstraints}
g^\dag g = e^{A},~~~~g^\dag P_i g=0,~~~ P_i = (\sigma_1\otimes \sigma_1,~\sigma_1\otimes \mathbb{I},~\mathbb{I}\otimes \sigma_1)_i 
\eeq 
where $g=(g_1,g_2,g_3,g_4)^T$. This gives 4 complicated relations that $g_{1},...,g_{4}$ need obey, however since $P_i$ are mutually commuting, they are simultaneously diagonalisable. Upon expanding $g$ in a basis of the eigenvectors of $P_i$ one finds that only 4 phases remain unfixed after solving \eqref{eq:eAconstraints}. Through these consideration we arrive at a general form for two independent Majorana spinors on S$^2\times$S$^4\times \mathbb{R}$:
\beq\label{eq:spinorS2S4}
\chi_1 = \frac{e^{\frac{A}{2}}}{\sqrt{2}}{\cal B}^T
_1 \Lambda {\cal X}+ \text{m.c},~~~\chi_2 = \frac{e^{\frac{A}{2}}}{\sqrt{2}} {\cal B}^T_2 \Lambda {\cal X}+ \text{m.c}
\eeq
where we define 
\beq
{\cal B}_p= \left(\begin{array}{c}e^{i\beta^p_{1}}\\e^{i\beta^p_{2}}\\ e^{i\beta^p_{3}}\\e^{i\beta^p_{4}}\end{array}\right),~~~{\cal X}=\left(\begin{array}{c}\xi^{\text{S}^2}\otimes \xi^{\text{S}^4}\\\sigma_3\xi^{\text{S}^2}\otimes \hat\Gamma^{(4)}\xi^{\text{S}^4}\\ \xi^{\text{S}^2}\otimes \hat\Gamma^{(4)}\xi^{\text{S}^4}\\\sigma_3\xi^{\text{S}^2}\otimes \xi^{\text{S}^4}\end{array}\right),~~~\Lambda=\frac{1}{2}\left(\begin{array}{cccc}1&1&1&1\\1&-1&-1&1\\1&1&-1&-1\\-1&1&-1&1\end{array}\right).
\eeq
We now automatically solve $|\chi_{1,2}|^2= e^{A}$ independent of the sphere directions, and have 8 phases left to solve for.\\
~~\\
We are now ready to compute the 7d bi-linears that \eqref{eq:spinorS2S4} imply, as $\chi_{1,2}$ consist of tensor products of 2 and 4-sphere Killing vectors, we anticipate that the 7d bi-linears will decompose as wedge products of bispinors on S$^2$, S$^4$ and the interval. Adapting \eqref{eq:evenbi} to the case at hand, we then define these as
\begin{align}\label{eq:s2s4bispinors}
\Phi^{\text{S}^{2(n+1)}}\pm \Phi_{\hat\Gamma}^{\text{S}^{2(n+1)}} &= \frac{1}{2^{n+1}}\Big(1\mp (1+i e^{C_n}d)\cos\alpha_n+ \nn\\
&e^{C_n}\sin\alpha_n V_{n}\wedge (\sin\alpha_{n} +  e^{C_n}d (i\sin\alpha_{n}\mp \cos\alpha_n) )\Big)\wedge e^{-i e^{2C_n}\sin^2\alpha_{n} J_{n}},\nn\\[2mm]
\tilde{\Phi}^{\text{S}^{2(n+1)}}\pm \tilde{\Phi}_{\hat\Gamma}^{\text{S}^{2(n+1)}} &=\frac{1}{2^{n+1}}\Big(e^{C_n}d\alpha_{n}\pm (i-e^{C_n}d)\sin\alpha_n\\
&+e^{C_n}\sin\alpha_{n} V_{n}\wedge(i( \cos\alpha_{n}\mp \sin\alpha_n)- e^{C_n}d\cos\alpha_{n})\Big)\wedge(e^{n C_n}\sin^{n}\alpha_{n}\Omega_{n}),\nn
\end{align}
for $n=0,1$ where $(V_n, J_n,\Omega_n)$ are related as in \eqref{eq:SEcond}, and in particular on S$^2$, $J_0=0$. The 7 dimensional bi-spinors that follow are rather long winded, they can be expressed most succinctly in terms of the matrix bilinear 
\begin{align}
 &\Xi_{\pm}(\Phi,\Psi)=\\
& \left(\begin{array}{cccc}
G_+\wedge\Phi_{P_{++}}\wedge\Psi_{P_{+\pm}} &G_+\wedge\Phi_{P_{+-}}\wedge\Psi_{P_{+\pm}}&G_+\wedge\Phi_{P_{+-}}\wedge\Psi_{P_{+\mp}}&-G_+\wedge\Phi_{P_{++}}\wedge\Psi_{P_{+\mp}}\\
G_-\wedge\Phi_{P_{--}}\wedge\Psi_{P_{+\pm}} &G_-\wedge\Phi_{P_{-+}}\wedge\Psi_{P_{+\pm}}&G_-\wedge\Phi_{P_{-+}}\wedge\Psi_{P_{+\mp}}&-G_-\wedge\Phi_{P_{--}}\wedge\Psi_{P_{+\mp}}\\
-G_+\wedge\Phi_{P_{--}}\wedge\Psi_{P_{-\mp}} &G_+\wedge\Phi_{P_{-+}}\wedge\Psi_{P_{-\mp}}&G_+\wedge\Phi_{P_{-+}}\wedge\Psi_{P_{-\pm}}&G_+\wedge\Phi_{P_{--}}\wedge\Psi_{P_{-\pm}}\\
-G_-\wedge\Phi_{P_{++}}\wedge\Psi_{P_{-\mp}} &G_-\wedge\Phi_{P_{+-}}\wedge\Psi_{P_{-\mp}}&G_-\wedge\Phi_{P_{+-}}\wedge\Psi_{P_{-\pm}}&G_-\wedge\Phi_{P_{++}}\wedge\Psi_{P_{-\pm}}\nn\\
\end{array}\right)
\end{align}
where we define
\beq
\Phi_{P_{++}}= (\Phi+ \Phi_{\hat\Gamma})_+,~~~\Phi_{P_{--}}= (\Phi- \Phi_{\hat\Gamma})_-,~~~\Phi_{P_{-+}}= (\Phi- \Phi_{\hat\Gamma})_+,~~~\Phi_{P_{+-}}= (\Phi+ \Phi_{\hat\Gamma})_-,~~~ G_{\pm}= 1\pm e^{k}dr\nn,
\eeq
so that the first subscript of P indicates a sum or difference of bispinors, and the second indicates a projection onto their even/odd form degree components. We then find that the 7d bispinors decompose as
\begin{align}\label{eq:s2s4bispinors}
\Psi_+ &= \frac{e^A}{8}\text{Re}\bigg({\cal B}^T_1 \Xi_+(\Phi^{\text{S}^4},\Phi^{\text{S}^2}) {\cal B}^*_2+{\cal B}^T_1 \Xi_+(\tilde{\Phi}^{\text{S}^4},\tilde{\Phi}^{\text{S}^2}){\cal B}_2\bigg),\nn\\[2mm]
\Psi_- &= \frac{e^A}{8}\text{Im}\bigg({\cal B}^T_1 \Xi_-(\Phi^{\text{S}^4},\Phi^{\text{S}^2}) {\cal B}^*_2+{\cal B}^T_1 \Xi_-(\tilde{\Phi}^{\text{S}^4},\tilde{\Phi}^{\text{S}^2}) {\cal B}_2\bigg),
\end{align}
which are clearly rather complicated. However it turns out that they are consistent with a single solution in massless IIA, we sketch its  derivation in the next section. The analysis of type IIB proceeds in the same fashion as IIA, but in this case leads to a system of differential and algebraic constraints admitting no AdS$_3\times$S$^4\times$S$^2$ solution. As the analysis is long winded and ultimately leads to a null result we shall omit details and focus only on the IIA case here\footnote{If the reader finds this unsatisfactory consider this short argument for the non existence of such IIB solutions: An AdS$_3\times $S$^4\times $S$^2$ solution in IIB would suggest that one can double wick rotate  on AdS$_3$ and S$^4$ arriving at an AdS$_4\times$S$^2\times$ S$^3$ solution. However  AdS$_4\times$S$^2\times$S$^3$ solution of type II are classified in \cite{Macpherson:2017mvu} - there does exist exactly one  such solution  (locally) but it is in IIA. This is consistent with our omitted IIB analysis and our finding for IIA presented in the next section.}.

\subsection{Unique Solution}\label{eq:sp2sp1IIA}
The bispinors \eqref{eq:s2s4bispinors} are rather complicated, however, since the bosonic fields for solutions we seek are are of the form \eqref{S2S4ansatz}, the supersymmetry conditions \eqref{eq:GeneralDiff} give many constraints, let us sketch how we solve them.\\
~\\
A large number of algebraic constraints follow almost immediately once one considers the fact that in IIA, any charged term in $\Psi_-$ that cannot be mapped to a term in $\Psi_+$ under d or wedge must be zero - otherwise the fluxes would not be $\mathfrak{sp}(2)\oplus \mathfrak{sp}(1)$ singlets.  Examining the forms appearing in \eqref{eq:s2s4bispinors} we observe that there is no zero form on S$^2$ or S$^4$ that can generate the 1-forms $(V_0,V_1,\text{Re}\Omega_1,\text{Im}\Omega_1)$ under d - given that each can appear multiplied  by the zero forms on S$^2$ and S$^4$, which the phases in ${\cal B}_{1,2}$ cannot depend on. This leads to 6 real constraints from just studying $\Psi_1$, namely
\begin{align}\label{eq:zeroforms1}
&\sin(\beta^1_3-\beta^2_2)+ \sin(\beta^1_2-\beta^2_3)=\sin(\beta^1_4-\beta^2_1)+ \sin(\beta^1_1-\beta^2_4)=0,\nn\\[2mm]
&\sin(\beta^1_4-\beta^2_3)+ \sin(\beta^1_3-\beta^2_4)=\sin(\beta^1_2-\beta^2_1)+ \sin(\beta^1_1-\beta^2_2)=0,\nn\\[2mm]
&e^{i(\beta^1_3+ \beta^2_1)}+ e^{i(\beta^1_4+ \beta^2_2)}= e^{i(\beta^1_1+ \beta^2_3)}+ e^{i(\beta^1_2+ \beta^2_4)},
\end{align}
likewise $\Psi_3$ contains the term $V_1\wedge J_1$ that must be set to zero imposing an additional 2 real constraints
\beq\label{eq:zeroforms2}
\cos(\beta^1_4-\beta^2_3)+ \cos(\beta^1_3-\beta^2_4)=\cos(\beta^1_2-\beta^2_1)+ \cos(\beta^1_1-\beta^2_2)=0.
\eeq 
There are several distinct ways to solve the 8 constraints derived so far, each leading to branching possible ways to solve the remaining conditions that follow from \eqref{eq:s2s4bispinors}. Following every branch to its conclusion is straightforward but extremely long winded and repetitive, we  will spare the reader an exercise in tedium - the punch line is that each way one can solve \eqref{eq:zeroforms1}-\eqref{eq:zeroforms2} either leads to no supergravity solution or truncates to a unique one once the rest of \eqref{eq:GeneralDiff} is considered. Allow us then to just paint the broad strokes of how this solutions arises from one simple tuning of the phases satisfying  \eqref{eq:zeroforms1}-\eqref{eq:zeroforms2}, namely 
\beq\label{eq:simpleansatz}
{\cal B}_1=(- e^{\frac{i}{2}\beta},e^{\frac{i}{2}\beta},-i e^{-\frac{i}{2}\beta},i e^{\frac{i}{2}\beta})^T,~~~{\cal B}_2=( e^{\frac{i}{2}\beta},e^{\frac{i}{2}\beta},-i e^{\frac{i}{2}\beta},-i e^{-\frac{i}{2}\beta})^T
\eeq
Substituting \eqref{eq:simpleansatz} into \eqref{eq:s2s4bispinors} leads to a pronounced truncation  - upon plugging the bispinors into \eqref{eq:GeneralDiff} one finds that all the S$^2\times$S$^4$ data factors out, leaving a set of algebraic and differential conditions to solve for. We find the following definitions for the functions appearing in \eqref{S2S4ansatz}
\begin{align}
h&=u_1= u_4=0,~~~ e^{C_1}=-\frac{1}{2}e^{A}\sin\beta,~~~e^{C_2}=\frac{1}{2}e^A\cos\beta\nn\\[2mm]
u_2&=e^{A-\Phi}\sin^2\beta\sec2\beta(2\sin\beta+ \cos\beta e^k(e^A)'),\nn\\[2mm]
u_3&= \frac{1}{4}e^{3A-\Phi}\cos^4\beta \sec2\beta(2\cos\beta+ \sin\beta e^k(e^A)')
\end{align}
and the following ODEs
\begin{align}\label{eq:S2S4bps}
&(e^{3A-\phi}\cos\beta)'+ 2 e^{2A+k-\phi} \sin\beta=0,\nn\\[2mm]
&e^{2A}\phi'- \frac{1}{2}(e^{2A}(2+  \cos 2\beta))'=0.
\end{align}
Remarkably, with these definitions \eqref{eq:GeneralPair} are implied - so all that remains is to solve the Bianchi identities for the fluxes. As the NS 3-form is necessarily zero, this amounts to imposing that $df_+=0$ away from the loci of possible sources, {\it i.e.} $u_2'=u_3'=0$. Using  \eqref{eq:S2S4bps}, $u_2'$ gives a second order ODE for $e^A$ that one can substitute into $u_3'$ to get a first order one - we find
\beq\label{eq:S2S4Bianchi}
\sin2\beta(e^A)'+ (2-\cos 2\beta) e^k=0 
\eeq
implies $u_2'=u_3'=0$ without loss of generality.\\
~~\\
All that is left is to solve \eqref{eq:S2S4bps} and \eqref{eq:S2S4Bianchi} - we can do this with a suitable choice of $e^{k}$, which just parametrises a coordinate transformation in $r$. We choose this such that our final result will be conveniently expressed - we thus define
\beq
e^{k}= L e^{\frac{1}{2}\phi},~~~~ \frac{1}{\cos\beta}= \cosh \tilde \beta,~~~ e^{A}= e^{\Delta+ \frac{1}{3}\phi}
\eeq
for $\tilde{\beta}, \Delta$  arbitrary functions of $r$, this reduces the ODEs we must solve to
\beq
e^{\Delta}= \frac{\cosh\tilde\beta}{\tilde{\beta}'},~~~(e^{-\frac{2}{3}\phi}\sinh\tilde\beta)'=0,~~~\tilde\beta''=0,
\eeq 
which are trivial to solve - indeed since we cannot set $\tilde\beta'=0$, we can without loss of generality simply fix
\beq
\tilde\beta = r.
\eeq
The resulting solution is then given by
\begin{align}\label{eq:S2S4sol}
ds^2&= e^{\frac{2}{3}\phi}L^2\bigg(\cosh^2 r ds^2(\text{AdS}_3)+\frac{1}{4}\sinh^2 r ds^2(\text{S}^2)+ dr^2 +\frac{1}{4}ds^2(\text{S}^4) \bigg),~~~ e^{\frac{4}{3}\phi}= \frac{L^2}{k^2} \sinh^2 r,\nn\\[2mm]
F_2&= \frac{k}{2}\text{Vol}(\text{S}^2),~~~~ F_4= \frac{3 L^3}{8}\text{Vol}(\text{S}^4), 
\end{align}
for $L,k$ constants and where $F_0=H=0$. Clearly this solution is non compact as $r=\infty$ is at infinite proper distance, but it is bounded from below at $r=0$ where the behaviour of $k$ D6 branes wrapped on AdS$_3\times$S$^4$ is reproduced. The solution is easy to interpret - indeed we have tried to write \eqref{eq:S2S4sol} in a suggestive manner. This solution can be lifted to 11 dimensions via the identities
\beq
ds^2_{11}= e^{-\frac{2}{3}\phi}ds^2+ e^{\frac{4}{3}\phi}(d\psi+ C_1),~~~~ G_4= F_4,~~~~ dC_1= F_2
\eeq
the resulting solution is AdS$_7/\mathbb{Z}_k\times$S$^4$, which is related to standard AdS$_7\times$S$^4$ as follows: One parameterises AdS$_7$ as a foliation of AdS$_3\times$S$^3$ over an interval then performs the orbifolding by restricting the period of the coordinate $\psi$, which parametrisies the Hopf fiber U(1) inside S$^3$. This solution preserves $\mathcal{N}=(8,0)$ from the AdS$_3$ perspective, which is enhanced to  $\mathcal{N}=(8,8)$ for $k=1$, where  AdS$_7\times$S$^4$ is recovered. Of course from the 11d perspective one could equally well perform an additional orbifolding on  AdS$_3$, and reduce along the Hopf fibere of AdS$_3$ to an $\mathcal{N}=8$ solution with AdS$_2$ in IIA - this can actually be mapped to the $\frak{sp}(2)\oplus \frak{sp}(1)$ preserving AdS$_2$ solution found in \cite{Dibitetto:2019nyz} via T and S duality.

\section{AdS$_3\times$S$^4$ in M-theory}
\label{sec:M-theory}
In this section we will find all $\mathcal{N}=(8,0)$ solutions of 11 dimensional supergravity containing round spheres in their internal space. As already argued this can be achieved by classifying supersymmetric solutions that are warped products of the form AdS$_3\times$S$^4\times$M$_4$. What remains beside this are certain squashing of the 7-sphere that shall be explored in section \ref{sec:squshed7spheres}. As in type II, we shall build our classification upon a set of geometric conditions that AdS$_3$ solutions with ${\cal N}=1$ supersymmetry must satisfy.

\subsection{Geometric supersymmetry conditions for AdS$_3\times$M$_8$}\label{sec:Gstructuremtheory}
In \cite{Martelli:2003ki} a set of geometric supersymmetry conditions where derived for warped AdS$_3\times$M$_8$ solutions preserving at least $\mathcal{N}=1$ supersymmetry. However, the objects appearing in (some of)  the conditions in \cite{Martelli:2003ki} are several steps removed from the bi-linears the spinor of such a solution give rise to. For this reason, we find it easier to work with an alternative set of conditions, involving directly the bi-linears, that we derive in appendix \ref{sec:Neq1AdS3mtheory}. We obtain these by imposing an AdS$_3$ factor on the general geometric constraints that any $\mathcal{N}=1$ solution in $d=11$ should obey presented in \cite{Gauntlett:2002fz,Gauntlett:2003wb}.\\
~\\  
We decompose the bosonic fields of 11d supergravity as
\beq\label{eq:ansatzMtheory}
ds^2= e^{2A} ds^2(\text{AdS}_3)+ ds^2(\text{M}_8),~~~~G= e^{3A}\text{vol}(\text{AdS}_3)\wedge F_1+ F_4
\eeq 
where $F_1,F_4,A$ have support on $\text{M}_8$ only and AdS$_3$ has radius $m$. Our geometric conditions for ${\cal N}=1$ supersymmetry are defined in terms of the following bi-linears on M$_8$
\begin{align}\label{eq:mtheoryforms}
2e^A&= |\chi|^2,~~~2e^Af= \chi^{\dag}\hat\gamma^{(8)}\chi,~~~~2 e^{A}K =\chi^{\dag}_+\gamma^{(8)}_a\chi_- e^a ,\nn\\[2mm]
2e^A\Psi_3& = \frac{1}{3!}\chi^\dag\gamma^{(8)}_{abc}\hat\gamma^{(8)}\chi e^{abc},~~~2e^A\Psi_4=\frac{1}{4!}\chi^\dag\gamma^{(8)}_{abcd}\chi e^{abcd}
\end{align}
where $\gamma^{(8)}_a$ are eight-dimensional flat space gamma matrices, $\hat\gamma^{(8)}=\gamma^{(8)}_{12345678}$ is the chirality matrix and $e^a$ is a vielbein on M$_8$. At least $\mathcal{N}=1$ supersymmetry is preserved by a background whenever the bi-spinors obey the following differential conditions
\begin{subequations}
\begin{align}
& d(e^{2A} K)=0,\label{eq:MtheoryBPS1}\\[2mm]
&d(e^{3A} f)- e^{3A} F_1-2 m  e^{2A} K=0,\label{eq:MtheoryBPS2}\\[2mm]
& d(e^{3A} \Psi_3)- e^{3A}(-\star_8 F_4+ f F_4)+2 m e^{2A} \Psi_4=0\label{eq:MtheoryBPS3},\\[2mm]
&d (e^{2A} \Psi_4)- e^{2A} K \wedge F_4=0\label{eq:MtheoryBPS4},\\[2mm]
&6 \star_8 dA-2 f \star_8 F_1+ \Psi_3\wedge F_4=0,\label{eq:MtheoryBPS5}\\[2mm]
&6 e^{-A} m \star_8 K - 6 f \star_8 dA+2 \star_8 F_1+\Psi_3 \wedge \star_8 F_4=0\label{eq:MtheoryBPS6}
\end{align}
\end{subequations}
where $\star_8$ is the hodge dual on the M$_8$. This is necessary and sufficient for supersymmetry, but this is not enough for all the equations of motion of 11 dimensional supergravity to necessarily follow. For this, in a region of the internal space away from the loci of potential sources, we must additionally impose
\beq\label{eq:BianchiMtheory}
d(F_4)=0,~~~~ d(\star_8 F_1)-\frac{1}{2} F_4\wedge F_4=0.
\eeq
When a geometry supports localised sources, the right hand side of at least one of these expression should be modified to include appropriate delta function sources. We will not concern ourselves with such sources a priori, as it is sufficient to solve \eqref{eq:BianchiMtheory} locally in a smooth region of the internal space, and then check whether the domain of this solution may be extended to singular loci (of physical origin) a posteriori. 

A result of \cite{Martelli:2003ki} is important to stress: all electric AdS$_3\times$M$_8$ solutions, {\it i.e.} those with $F_4=0$, are diffeomorphic to AdS$_4$ solutions with an internal 7-manifold that is conformally of weak G$_2$ holonomy. Upon parameterising $(\tilde{K}=\frac{1}{\cos\theta}K,~ f= \sin\theta)$ one has in general that 
\beq\label{eq:elecconds}
\tilde{K}=\frac{e^A}{m}d\theta,~~~F_1= 3 \sec\theta d\theta,~~~ e^{-A}=\cos\theta
\eeq
and AdS$_4$ of radius $m$  becomes manifest upon identifying
\beq
\sec\theta = \cosh(m r).
\eeq
Thus whenever we are forced to set $F_4=0$ in the following classification, we know that we cannot have an AdS$_3$ solution with compact internal space.\\
~\\
In the next section we will derived reduced 4-dimensional geometric conditions that imply supersymmetry when M$_8=$S$^4\times$M$_4$.

\subsection{Reducing the 8 dimensional conditions on S$^4$} 
To realise an S$^4$ factor in solutions we decompose the internal space and flux in \eqref{eq:ansatzMtheory} as
\beq
ds^2(\text{M}_8) = e^{2C} ds^2(\text{S}^4)+ ds^2(\text{M}_4),~~~ F = g_4+e^{4C}g_0 \text{vol}(\text{S}^4)
\eeq
with $e^{2C},e^{2A },g_0,g_4,f$ defined on M$_4$ only. We decompose the 8d gamma matrices consistently with appendix \ref{eq:evenspherebilinears}   as
\beq
\gamma^{(8)}_a= e^{C}\gamma^{\text{S}^4}_a\otimes \mathbb{I},~~~~ \gamma^{(8)}_i= \hat\gamma^{\text{S}^4}\otimes \gamma_i,~~~ \hat\gamma^{(8)}=\hat\gamma^{\text{S}^4}\otimes \hat\gamma
\eeq
where $a=1,...,4$ are indices on S$^4$, $i=1,...,4$ indices  on M$_4$ and $\hat\gamma^{\text{S}^4}$, $\hat\gamma$ are the corresponding chirality matrices. The 8d intertwiner is $B^{(8)}=  B^{\text{S}^4}\otimes B$, where $B\gamma_i B^{-1}= \gamma_i^*$, $B^{\text{S}^4}\gamma_i^{\text{S}^4} (B^{\text{S}^4})^{-1}= (\gamma^{\text{S}^4}_i)^*$  and $BB^*=B^{\text{S}^4}(B^{\text{S}^4})^*=-\mathbb{I}$. Likewise we decompose the spinors on M$_8$ as a product of the Killing spinors on S$^4$, $\xi^{\text{S}^4}$. We choose to do this by decomposing $\chi=e^{A/2}(\chi_+ +\chi_-)$ for 
\beq
\chi_+ = P_+(\xi^{\text{S}^4}\otimes \eta_1)+ \text{m.c},~~~\chi_- = P_-(\xi^{\text{S}^4}\otimes \eta_2)+ \text{m.c},~~~~ P_{\pm}=\frac{1}{2}(\mathbb{I}\otimes \mathbb{I}\pm \hat\gamma^{\text{S}^4}\otimes \hat\gamma).
\eeq 
As the 8 dimensional spinors decompose as a product of S$^4$ and M$_4$ spinors, the 8 dimensional forms in \eqref{eq:mtheoryforms} will likewise decompose as in terms of wedge products of forms on these spaces. To proceed it is helpful to define some bilinears - those on S$^4$ are defined as in \eqref{eq:s2s4bispinors} (albeit with $\alpha_n\to\alpha$ and $C_2\to C$) so decompose in terms of an angle $\alpha$ and $(V,J,\Omega)$ which here define a Sasaki--Einstein structure on S$^3$. On M$_4$ we define the following set of forms
\begin{align}
&\Psi^{\alpha\beta}_n = \eta^{\dag}_{\beta}\gamma_{a_1...a_n} \eta_{\alpha} e^{a_1...a_n},~~~\Psi^{\alpha\beta}_{\hat\gamma n} = \eta^{\dag}_{\beta}\gamma_{a_1...a_n} \hat\gamma\eta_{\alpha} e^{a_1...a_n},\\[2mm]
&\tilde\Psi^{\alpha\beta}_n = \eta^{c\dag}_{\beta}\gamma_{a_1...a_n} \eta_{\alpha} e^{a_1...a_n},~~~\tilde\Psi^{\alpha\beta}_{\hat\gamma n} = \eta^{c\dag}_{\beta}\gamma_{a_1...a_n} \hat\gamma\eta_{\alpha} e^{a_1...a_n},
\end{align}
for $e^{a}$ a veilbein on M$_4$, to easy presentation we also use the shorthand notation $\Psi^{\pm}=\frac{1}{2}(\Psi^{11}\pm \Psi^{22})$.  Plugging this ansatz into the definition of $\sin\zeta$ in \eqref{eq:mtheoryforms} we find we must impose
\beq\label{eq:sindef}
\Psi^{+}_0=1,~~~\Psi^{-}_{\hat\gamma 0}=0,~~~f=\Psi^{-}_0- \cos\alpha \Psi^{+}_{\hat\gamma 0}
\eeq
where $\alpha$ is an angle in terms of which  S$^4$ is expressed as a foliation of $\text{S}^3$ over an interval - see \eqref{eq:metric_even}. For the 1-form in \eqref{eq:s2s4bispinors} we find
\begin{align}\label{eq:Kexpanded}
 K &= \text{Re}\Psi^{12}_{\hat\gamma 1}+e^C d(\cos\alpha) \text{Im}\Psi^{12}_{\hat\gamma0}-\cos\alpha \text{Re}\Psi^{12}_1\nn\\[2mm]
&+ e^{C} \sin^2\alpha V \text{Re}\Psi^{12}_0-e^{C}\sin^2\alpha \text{Im}\big(\Omega_1\tilde\Psi^{12}_{\hat\gamma 0}\big),
\end{align}
where the constraint \eqref{eq:MtheoryBPS1} will force us to set each terms in the second line of this expression to zero, as they are not closed and do not mix under exterior differentiation. For the 3-form we find 
\begin{align}
\Psi_3&=\text{Re}(\Psi^{12}_{\hat\gamma3}- \cos\alpha \Psi^{12}_3)+e^Cd(\cos\alpha)\wedge\text{Im}\Psi^{12}_{\hat\gamma2}-e^{2C} d(\cos\alpha)\wedge V\wedge\text{Im}(\Psi^{12}_1-\cos\alpha \Psi^{12}_{\hat\gamma1})\nn\\[2mm]
&-e^{3C}\sin^3\alpha d\alpha\wedge J\text{Re}\Psi^{12}_{\hat\gamma0}-e^{2C}\sin^2\alpha J\wedge\text{Im}(\Psi^{12}_{\hat\gamma1}+\cos\alpha \Psi^{12}_1)- e^{3C}\sin^4\alpha V\wedge J\text{Im}\Psi^{12}_0\nn\\[2mm]
&-e^{2C}d(\cos\alpha)\wedge\text{Re}\big(\Omega_1\wedge (-\tilde{\Psi}^{12}_{\hat\gamma1}+\cos\alpha \tilde{\Psi}^{12}_{1})\big)- e^{2C}\sin^2\alpha V\wedge \text{Im}\big(\Omega_1\wedge (\tilde{\Psi}^{12}_1-\cos\alpha \tilde{\Psi}^{12}_{\hat\gamma 1}\big)\nn\\[2mm]
&+ e^{C}\sin^2\alpha V\wedge\text{Re}\Psi^{12}_2+ e^{3C}\sin^3\alpha d\alpha\wedge V\wedge\text{Re}(\Omega_1 \tilde{\Psi}^{12}_0)- e^{C} \sin^2 \alpha \text{Im}(\Omega_1\wedge\tilde{\Psi}^{12}_{\hat\gamma 2}).
\end{align}
Finally the 4-form is 
\begin{align}\label{eq:Ydef}
\Psi_4&=\text{Re}\Psi^+_4-\cos\alpha \text{Re}\Psi^-_{\hat\gamma 4}-e^{4C}\text{vol}(\text{S}^4)\text{Re}(\Psi^-_{\hat\gamma0}+ \cos\alpha \Psi^+_0)+ e^{C}d(\cos\alpha)\wedge \Psi^+_3+e^{C}\sin^2\alpha V\wedge\text{Re}\Psi^-_{\hat\gamma 3}\nn\\[2mm]
&- e^{2C}d(\cos\alpha)\wedge V\wedge\text{Im}\big(\Psi^-_{\hat\gamma2}- \cos\alpha \Psi^+_2\big)+e^{2C}\sin^2\alpha J\wedge\text{Im}\big(- \Psi^+_2+\cos\alpha\Psi^-_{\hat\gamma 2}\big)\\[2mm]
&-e^{3C}\sin^2\alpha d\alpha\wedge J_2\wedge \text{Re}\Psi^+_2-e^{3C}\sin^4\alpha J\wedge V\wedge\text{Im}\Psi^-_{\hat\gamma1}+e^{2C}d(\cos\alpha)\wedge\text{Re}\big(\Omega\wedge\big(\tilde\Psi^+_2-\cos\alpha \tilde\Psi^-_{\hat\gamma2}\big)\big)\nn\\[2mm]
&+e^{3C}\sin^3\alpha d\alpha\wedge V\wedge\text{Re}\big(\Omega\wedge\tilde\Psi^-_{\hat\gamma1}\big)-e^{C}\sin^2\alpha\text{Im}\big(\Omega\wedge \tilde\Psi^+_{3}\big)-e^{2C}\sin^2\alpha V\wedge \text{Im}\big(\Omega\wedge \big(\tilde\Psi^-_{\hat\gamma2}+ \cos\alpha \tilde\Psi^+_2\big)\big)\nn.
\end{align}
Plugging \eqref{eq:sindef}-\eqref{eq:Ydef} into \eqref{eq:MtheoryBPS1}- \eqref{eq:MtheoryBPS6} leads to a large but highly redundant  set of conditions on M$_4$, once the S$^4$ forms are factored out. To make progress it is helpful to study just the 0 and 1-form constraints following from \eqref{eq:MtheoryBPS1}- \eqref{eq:MtheoryBPS4} first, 
\begin{subequations}
\begin{align}
&\Psi^+_0=1,~~~\text{Re}\Psi^-_{0}g_0=\star_4g_4,~~~ \Psi^-_{\hat\gamma_0}=\tilde{\Psi}^{12}_{\hat\gamma0}= \text{Re}\Psi^{12}_0=\text{Re}\Psi^{12}_{\hat\gamma_1}g_0=0,\label{eq:mbps1}\\[2mm]
&2 m e^{C}\text{Im}\Psi^{12}_{\hat\gamma0}= -\text{Re}\Psi^+_{\hat\gamma0},~~~ e^{C}(-2m+e^{A}\text{Re}\Psi^+_{\hat\gamma 0}g_0)=4e^{A}\text{Im}\Psi^{12}_0,\label{eq:mbps2}\\[2mm]
&d(e^{2A+C}\text{Im}\Psi^{12}_{\hat\gamma 0})+ e^{2A}\text{Re}\Psi^{12}_1=d(e^{3A+3C}\text{Im}\Psi^{12}_0)-2m e^{2A+3C}\text{Im}\Psi^-_{\hat\gamma 1}\nn\\[2mm]
&=d(e^{3A}\text{Re}\Psi^-_0)- 2m e^{2A}\text{Re}\Psi^{12}_{\hat\gamma1}-e^{3A}f= d(e^{2A+4C})+ e^{2A+3C}(4\text{Im}\Psi^-_{\hat\gamma1}- e^C \text{Re}\Psi^{12}_1g_0)=0,\label{eq:mbps3}\\[2mm]
&d(e^{3A+3C}\tilde{\Psi}^{12}_0)-e^{2A+2C}(3 i \tilde{\Psi}^{12}_{\hat\gamma1}+ 2me^{C}\tilde{\Psi}^-_{\hat\gamma 1})= d(e^{3A+3C}\text{Re}\Psi^{12}_{\hat\gamma0})-e^{2A+2C}(3  \text{Im}\Psi^{12}_{1}+ 2m e^{C}\text{Re}\Psi^+_{1}) =0\label{eq:mbps4}.
\end{align}
\end{subequations}
These conditions are restrictive - indeed from \eqref{eq:mbps1} it follows that either $\text{Re}\Psi^{12}_{\hat\gamma_1}=0$ or $g_0=g_4=0$ - in which case we have a purely electric 4-form flux and so local AdS$_4$.

To make progress we decompose the M$_4$ spinors in a basis of a single unit norm spinor $\eta$, which satisfies $\eta^\dag\hat\gamma\eta=0$. The most general decomposition compatible with \eqref{eq:mbps1} may be parameterised as
\begin{align}\label{eq:mspinors}
\eta_1&= \cos\lambda(\cos\frac{\beta}{2}\eta+ \sin\frac{\beta}{2}\hat\gamma \eta),\nn\\[2mm]
\eta_2&=\sin\lambda( ( c x_3\sin\frac{\beta}{2} +i a)\eta+ (-c x_3 \cos\frac{\beta}{2}+ i b)\hat\eta+ c(- x_1 +i x_1 )\cos\frac{\beta}{2}  \eta^c+ c(x_2-i x_1 )\sin\frac{\beta}{2} \hat\gamma\eta^c),\nn\\[2mm]
&a^2+ b^2+ c^2= x_1^2+ x_2^2+ x_3^2=1,~~~~(ab-\frac{c^2}{2}\sin\beta)\sin^2\lambda= \frac{1}{2}\sin\beta \cos^2\lambda,
\end{align}
for $(a,b,c,x_i)$ real, which solves the conditions not involving fluxes - these yield two cases we will solve in the next sections
\subsubsection{Case I}
We have two conditions in \eqref{eq:mbps1} left to solve for, in this section we solve $\text{Re}\Psi^{12}_{\hat\gamma_1}g_0=0$ by setting the first factor to zero, which implies we can without loss of generality fix
\beq
c=0,~~~ a=\sin\frac{\beta}{2},~~~b=\cos\frac{\beta}{2}
\eeq
which makes $x_i$ drop out of the ansatz and also implies $\text{Re}\Psi^-_{0}=0$,  leading to $g_4=0$ - the final constraint in \eqref{eq:mspinors} then reduces to 
\beq
\sin\beta \cos2\lambda=0.
\eeq
This appears to suggest two cases, however only $\cos2\lambda=0$ is consistent with the first of \eqref{eq:mbps2}, which demands $\sin\beta\neq0$ when $m\neq0$, and we may solve this and whole of \eqref{eq:mbps2} by fixing
\beq
\sin\lambda=\cos\lambda= \frac{1}{\sqrt{2}},~~~ e^C=e^{A}\frac{\sin\beta}{2m},~~~ e^{A}g_0=- \frac{6m}{\sin\beta},
\eeq
without loss of generality. At this stage what remains non trivial in \eqref{eq:mbps3}-\eqref{eq:mbps4} then imposes 
\beq
F_1=0,~~~ d(e^{3A}\sin\beta)+ 2m e^{2A}\cos\beta e^2,~~~ d(e^{A}\sin\beta)=0,
\eeq
which we can solve locally as
\beq
e^{A}=\frac{L}{\sin\beta},~~~e^2=\frac{e^{A}}{m} d\beta,
\eeq
for $L$ a constant, and where $\beta$ has become a coordinate. 
Substituting what has been derived thus far into \eqref{eq:MtheoryBPS1}-\eqref{eq:MtheoryBPS6} we find just 3 real constraints imply the entire system, namely
\beq\label{eq:mc1remains}
d(e^{3A+2C}\tilde{\Psi}^{12}_1)+ 2e^{2A+C}(i e^{A}\tilde{\Psi}^{12}_{\hat\gamma2}+ m e^C\tilde{\Psi}^-_{\hat\gamma2})=d(e^{3A+2C}\text{Im}\Psi^{12}_{\hat\gamma1})- 2e^{2A+C}(e^{A}\text{Re}\Psi^{12}_{ 2}- m e^C\text{Im}\Psi^+_{2})=0,\nn
\eeq
which all follow from \eqref{eq:MtheoryBPS3}. These can be put into a simple form by defining
\beq
(e^{1},e^3,e^4)=\frac{\cot\beta}{2m}(\omega^1,\omega^2,\omega^3),
\eeq
which reduces the 3 real constraints to $d\omega^i = \frac{1}{2}\epsilon^i_{~jk}\omega^j\wedge \omega^k$, so that these directions locally define an S$^3\to$ S$^3/\mathbb{Z}_k$ globally.\\
~\\
Having solved all the supersymmetry constraints we arrive at a single solution, defining $\sin\beta= \frac{1}{\cosh r}$ and fixing $m=1$, we find
\beq\label{eq:mtheorysol1}
ds^2=L^2\bigg[\cosh^2r ds^2(\text{AdS}_3)+ \sinh^2r ds^2(\text{S}^3/\mathbb{Z}_k)+ dr^2+ \frac{1}{4}ds^2(\text{S}^4) \bigg],~~~ G= \frac{3 L^2}{8}\text{vol}(\text{S}^4),
\eeq
which is just AdS$_7/\mathbb{Z}_k\times$S$^4$, {\it i.e.} the lift of the unique $\mathfrak{sp}(2)\oplus \mathfrak{sp}(1)$ solution in IIA derived in section \ref{eq:sp2sp1IIA}.
\subsubsection{Case II}
In this section we solve $\text{Re}\Psi^{12}_{\hat\gamma_1}g_0=0$ by fixing $g_0=0$, which implies $g_4=0$ as well so as explained earlier we know that for such solutions AdS$_3$ is enhanced to AdS$_4$ - there is one such solution, let us briefly describe how it is derived.\\
~\\
The final condition in \eqref{eq:mspinors} combined with  \eqref{eq:mbps2} impose
\beq
\beta= b=0,~~~~ e^{C}=\frac{2}{m} a\sin2\lambda,~~~ a^2+c^2=1,
\eeq
so that we necessarily have $a\neq0$ and $\sin2\lambda\neq 0$. With \eqref{eq:mbps1} and \eqref{eq:mbps2} solved, what remains non trivial in  \eqref{eq:mbps3}-\eqref{eq:mbps4} then serves to define the vielbein on M$_4$ as
\begin{align}
&(m e^3,m e^4,-m e^1)=(u_1,u_2,u_3),~~~u_i=e^{A}\bigg( x_i\big(2\sin2\lambda dc+c(4 \lambda+ 3\tan 2\lambda dA )\big)+2\sin2\lambda c dx_i\bigg),\nn\\[2mm]
&m e^4=-e^{A}\big( a(2d(2 \lambda)+ 3\tan 2\lambda dA)+ 2\sin2\lambda da\big),
\end{align}
and yield the electric flux component
\beq\label{eq:mfluxel}
e^{A}f=3\left(\tan2\lambda d(e^{A}\sin2\lambda)+ \frac{de^{A}}{\cos2\lambda} \right) .
\eeq
Given the definition of the veilbein it should be clear that $x_i$ have become embedding coordinates for a round S$^2$ within M$_4$, while $a,c$ are embedding coordinates for an S$^1$ over which the S$^2$ and the  S$^4$ are foliated. The remaining coordinate on M$_4$ is $\lambda$, which $A$ must be an exclusive function of $\lambda$ for $d(e^{3A}f)=0$ to hold. All that remains is to determine $e^{3A}$ as a function of $\lambda$ - this follows from the remaining conditions in \eqref{eq:MtheoryBPS1}-\eqref{eq:MtheoryBPS6} and is actually fixed uniquely for all electric solution \eqref{eq:elecconds}, we find
\beq
e^{A}= \frac{1}{\cos\theta}= \frac{1}{ \sin2\lambda},
\eeq
we have thus derived a single solution. Upon identifying $\sec\theta = \cosh(m r)$ and $(a,c)=(\cos\gamma,\sin\gamma)$ this may be expressed as
\begin{align}\label{eq:mtheorysol2}
ds^2&= \cosh(m r)ds^2(\text{AdS}_3)+ dr^2+\frac{4}{m^2}\bigg[d\gamma^2+\sin^2\gamma ds^2(\text{S}^2)+ \cos^2\gamma ds^2(\text{S}^4)\bigg],\nn\\[2mm]
G&= 3m \cosh^3(m r)\text{vol}(\text{AdS}_3)\wedge dr,
\end{align}
which is nothing more than AdS$_4\times$S$^7$, with S$^7$ expressed as a foliation of S$^2\times$S$^4$ over an interval. The appearance of the 7-sphere is consistent with $\mathcal{N}=(8,0)$ with algebra $\mathfrak{osp}(8|2)$, however the AdS$_4$ factor enhances this to $\mathcal{N}=(8,8)$ from the AdS$_3$ perspective.\\
~\\
This concludes our analysis of  AdS$_3\times$S$^4$ solutions in M-theory.

\section{AdS$_3$ with squashed 7-spheres}\label{sec:squshed7spheres}
In this final section we will consider the possibility of realising $\mathcal{N}=(8,0)$ on 3 distinct squashed 7-spheres realising the R-symmetries U(4), Sp(2)$\times$ Sp(1) and Spin(7) in sections \ref{sec:u4cases}, \ref{sec:sp2sp1cases} and \ref{sec:spin7cases} respectively. These differ from the round 7-sphere both in terms of the squashing which introduces an additional function in the metric, but also in terms of the fluxes which can  depend on additional invariant forms. At the level of the equations of motion it is a simple matter to establish that no such solution can exist in type II supergravity so our focus here will be in M-theory. We will again make uses of geometric supersymmetry conditions reviewed in \eqref{sec:Gstructuremtheory}. 

\subsection{U(4) preserving squashed S$^7$}\label{sec:u4cases}
In this section will address the possibility of realising $\mathcal{N}=(8,0)$ on a U(4) preserving squashed 7-sphere in M-theory, as such we should refine the ansatz of \eqref{eq:ansatzMtheory} as
\begin{align}\label{eq:U4ansatz}
ds^2(\text{M}_8)&= e^{2B} ds^2(\mathbb{CP}^3)+ e^{2C}(V^0)^2+e^{2k} dr^2,~~~(V^0)^2=(d\psi+ \eta)^2~~~ d\eta=2J^0,\nn\\[2mm]
e^{3A}F_1&= f_1dr+g_0 V^0,~~~~ F_4= \frac{1}{2}f_2 J^0\wedge J^0+ \frac{1}{4}f_3 dr\wedge V^0\wedge J^0
\end{align}
where $(e^{2B}, e^{2C}, f_i,e^{2k})$ and the AdS warp factor $e^{2A }$ are all functions of $r$ only and we are free to choose an $e^{2k}$ that suits us.  Here $V^0$ and $J^0$ are the U(4) invariant 1 and 2 forms one can define on the 7-sphere - there is also an SU(4)$\subset$ U(4) invariant 3-form $\Omega^0$, but this is charged under the U(1) $\partial_{\psi}$ so cannot appear in the flux. Together $(V^0,J^0,\Omega^0)$ span a Sasaki--Einstein structure and obey the relation \eqref{eq:SEcond}. Explicit expressions for these forms are given in terms of complex coordinates $Z^I$, that embed unit radius S$^7$ in $\mathbb{C}^4$, as
\beq\label{eq:SU(4)invariants}
V^0=\frac{i}{2}(\overline{Z^I}d Z^I-Z^I d\overline{Z^I}),~~~J^0= \frac{i}{2} d\overline{Z^I}\wedge dZ^I,~~~ V^0\wedge \Omega^0= -i d\overline{Z^1}\wedge d\overline{Z^2}\wedge d\overline{Z^3}\wedge d\overline{Z^4},
\eeq 
the metric on $\mathbb{CP}^3$ is simply $dZ^i \overline{dZ^i}-(V^0)^2$. In what follows we will make use of topological joint coordinates that express (the round) S$^7$ as a foliation of S$^3\times$S$^3$ over an interval
\beq\label{eq:U4emedding}
Z^I= (e^{\frac{i}{2}(\phi_1+\psi_1)}\cos\alpha \cos\frac{\theta_1}{2},e^{-\frac{i}{2}(\phi_1-\psi_1)}\cos\alpha \sin\frac{\theta_1}{2},e^{\frac{i}{2}(\phi_2+\psi_2)}\sin\alpha \cos\frac{\theta_2}{2},e^{-\frac{i}{2}(\phi_2-\psi_2)}\sin\alpha \sin\frac{\theta_2}{2})^I\nn,
\eeq
we  shall take the U(1) portion of the R-symmetry to be
\beq
\partial_{\psi}= 2(\partial_{\psi_1}+\partial_{\psi_2})
\eeq
which fixes an arbitrary choice of overall sign.\\
~\\
Taking the bosonic fields as in \eqref{eq:U4ansatz} ensures that they are singlets under U(4), but to have a U(4) R-symmetry the spinors on M$_8$ should transform in the $\textbf{4}\oplus \overline{\textbf{4}}$. General spinors of this type may be constructed from the Killing spinors on the round S$^7$, transforming in the $\textbf{8}$ of SO(8), as follows. Depending on the embedding of U(4) into SO(8) the $\mathbf{8}$ can branch as $\textbf{6}_0\oplus \textbf{1}_{-2}\oplus\textbf{1}_{2}$ or $\textbf{4}_{-1}\oplus\overline{\textbf{4}}_{1}$ - where the subscript is the U(1) charge under $\partial_{\psi}$. The portion of the  S$^7$ Killing spinors that transform in these reps may be defined via the following operators relations\footnote{More concretely, with respect to the embedding \eqref{eq:U4emedding}, we can define the frame
\begin{align}
e^1&=d\alpha,~e^2=\frac{1}{2}\cos\alpha d\theta_1,~e^3=\frac{1}{2}\cos\alpha \sin\theta_1d\phi_1,~e^4=\frac{1}{2}\sin\alpha d\theta_2,~e^5=\frac{1}{2}\sin\alpha \sin\theta_2d\phi_2\nn\\
e^6&= \frac{1}{2}\cos\alpha(-\cos\theta_1d\phi_1+ \cos\theta_2d\phi_2+ d\psi_2),~ e^7=-V^0= \frac{1}{2}(\cos^2\alpha(d\psi_1+ \cos\theta_1 d\phi_1)+ \sin^2\alpha(d\psi_2+ \cos\theta_2 d\phi_2)).\nn
\end{align}
The Killing spinor equations are then solved in terms of arbitrary constant spinors $\xi^0_{\pm}$ as
\beq
\xi_{\pm}={\cal M_{\pm}}\xi^0_{\pm},~~~{\cal M_{\pm}}=e^{\frac{\alpha}{2}(\pm i\gamma_1-\gamma_{67})}e^{\frac{\psi_1}{4}(\pm i\gamma_7-\gamma_{23})}e^{\frac{\theta_1}{4}(\pm i \gamma_2+ \gamma_{37})}e^{\frac{\phi}{4}(\pm i\gamma_7+\gamma_{23})}e^{\frac{\psi_2}{4}(\gamma_{16}-\gamma_{45})}e^{\frac{\theta_2}{4}(\gamma_{14}+\gamma_{56})}e^{\frac{\phi_2}{4}(\gamma_{16}+\gamma_{45})}.\nn
\eeq
}
\beq
\textbf{6}_0:~ (V^0-i J^0)\xi_+=0,~~~\textbf{1}_{-2}:  \left(V^0+\frac{i}{3}J^0 \right)\xi_+=0,~~~ \textbf{4}_{-1}:~ (V^0+i J^0)\xi_-=-2\xi_-
\eeq
where $\xi_{\pm}$ solves $\nabla_{a}\xi_{\pm}=\pm\frac{i}{2}\gamma^{(7)}_a \xi_{\pm}$ on the round sphere, and the  $\textbf{1}_{2}$ and $\overline{\textbf{4}}_{1}$ can be taken to be the Majorana conjugates of  the $\textbf{1}_{-2}$ and $\textbf{4}_{-1}$. Thus the Killing spinors on the round sphere furnish us with a spinor in the $\textbf{4}_{-1}$,  but this is not all that we have at our disposal. The embedding coordinate transform in the $\textbf{4}_1$, so we can construct another spinor in the $\textbf{4}_{-1}$ by dressing the $\textbf{1}_{-2}$  part of the 7-sphere Killing spinors by them. Let us denote the portions of the round sphere Killing spinors in the $\textbf{4}_{-1}$ and $\textbf{1}_{-2}$ as
\beq
\textbf{4}_{-1}:~ \xi^I_4~~~ \text{s.t.}~~ \xi_4^{\dag I}\xi_4^J= \delta^{IJ},~~~~ \textbf{1}_{-2}:~ \xi^0,~~~ \text{s.t.}~~|\xi^0|^2=1,~~~ \xi^{0\dag}\xi_4^I= i Z^I,~~~\xi^{0 c\dag}\xi_4^I=0,
\eeq
where the inner product conditions can be achieved without loss of generality, and made consistent with \eqref{eq:U4emedding}, by suitably defining the spinors\footnote{Specifically, in the notation of the previous footnote, one first defines two spinors in the \textbf{8} of SO(8) as $
\xi^A_{\pm}= {\cal M_{\pm}} \hat\eta^A,~~~ \hat\eta^A=\sum_B\delta^{AB}$ one then has 
\beq
(\xi^I_4)^T=\big(\frac{i}{2}(\xi_-^2+\xi^5_-),~\frac{i}{2}(\xi_-^3-\xi^5_-),~-\frac{1}{2}(\xi^1_-+ \xi^6_-),~\frac{i}{2}(\xi^1_--\xi^6_-)\big)^I,~~~ \xi^0= \frac{1}{\sqrt{2}}(\xi^2_++ \xi^5_+).
\eeq}. Then since there are exactly 2 independent  spinors in the $\textbf{4}_{-1}$, as is shown  in section \ref{sec:squashS7},  any such spinor may be expanded in a basis of $(\xi^I, Z^I \xi^0)$, we also have that on the unit round sphere
\beq
 V^0 \xi^0=-\frac{i}{3} J^0\xi^0=\xi^0,~~~V^0 \xi^I_4=-\xi^I_4+2 i Z^I \xi^0,~~~J^0 \xi^I_4=i \xi^I-2 Z^I \xi^0,
\eeq
consistent with this claim. A spinor on S$^7$ manifestly transforming in the  $\textbf{4}_{-1}\oplus \overline{\textbf{4}}_{1}$ is then given by
\beq
\hat\xi^{\cal I}= \left(\begin{array}{c}\xi^I_4\\(\xi^I_4)^c\end{array}\right), ~~~{\cal I}=1,...,8
\eeq
with a second given in terms of $Z^I\xi^0$. However we need 8d spinors which are also Majorana so this is not a very sensible basis.  We split the flat space gamma matrices as $\gamma^{(8)}_a=\sigma_1\otimes \gamma^{(7)}_a,~\gamma^{(8)}_r= \sigma_2\otimes \mathbb{I}_8$ with chirality matrix and intertwiners $\hat\gamma^{(8)}= \sigma_3\otimes \mathbb{I}_8$ and $B^{(8)}= \sigma_3\otimes B^{(7)}$. Two independent sets of Majorana-Weyl 8d spinors transforming in the $\textbf{4}_{-1}\oplus \overline{\textbf{4}}_{1}$ are then given by\footnote{One can show that the $\mathfrak{u}(4)$ reps of $(\chi^{\cal I}_{\pm},\tilde\chi^{\cal I}_{\pm})$ and $\hat\xi^{\cal I}$  are related by the similarity transformation
\beq
{\cal S}=\left(\begin{array}{cc}\mathbb{I}_{4\times 4}&\mathbb{I}_{4\times 4}\\
i \mathbb{I}_{4\times 4}&-i\mathbb{I}_{4\times 4}\end{array}\right)
\eeq
 so clearly both are in the $\textbf{4}\oplus \overline{\textbf{4}}$.}
\beq
\chi_{\pm}^{\cal I}= \left(\begin{array}{c}\theta_{\pm}\otimes \xi_4^I+\theta^c_{\pm}\otimes (\xi_4^I)^c\\
i (\theta_{\pm}\otimes \xi_4^I-\theta^c_{\pm}\otimes (\xi_4^I)^c)\end{array}\right)^{\cal I},~~~\tilde\chi_{\pm}^{\cal I}= \left(\begin{array}{c}\theta_{\pm}\otimes Z^I\xi^0+\theta^c_{\pm}\otimes (Z^I\xi^0)^c\\
i (\theta_{\pm}\otimes Z^I\xi^0-\theta^c_{\pm}\otimes (Z^I\xi^0)^c)\end{array}\right)^{\cal I},
\eeq
where $\theta_{\pm}$ are eigenvectors of $\sigma_3$. A general $\textbf{4}_{-1}\oplus \overline{\textbf{4}}_{1}$ spinor will contain both these terms coupled to arbitrary complex functions of the interval.  As explained in section \ref{sec:arecipe}, we only need to explicitly solve for a single $\mathcal{N}=1$ subsector of this general spinor, the rest is implied by acting with the spinoral Lie derivative along the U(4) Killing vectors by construction. We take a representative $\mathcal{N}=1$ sub-sector of the full  $\textbf{4}\oplus \overline{\textbf{4}}$ to be $\chi=e^{A/2}(\chi_+ +\chi_-)$ for
\begin{align}\label{eq:u4repneq1}
\chi_+&= \frac{1}{2}\left(\begin{array}{c}1\\0\end{array}\right)  \otimes ( (a_1+b_1)\eta_++(a_2+b_2)\eta_-) +\text{m.c},~~~\chi_-=\frac{1}{2}\left(\begin{array}{c}0\\i\end{array}\right)\otimes ( (a_1-b_1)\eta_++ (a_2-b_2)\eta_-)+\text{m.c},\nn\\[2mm]
\eta_{\pm} &=\frac{1}{2}(1\pm V^0)\xi^1_4,~~~ |a_1|^2+|b_1|^2=|a_2|^2+|b_2|^2=2,
\end{align}
where $(a_1,a_2,b_1,b_2)$ are otherwise\footnote{The restriction on their norms is required by \eqref{eq:mtheoryforms}} arbitrary complex functions of $r$ taken in this combination for later convenience, $\xi^1_4$ is the 1st component of $\xi^I_4$.\\
~\\
We would now like to compute the bi-linears appearing in \eqref{eq:mtheoryforms} that follow from \eqref{eq:u4repneq1}. These decompose as wedge products of terms on the interval and the $d=7$ bi-linears
\beq\label{eq:squashedbilinearsu4}
\Phi_{ab}= 8\eta_a\otimes \eta_b^{\dag},~~~\tilde{\Phi}_{ab}= 8\eta_a\otimes \eta_b^{c\dag},~~a,b=\pm,
\eeq
which are the non trivial part. To compute $(\Phi_{ab},\tilde{\Phi}_{ab})$ we need to define some forms that respect the squashing:  As $\eta_{\pm}$ are spanned by $(\xi^1,\xi^0)$ we expect to see deformations of two Sasaki--Einstein structures that follow from the round sphere versions of $\xi^0\otimes \xi^{0\dag}$ and $\xi^1\otimes \xi^{1\dag}$ (as explained in appendix \ref{sec:BispSphere}). With our definitions  $\xi^0\otimes \xi^{0\dag}$ is spanned by $(V^0,J^0,\Omega^0)$ as defined in \eqref{eq:SU(4)invariants} on both the squashed and round sphere. On round S$^7$ $\xi^1\otimes \xi^{1\dag}$ is spanned by  $(V^1,J^1,\Omega^1)$ obtained as in \eqref{eq:SU(4)invariants}, but for the embedding $(-\overline{Z^1},Z^2,...,Z^4)$ - however these expressions  get deformed by the squashing. It is thus helpful to decompose $(V^1,J^1,\Omega^1)$  into parts preserved by the squashing, i.e.  those parallel and orthogonal to $V^0$, we find
\begin{align}
V^1&=(1-2|Z^1|^2) V^0+2\text{Im}U_1,~~~~ J^1=J^0+\frac{i}{|Z^1|^2}U_1\wedge \overline{U_1}-2\text{Re}U_1\wedge V^0 ,\nn\\[2mm]
\Omega^1&=2i |Z^1|^2V^0\wedge U_2-U_1\wedge U_2-(Z^1)^2 \Omega^0, 
\end{align}
where we define the one and two forms $U_{1,2}$ via
\beq
\overline{Z^1} dZ^1= U_1- i |Z^1|^2 V^0,~~~\iota_{dZ^1}\Omega^0= 2 \overline{Z^1} U_2
\eeq
they are both orthogonal to $V^0$ and so respect the squashing. One can show that they obey the additional differential 
\beq
dU_1= i d\text{Im}U_1=2i|Z^1|^2J^0-\frac{1}{|Z^1|^2}U_1\wedge\overline{U_1},~~~ d(\overline{Z^1}U_2)=3(- Z^1 \Omega^0+i \overline{Z^1} V^0\wedge U^2),
\eeq
and exterior product
\begin{align}
&\overline{Z^1}\overline{U_1}\wedge U_2= (1-|Z^1|^2)Z^1\Omega^0,~~(U_2)^2=0,~~ |Z^1|^4U_2\wedge\overline{U_2}=2(|Z^1|^2(1-|Z^1|^2)J^0+ i U_1\wedge \overline{U_1})\wedge J^0,\nn\\[2mm]
& (1-|Z^1|^2)|Z^1|^2 J^0\wedge U_2+\frac{i}{2}U_1\wedge \overline{U_1}\wedge U_2=0,
\end{align}
 constraints (where we abuse notation such that $U_2^2=U_2\wedge U_2$ and so on) so together with $(V^0,J^0,\Omega^0)$ which obey the Sasaki--Einstein conditions
\beq
dV^0=2J^0,~~~~d\Omega^0= 4 i V^0\wedge \Omega^0,~~~J^0\wedge \Omega^0=0,~~~ (J^0)^3=\frac{3i}{4}\Omega^0\wedge\overline{\Omega^0}
\eeq 
and the embedding coordinate $Z^1$, $(U_1,U_2)$ form a closed set of forms under $d$ and $\wedge$. This is actually all that appears in \eqref{eq:squashedbilinearsu4}, we find
\begin{align}
\Phi_{++}&=|Z^1|^2(1+e^C V^0)\wedge e^{-i e^{2B}J^0},~~~~\tilde{\Phi}_{++}= - e^{3B}(Z^1)^2(1+e^{2C}V^0)\wedge\Omega^0,\\[2mm]
\Phi_{--}&=(1-e^{C}V^0)\wedge\big(e^{ \frac{e^{2B}}{|Z^1|^2} U_1\wedge \overline{U_1}}-|Z^1|^2\big)\wedge e^{-i e^{2B} J^0}  ,~~~\tilde{\Phi}_{--}=-e^{3B}(1-e^{C} V^0)\wedge U_1\wedge U_2,\nn\\[2mm]
\Phi_{-+}&=i e^{B}(1-e^{C}V^0)\wedge U_1\wedge e^{-i e^{2B}J^0},~~~\tilde{\Phi}_{-+}=-i e^{2B}|Z^1|^2 (1-e^C V^0)\wedge U_2\wedge e^{i e^{2B} J^0}.\nn
\end{align} 
It is then simple to use these expression to calculate \eqref{eq:mtheoryforms}, we find
\begin{subequations}
\begin{align}
f &= \text{Re}(|Z^1|^2 \overline{a_1} b_1+(1-|Z^1|^2) \overline{a_2} b_2),\\[2mm]
 K&=\frac{1}{2}(|a_1|^2-|b_1|^2)|Z^1|^2e^{C}V^0+\frac{1}{2}(|a_2|^2-|b_2|^2)(|Z^1|^2-1)V^0\nn\\[2mm]
&- \text{Im}((\overline{a_1}a_2-\overline{b_1} b_2)U_1)- e^{k}\text{Im}(\overline{a_1}b_1|Z^1|^2+\overline{a_2}b_2(1-|Z^1|^2)) dr,\\[2mm]
\Psi_3&=-\frac{e^{2B}}{2}\text{Re}\bigg[(a_1^2-b_1^2)(Z^1)^2\Omega^0+(a_2^2-b_2^2)U_1\wedge U_2-2(\overline{a_1}b_2-\overline{b_1}a_2)U_1\wedge J^0\bigg],\nn\\[2mm]
&- e^{2B+C}V^0\wedge \text{Im}\bigg[|Z^1|^2\big((a_1a_2-b_1b_2)U_2+ a_1 b_1 J^0)+ a_2 b_2\big((1-|Z^1|^2)J^0+ \frac{i}{|Z^1|^2} U_1\wedge \overline{U_1}\big)\bigg],\nn\\[2mm]
&-e^{k}dr\wedge\text{Re}\bigg[e^{2B+C}V^0\wedge(\overline{a_1}a_2-\overline{b_1}b_2) U_1- e^{2B}|Z^1|^2(a_1 b_2-a_2 b_1)U_2\bigg]+...,\\[2mm]
\Psi_4&=\frac{e^{4B}}{2}\bigg[\text{Im}\big((a_1 b_1+a_2 b_2)\frac{Z^1}{\overline{Z^1}}U_1\wedge \Omega^0\big)-(J_0+2 i\frac{1}{|Z^1|^2} U_1\wedge \overline{U_1})\wedge J^0 \bigg]\nn\\[2mm]
&+e^{3B+C}V^0\wedge \text{Re}\bigg[a_2 b_2 U_1\wedge U_2- a_1 b_1 (Z^1)^2 \Omega^0- (a_1 a_2+b_1 b_2)U_1\wedge J^0\bigg]\nn\\[2mm]
&+e^{2k}dr\wedge\bigg(e^{2B+C}V^0\wedge \text{Re}\bigg[|Z^1|^2\big(a_1 b_1 J^0-(a_1a_2+b_1 b_2)U_2\big)- a_2 b_2\big((1-|Z^1|^2)J^0+ i \frac{1}{|Z^1|^2} U_1 \wedge \overline{U_1}\big) \bigg]\nn\\[2mm]
&\frac{e^{3B}}{2}\text{Im}\bigg[(a_1^2+b_1^2) (Z^1)^2\Omega^0+(a_2^2+b_2^2 )U_1\wedge U_2-2\text{Im}(\overline{a_1}b_2+\overline{b_1} a_2)U_1 \wedge J^0\bigg]\bigg)
\end{align}
\end{subequations}
where ... are terms proportional to $(|a_1|^2-|b_1|^2)$ and $(|a_2|^2-|b_2|^2)$ which must vanish for $e^{2A}K$ to be closed as \eqref {eq:MtheoryBPS1} demands. Although these appear rather complicated, things truncate rather quickly once \eqref{eq:MtheoryBPS1} and \eqref{eq:MtheoryBPS2} are considered, we find these impose
\begin{align}\label{eq:u4bpscondtions1}
&|a_1|=|a_2|=|b_1|=|b_2|=1,~~~g_0=0,~~~ e^{\Delta}\text{Re}(\overline{a_1}b_1-\overline{a_2}b_2)+ i e^{B}m (\overline{a_1} a_2-\overline{b_1} b_2)=0,\\[2mm]
&e^{3A}f_1- \partial_{r}(e^{3A}\text{Re}(\overline{a_2}b_2))-2 m e^{2A+k}\text{Im}(\overline{a_2}b_2)=0,~~~\partial_r(e^{2A+B}\text{Im}(\overline{a_1}a_2-\overline{b_1}b_2))+2m e^{3\Delta+k}\text{Im}(\overline{a_1} b_1-\overline{a_2}b_2)=0.\nn
\end{align}
So the functions of the spinor ansatz are necessarily pure phases, we parameterise them as
\beq
(a_1~,b_1,~ a_2,~b_2)=(e^{i \beta_1}~,e^{i \beta_2}~,e^{i \beta_3}~,~e^{i \beta_4})
\eeq
 which reduces the complex algebraic constraint to the two real constraints
\begin{align}
&\cos(\beta_1-\beta_3)=\cos(\beta_2- \beta_4),\\[2mm]
&\sin\frac{1}{2}\big(\beta_1-\beta_3-(\beta_2-\beta_4)\big)\big(m e^{B}\cos\frac{1}{2}\big(\beta_1-\beta_3-(\beta_2-\beta_4)\big)+e^{A}\sin\frac{1}{2}(\beta_1-\beta_2+\beta_3-\beta_4)\big).\nn
\end{align}
These can be solved in two physically distinct ways - either $\beta_1-\beta_3= \beta_2-\beta_4$, which automatically solves both real constraints, or $\beta_1-\beta_3=-( \beta_2-\beta_4)$, which then fixes a warp factor. One can show that the first option ultimately leads uniquely to AdS$_4\times$S$^7$, which we have no interest in rederiving here, so let us instead take the second option and fix
\beq
\beta_1=\beta_3-\beta_2+\beta_4,~~~~ e^{B}=\frac{e^{A}}{m}\sin(\beta_2-\beta_4).
\eeq
Plugging this into the rest of \eqref{eq:u4bpscondtions1} and the conditions following \eqref{eq:MtheoryBPS3} then fixes the majority of \eqref{eq:U4ansatz} as
\beq
\beta_3=\beta_4,~~~e^{C}= \frac{e^A}{2m}\sin 2(\beta_2-\beta_4),~~~f_2= -\frac{4}{m^2}e^{3A} \sin^4(\beta_2-\beta_4),~~~ f_3=f_2',~~~e^{A}f_1=6m e^{k}\cot \beta_2 dr\nn
\eeq
and furnishes us with three ODEs to solve
\beq\label{eq:u4odes}
\partial_{r}(e^{A})=2m e^{k}\cot(\beta_2-\beta_4),~~~e^{A} \partial_r\beta_2=-2m e^{k},~~~ \beta_4'=0.
\eeq
Upon plugging these definitions into \eqref{eq:MtheoryBPS4}-\eqref{eq:MtheoryBPS6} we find that they and the Bianchi identity and equation of motion of the flux are implied - so we indeed have a solution if we can solve \eqref{eq:u4odes}. The ODEs are basically trivial and lead to
\beq
e^{A}= \frac{L}{\sin(\beta_2-\beta_4)},~~e^{k}=-\frac{L}{2m \sin(\beta_2-\beta_4)}\partial_r\beta_2,~~\beta_4=\text{constant}.
\eeq
At this point we find that all the bosonic fields are functions of $\beta_2$, so we can in fact fix this such that our solution take a concise form - we take
\beq
\sin(\beta_2-\beta_4)=\frac{1}{\cosh (2mr)}.
\eeq
The unique 11 dimensional (besides AdS$_4\times$ S$^7$) solution  realising the algebra $\mathfrak{su}(1,1|4)$ then takes the form
\begin{align}
ds^2&=L^2\bigg[\cosh^2(2mr)ds^2(\text{AdS}_3)+ dr^2+ \frac{1}{m^2}\bigg( ds^2(\mathbb{CP}^3)+ \tanh^2(2mr)(V^0)^2\bigg)\bigg],\label{eq:squshed1}\\[2mm]
G&= 6m\tanh(2mr)\text{vol}(\text{AdS}_3)\wedge dr- \frac{2 L^2}{m^3\cosh(2m r)}  J^0\wedge J^0-\frac{2 L^2\sinh(2m r)}{m^2\cosh^2(2m r)} dr\wedge V^0\wedge J^0\nn.   
\end{align}
Clearly this solution is non compact - however it does still have a physical interpretation. As $r \to \infty$  the $(\text{AdS}_3,r)$ directions ( weighted by $L^{-2}$) become an AdS$_4$ factor of radius $2m$ while $(V^0,\mathbb{CP}^3)$ become a round S$^7$, meanwhile the magnetic component of $G$ is exponentially suppressed tending to zero. Hence as $r \to \infty$ the solution becomes AdS$_4\times$S$^7$ with the correct electric flux and supersymmetry is consequentially enhanced. The other end of the space is at $r=0$, where the warp factor of the U(1) fiber goes to zero, but does so in a regular fashion leaving the warping of AdS$_3\times \mathbb{CP}^3$ constant in this limit. The  flux becomes purely magnetic, independent of $r$  and orthogonal to the U(1) fiber. However, unlike the $r\to \infty$ limit, truncating the expansion of \eqref{eq:squshed1} about $r=0$ to leading order does not give a solution on its own. It is natural to interpret this solution as the holographic dual of a $\mathfrak{su}(1,1|4)$ preserving conformal defect in $\mathcal{N}=8$ Chern-Simons matter theory. It would be interesting to study the consequences of this, but we shall not attempt that here.

\subsection{Spin(7) preserving ``squashed" S$^7$}\label{sec:spin7cases}
In this section we consider solutions realising the algebra $\mathfrak{f}_4$ on a deformed 7-sphere - in this  case the metric is actually round, however the flux can depend on more than just $\text{vol}(\text{S}^7)$. The most general way to preserve a Spin(7) isometry is to specialise the ansatz of \eqref{eq:ansatzMtheory} to
\begin{align}
ds^2(\text{M}_8)&= e^{2B}\bigg[d\alpha^2+ \sin^2\alpha ds^2(\text{S}^6)\bigg]+ e^{2k}dr^2,\nn\\[2mm]
e^{3A}F_1&= f_1 dr,~~~F_4=  4f_2 \star_{7}\Phi^0+ f_3 dr\wedge\Phi^0.  
\end{align}
where $(e^{2A},e^{2B},e^{2k},f_i)$ have support on $r$ only and $(\Phi^0,\star_7 \Phi^0)$ are Spin(7) invariant 3-form and 4-forms defined in terms of the nearly Kahler G$_2$ invariants on S$^6$ as
\begin{align}
\Phi^0&= \sin^2\alpha d\alpha\wedge J_{\text{G}_2}+\sin^3\!\alpha\,\text{Re}(e^{-i \alpha}\Omega_{\text{G}_2}),~~~\star_7\Phi^0=-\frac{1}{2}\sin^4\!\alpha\,J_{\text{G}_2}\wedge J_{\text{G}_2}-\sin^3\!\alpha \,d\alpha\wedge\text{Im}(e^{-i \alpha}\Omega_{\text{G}_2}),\nn\\[2mm]
J_{\text{G}_2}&=\frac{1}{2}{\cal C}_{ijk}Y_{\text{S}^6}^{i}dY_{\text{S}^6}^{j}\wedge dY_{\text{S}^6}^{k},~~~\Omega_{\text{G}_2}=\frac{1}{3!}(1-i \iota_{d\alpha}\star_6){\cal C}_{ijk}dY_{\text{S}^6}^{i}\wedge dY_{\text{S}^6}^{j}\wedge dY_{\text{S}^6}^{k}
\end{align}
where ${\cal C}_{ijk}$ are the structure constants defining the product between the octonions imaginary units\footnote{{\it i.e.} the product between the octonions imaginary units $o^i$ is defined as $o^io^j=-\delta^{ij}+\mathcal{C}^{ijk}o_k$.}  and $Y_{\text{S}^6}^{i}$ are embedding coordinates for the unit radius 6-sphere.  They obey 
\beq
d\Phi^0=4 \star_7\Phi^0,~~~\Phi^0\wedge \star_7\Phi^0=7 \text{vol}(\text{S}^7),
\eeq
which states that the seven-sphere metric has weak $\text{G}_2$ holonomy. These conditions follow from the nearly Kahler conditions on $\{J_{\text{G}_2},\Omega_{\text{G}_2}\}$  namely
\begin{equation}
d[J_{\text{G}_2}]=3\text{Re}\Omega_{\text{G}_2}\,,\quad d[\text{Im}\Omega_{\text{G}_2}]=-2 J_{\text{G}_2}\wedge J_{\text{G}_2}\,.
\end{equation}
We can define general Killing spinors on S$^7$ as $\xi_{\pm}= \sum_{I=1}^8\xi^{I}_{\pm} c^I_{\pm}$ where 
\beq
\xi^I_{\pm}= e^{(-\frac{i}{4}\pi\pm \frac{i}{2} \alpha) \gamma_{\alpha}}{\cal M}^{\text{S}^6}\hat\eta^I,~~~ \hat \eta^I = \sum_{J=1}^8 \delta^{IJ}.
\eeq
$c^I$ are constants and ${\cal M}^{\text{S}^6}\hat\eta^I$ span a general Killing spinor on S$^6$. We can take 7-sphere embedding coordinates to be
\beq
Y_1= \cos\alpha,~~~ Y_{i+1}=\sin\alpha Y_{\text{S}^6}^{i},~~~ i=1,..7
\eeq
SO(8) branches as $\textbf{7}\oplus \textbf{1}$ and $\textbf{8}$ under Spin(7) which are realised by the portions of Killing spinors satisfying
\beq
\textbf{1}:~(\Phi^0+\frac{i}{7})\xi_+=0,~~~\textbf{7}:~(\Phi^0- i)\xi_+=0,~~~ \textbf{8}:~\xi_-
\eeq
We denote the $\mathbf{8}$ and $\textbf{1}$ of Spin(7) as $\xi^I_8$, $\xi^0$ respectively and take their components to be Majorana\footnote{Specifically
\beq
\xi^0= \frac{1}{\sqrt{8}}(-\xi_+^1+\xi^2_++\xi^3_++\xi^4+i \xi^5_+-i \xi^6_+-i \xi^7_+-i \xi^8_+),~~~
\eeq
and $\xi_8^1$ (the representative $\mathcal{N}=1$)  defined similarly, but with $\xi^I_{+}\to- \xi^I_-$.}, they obey
\beq
\Phi^0\xi^I_8= i (\xi^I_8+ 8 Y^I\xi^0),~~~ |\xi^0|=1,~~~ \xi^{I \dag}_8\xi^J_8=\delta^{IJ},~~~\xi^{0\dag} \xi^I_8=-Y^I
\eeq
We take an $\mathcal{N}=1$ representative of the most general multiplet transforming spinor transforming in the $8$-representation of $\text{Spin}(7)$ to be of the form $\chi=e^{A/2}(\chi_+ +\chi_-)$, for
\begin{equation}
\begin{split}
\chi_{+}&=\left(\begin{matrix}\sqrt2\\ 0 \end{matrix}\right) \otimes\left(\cos\left(\frac{\rho+\lambda}{2}\right)\,\xi_8^1\,+\,2 \cos\left(\frac{\rho}{2}\right)\cos\left(\frac{\lambda}{2}\right)\,Y_1\xi^0\right)\,\,,\\ 
\chi_{-}&=\left(\begin{matrix}0\\i\sqrt2\end{matrix}\right) \otimes\left(\sin\left(\frac{\rho+\lambda}{2}\right)\,\xi_8^1\,+\,2 \cos\left(\frac{\rho}{2}\right)\sin\left(\frac{\lambda}{2}\right)\,Y_1\xi^0\right)\,,
\end{split}
\end{equation}
where the angles $\rho,\lambda$ are functions of the interval coordinate $r$ and the coefficients $\xi^1_8$ and $Y_1\xi^0$ have been chosen in such a way to solve the constraint $\chi^\dagger\chi=2e^{A}$. 
 In the following, we will also need the forms $f,K,\Psi_3,\Psi_4$ defined as in \eqref{eq:N=1formsMtheory}. The zero-form $f$ and the one-form $K$ admits the following simple form:
 \begin{equation}
 \begin{split}
 f\,&=\,\cos\left(\lambda+\rho\right)+2 \sin\lambda\sin\rho\,Y_1^2\,,\\ 
 K\,&=\, \left(\sin(\lambda+\rho)-2\sin\lambda\cos\rho\,Y_1^2\right)e^{k}dr-2e^B \sin\lambda\,d(Y_1^2)\,.
 \end{split}
 \end{equation}
 Observe in particular that we have three independent one-forms, $\{e^kdr,\,Y_1^2\,e^kdr,\,dY_1^2\}$. In order to write down the forms $\Psi_{3,4}$ in a compact way, it is useful to introduce an appropriate basis of forms that can be defined starting from the G$_2$ structure on S$^7$. Let us define the basis of three-forms:
 \begin{equation}
 \omega^{(3)}_i=\left\{\Phi^0,\,\,\,Y_1^2\Phi^0,\,\,\,dY_1\wedge \iota_{dY_1}\Phi^0,\,\,\,Y_1\,\iota_{dY_1}\!\star_7\!\Phi^0,\,\,\,Y_1\,e^kdr\wedge\iota_{dY_1}\Phi^0\right\}\,,
 \end{equation}
  where the contraction $\iota_{dY_1}$ has to be understood with respect to the un-warped metric on S$^7$. Observe that the unique invariant three-form is $\omega^{(3)}_1=\Phi^0$ while all the other forms belong to the traceful rank-two symmetric representation of Spin$(7)$. In a similar way, we can write down all the appropriate basis of four-forms as:
  \begin{equation}
  \omega^{(4)}_i\,=\,\left\{ \star_7\Phi^0,\,\,\,Y_1^2\star_7\Phi^0,\,\,\, dY_1\wedge \iota_{dY_1}\!\star_7\Phi^0,\,\,\,Y_1 dY_1\wedge \Phi^0,\,\,\,e^k dr\wedge \omega^{(3)}_{1,2,3,4}\right\}
 \end{equation}
 Among the eight possible four-forms, only $\omega_1^{(4)}=\star_7\Phi^0$ is invariant, while all the others are elements of traceful rank-two symmetric multiplets of Spin$(7)$. It is quite straightforward to show that the  exterior derivative of the three-form $d(\omega^{(3)}_i)$ can be expressed in terms of the introduced basis of four-forms. In the following, it will be useful to observe that the five-forms can be all written in terms of the Hodge duals of the three-forms and analogously the seven-forms can be written in terms of the Hodge duals of the one-forms. Finally, let us write down the explicit form of $\Psi_3$ and $\Psi_4$ in terms of the previous basis:
\begin{align}
e^{-3B} \Psi_3\,=\,&\sin(\lambda+\rho)\omega^{(3)}_1-2 \cos\lambda\,\sin\rho\,\omega^{(3)}_2-2 \sin(\lambda+\rho)\omega^{(3)}_3+2 \sin\rho\,\omega^{(3)}_4+2 e^{-B}\sin\lambda\,\omega^{(3)}_5\,,\nn\\
e^{-4B}\Psi_4\,=\,&\omega^{(4)}_1-2\omega^{(4)}_3+2 \cos\lambda\,\omega^{(4)}_4+e^{-B}\cos(\lambda+\rho)\omega^{(4)}_5-2e^{-B}\cos\lambda\,\cos\rho\,\omega^{(4)}_6+\nn\\
&-2e^{-B}\cos(\lambda+\rho)\omega^{(4)}_7+2e^{-B}\cos\rho\,\omega^{(4)}_8\,.
\end{align}
 Having introduced some appropriate forms, we are now in a position to solve the supersymmetry equations. First, we consider the equation \eqref{eq:MtheoryBPS2}: this is a one-form equation and thus can be written in the form $(a_1+a_2 Y_1^2) e^kdr+a_3\,d(Y_1^2)=0$. In particular, focusing on the component proportional to $d(Y_1^2)$ and requiring it to vanish implies that:
 \begin{equation}
 \label{eq:firstEqSpin(7)}
 \sin\lambda\,\left(e^Bm\,-\,e^A\sin\rho\right)\,=\,0\,.
 \end{equation}
 This equation admits two different solutions and thus two branches. The first one is $\sin(\lambda)=0$. This condition simplifies a lot the equations and it is quite straightforward to show that only when $\lambda=\pi$ all the supersymmetry equations can be solved. However this solution again coincide with the well-known $\text{AdS}_4\times \text{S}^7$ background of M-theory and thus we prefer to move directly to the next more interesting branch. In this case, we will assume $\sin\lambda\neq 0$ and \eqref{eq:firstEqSpin(7)} fixes the seven-sphere warping to be $e^B=-\frac{1}{m}\sin\rho\,e^A$. Since by definition the warpings must be non-vanishing, also $\sin\rho$ is constrained to be non-vanishing. The component of \eqref{eq:MtheoryBPS4} proportional to $Y_1dY_1\wedge\star_7\Phi^0$ together with the components of \eqref{eq:MtheoryBPS5} proportional to $\text{vol}_{\text{S}^7}$ and $e^kdr\wedge \iota_{dY_1}\text{vol}_{\text{S}^7}$ completely fix the fluxes:
 \begin{equation}
 \begin{split}
 &e^{-3A}f_1=2e^k-B\cot\left(\frac{\lambda}{2}\right)\tan(\lambda+\rho)+\frac{3}{\sin(\lambda+\rho)}A' \,,\\
& f_2=-e^{3B}\cot\left(\frac{\lambda}{2}\right)\,,\quad f_3=-3e^{k+2B}\frac{1+\cos\lambda}{\sin\rho}\,.
\end{split}
 \end{equation}
 where $A'\equiv \partial_rA$. In this way, all the remaining BPS conditions consists of algebraic and differential constraints on the warping $A$ and the angles $\rho,\lambda$, while the warping $e^k$ is a gauge redundancy to be conveniently fixed. Let us now consider the equation \eqref{eq:MtheoryBPS1}\footnote{This is a two-form equation proportional to $d(Y_1^2)\wedge dr$.} and the equation \eqref{eq:MtheoryBPS3}, focusing in particular on the components proportional to $\omega^{(4)}_6$ and $\omega^{(4)}_8$. Such equations are differential constraints that allow us to get rid of all the derivatives of $A$ and the angles $\rho,\lambda$:
 \begin{equation}
 \begin{split}
 \label{eq:Spin7diff}
\lambda'&=\,-6m\,e^{k-A}\frac{1+\cos\lambda}{\sin\rho}\,,\\
\rho'&=\,\frac{e^{k-A}}{\sin(2\rho)}\left(12 \cot\left(\frac{\lambda}{2}\right)-9\sin\lambda-\sin(2\rho)\right)\,,\\
A'&=\,-\frac{e^{k-A}}{2m\sin^2\rho}\left(8\cot\left(\frac{\lambda}{2}\right)-7\sin\lambda-\sin(2\rho)\right)\,.
 \end{split}
 \end{equation}
 Substituting all the derivatives using the previous conditions, all the other BSP equations boil down to a unique algebraic constraint:
 \begin{equation}
 \sin\rho+3\sin(\lambda+\rho)=0\,.
 \end{equation}
This can be solved without loss of generality as:
 \begin{equation}
 \rho=-\arccos\left( \frac{1+3 \cos\lambda}{\sqrt{10+6\cos\lambda}} \right)\,,
 \end{equation}
 with the further requirement imposed by the BPS equations that $\lambda>0$. We can now use the gauge redundancy to fix the function $k$ and easily solve the differential equations \eqref{eq:Spin7diff}. Fixing:
 \begin{equation}
 e^k=-\frac{e^A}{6m}\,\frac{\sin^2\rho}{1+\cos\lambda}
 \end{equation} 
 the leftover differential constraints are solved by:
 \begin{equation}
 \lambda=r\,,\quad e^{A}=L\frac{\sqrt{5+3\cos r}}{\cos(r/2)\,\sin(r/2)^{2/3}}\,,
 \end{equation}
where $L$ is an integration constant. Observe that given the previous expressions, also the Bianchi identity $d G=0$, that boils down to $f_3'=f_2$ is satisfied - the flux equation of motion also holds.\\
~\\
In summary we find one solution preserving $\mathfrak{f}_4$ besides AdS$_4\times$S$^7$:
\begin{align}
ds^2\,&=\,\frac{L^2}{\cos^2(r/2) \sin^{4/3}(r/2)}\left( (5+3\cos r)ds^2(\text{AdS}_3)+\frac{9}{4 m^2}\frac{\sin^4(r/2)}{5+3\cos r}dr^2+\frac{9}{2 m^2}\sin^2\!r\,ds^2(\text{S}^7)\right)\,,\nn\\
G&=-L^3\frac{(5+3\cos r)(1+7\cos r)}{2\sqrt{2}\sin^3(r/2)\cos^4(r/2)} \text{vol}(\text{AdS}_3)\wedge dr\,+\frac{54 \sqrt2 L^3}{m^3}\,d\left( \cos(r/2)\Phi^0\right)\,.\label{eq:squshed2}
\end{align}
Although $0 <r<\pi$ the solution is non compact, however it still has an attractive physical interpretation. Close to $r=0$ the metric reproduces the singular behaviour of an O2 plane wrapping AdS$_3$ and the internal flux reduce to the invariant 4-form component with no $r$ dependence. As $r\to \pi$, the warping of S$^7$ becomes constant and the internal flux tends to zero. The warp factor of AdS$_3$ blows up, but this is not signaling a singularity, indeed the solution is regular at this point. Computing the curvature tensors of the $(\text{AdS}_3,r)$ directions (weighted by $L^{-2}$) shows that they become those of AdS$_4$ with radius $\frac{\sqrt{2}m}{3}$. So the solution interpolates between an O2 plane on AdS$_3$ and AdS$_4\times$S$^7$. Again we have derived a candidate holographic dual to a conformal defect in ${\cal N}=8$ Chern-Simons matter theory, this time preserving the algebra $\mathfrak{f}_4$.

\subsection{Sp(2)$\times$Sp(1) preserving squashed S$^7$}\label{sec:sp2sp1cases}
In this section we study the possibility of realising $\mathfrak{osp}(4^*|4)$ on the Sp(2)$\times$Sp(1) preserving squashed 7-sphere. In many ways this is the most complicated of the 3 squashings so we supplement this section with appendix \ref{Appendix C} which goes into greater detail on many of the technical points. In this case we should refine \eqref{eq:ansatzMtheory} as
\begin{align}\label{eq:sp2sp1ansatz}
ds^2(\text{M}_8)&=\frac{1}{4}\bigg[e^{2B}(d\alpha^2+ \frac{1}{4}\sin^2 \alpha(L_1^i)^2)+ e^{2C}(L_2^i+{\cal A}^i)^2\bigg]+e^{2k} dr^2,~~~{\cal A}^i =- \cos^2 \left(\frac{\alpha}{2}\right) L_1^i,\nn \\[2mm]
e^{3A}F_1&= f_1dr,~~~~ F_4= (f_2-f_3)\Lambda^0_3\wedge dr+ f_3\tilde{\Lambda}^0_3\wedge dr-12f_4\Lambda^0_4-2 f_5\tilde{\Lambda}^0_4
\end{align}
where $(e^{2A},e^{2B},e^{2C},f_i,e^{2k})$ are functions of the interval only and  $L^i_{1,2}$ are SU(2) left-invariant forms. The 3 and 4 forms $(\Lambda^0_3,\tilde\Lambda^0_3)$, $(\Lambda^0_4,\tilde\Lambda^0_4)$ are Sp(2)$\times$Sp(1)  invariants one can define on this squashed S$^7$; they take the form
\beq
\Lambda^0_3=\frac{1}{8}(L_2^1+{\cal A}^1)\wedge (L_2^2+{\cal A}^2)\wedge (L_2^3+{\cal A}^3),~~~\tilde{\Lambda}^0_3= \frac{1}{8}(L_2^i+{\cal A}^i)\wedge{\cal F}^i
\eeq
where ${\cal F}^i= d{\cal A}^i+\frac{1}{2}\epsilon^{i}_{~jk}A^j\wedge A^k$ and on the unit round sphere $\Lambda^0_4=\star_7 \Lambda^0_3$, $\tilde\Lambda^0_4=\star_7 \tilde\Lambda^0_3$. They obey
\beq
d\Lambda^0_3=-2 \tilde{\Lambda}^0_4,~~~d\tilde{\Lambda}^0_3=-12\Lambda^0_4-2 \tilde{\Lambda}^0_4,~~~  \Lambda^0_3\wedge \Lambda^0_4=\frac{1}{6}\tilde\Lambda^0_3\wedge\tilde \Lambda^0_4=\text{vol}(\text{S}^7),
\eeq
with the other wedge products yielding zero.\\
~~\\
Under $\mathfrak{sp}(2)\oplus \mathfrak{sp}(1)$, $\mathfrak{so}(8)$ branches as either $\textbf{3}\oplus\textbf{5}$ or $(\textbf{4},\textbf{2})$. These representations are all realised by the unit radius round sphere Killing spinors $\xi_{\pm}$ obeying the operator relations
\beq
\textbf{3}:~ (i+\Lambda^0_3)\xi_+=0,~~~\textbf{5}:~ (1+i\Lambda^0_3-\frac{i}{3}\tilde{\Lambda^0}_3)\xi_+=0,~~~(\textbf{4},\textbf{2}):~ \xi_-
\eeq
In addition there is also a singlet spinor $\xi_0$, that is not a Killing spinor for the round sphere, which obeys
\beq
(i+ \Lambda^0_3)\xi^0=(i+\frac{1}{6} \tilde{\Lambda}^0_3)\xi^0=0,
\eeq
note it is this spinor that the $\mathcal{N}=1$ AdS$_4$ solution with squashed 7-sphere in \cite{Awada:1982pk} preserves. Let us define our $(\textbf{4},\textbf{2})$ Killing spinor such that it is Majorana, namely
\beq
\xi^{\cal I}_{(4,2)}=\frac{1}{\sqrt{2}}\big(i(\xi^1_-+\xi^4_-),-(\xi^1_--\xi^4_-),\xi^2_-+\xi^3_-,i(\xi^2_--\xi^3_-),\xi^5_-+\xi^8_-,-i(\xi^5_-+\xi^8_-),i(\xi^6_-+\xi^7_-),-\xi^6_-+\xi^7_-\big)^T\nn
\eeq
where the spinor components are defined in  appendix \ref{eq:sp2sp1killingspinor}. We can then generate another two independent $(\textbf{4},\textbf{2})$ spinors by acting on this with $\Lambda^0_4$ and $\tilde{\Lambda}^0_4$, which together span the most general spinor of this type (as explained around \eqref{eq:branchSpin8-7}). One can additionally show on the unit round sphere that
\beq
(1+\Lambda^0_4+ \tilde{\Lambda}^0_4)\xi^{\cal I}_{(4,2)}=8 Y^{\cal I}\xi^0 
\eeq
where $Y^{\cal I}$ are real embedding coordinates defined in terms of the complex coordinates of \eqref{eq:sp4sp2emeddingcomplex} as $Z^1=Y^1+i Y^2, \dots , Z^4=Y^7+i Y^8$ - clearly consistency demands these also transform in the \textbf{(4,2)}.\\ 
~\\
We  take our representative $\mathcal{N}=1$ sub sector  to be $\chi = e^{A/2} (\chi_+ + \chi_-)$ for
\begin{equation}
\chi_{+}=\left(\begin{matrix}1 \\ 0 \end{matrix}\right) \otimes\left(a_1 \eta_+^1 + a_2 \eta_-^1 + (a_3-a_1) Y^1\xi^0\right),~~~\chi_{-}=\left(\begin{matrix}0\\i \end{matrix}\right) \otimes\left(b_1 \eta_+^1 + b_2 \eta_-^1 + (b_3-b_1) Y^1\xi^0\right),\nn
\end{equation}
where $(a_{1,2,3},b_{1,2,3})$ are real functions of $r$ and  we define 
\begin{equation}
\eta_\pm = \frac{1 \pm \Lambda^0_4}{2} \xi^{1}_{(4,2)}~~ \Rightarrow ~~\eta_+^\dag \eta_+ = 1-\eta_-^\dag \eta_- = \sin^2 \frac{\alpha}{2}, \qquad \eta_+^\dag \xi^0 = Y^1 , \qquad \eta_-^\dag \xi^0 = 0.
\end{equation}
One can easily see that the constraint $\chi^\dagger\chi=2e^{A}$ imposes
\begin{equation}
a_1^2 + b_1^2 = a_2^2 + b_2^2 = a_3^2 + b_3^2 = 1
\end{equation}
and therefore we can set $a_i = \cos \beta_i$ and $b_i = \sin \beta_i$ without loss of generality. In terms of these angles we have that
\begin{equation}
f= \cos ^2\left(\frac{\alpha }{2}\right) \cos \left(2 \beta _2\right)+\sin ^2\left(\frac{\alpha }{2}\right) \cos \left(2 \beta _1\right)+(Y^1)^2 \left(\cos \left(2 \beta _3\right)-\cos \left(2 \beta _1\right)\right) .
\end{equation}
Notice that by decomposing $d Y^1$ along the fiber and base direction
\begin{equation}
v_B = e^B d Y^1 |_B \, , \quad v_F= e^C d Y^1 |_F\, ,\qquad d Y^1=d Y^1 |_B+d Y^1 |_F \, ,
\end{equation}
it is possible to rewrite the spinors on the squashed sphere as following:
\begin{equation}
\eta_+= i v_F \xi^0 + Y^1 \xi^0 \, , \qquad \eta_-= - i v_B \xi^0 \, .
\end{equation}
This gives us a simple way of rewriting all the bilinears just in terms of
\begin{equation}
\Psi_0 =\xi^0 \otimes \xi^{0 \, \dagger}= 1 + i (e^{3C}\Lambda^0_3 + e^{2B + C} \tilde\Lambda^0_3 )+( e^{4B}\Lambda^0_4 +  e^{2B + 2C}\tilde\Lambda^0_4)+ i e^{4B + 3 C}\text{vol}(S^7) \, ,
\end{equation}
Indeed, using the usual gamma-matrix representation acting on differential forms\footnote{Namely, we can express the action of a gamma matrices on a $k$-form as $\gamma^\mu \Omega_k= (d x^\mu + \iota^\mu) \Omega_k$ and $\Omega_k \gamma^\mu = (-)^k (d x^\mu - \iota^\mu) \Omega_k$.} one can compute all the spinor bilinears
\begin{align}
\Psi_{+0}=\eta_+ \otimes \xi^{0 \, \dagger}  &= Y^1\Psi_0 + i (v_F \wedge + \iota_{v_F}) \Psi_0  , \qquad \qquad \Psi_{-0}=\eta_- \otimes \xi^{0 \, \dagger}  =  -i (v_B \wedge + \iota_{v_B}) \Psi_0 , \\[2mm]
\Psi_{+-}=\eta_+ \otimes \eta_-^{\dagger}  &= (i Y^1 -v_F \wedge-\iota_{v_F} )(v_B \wedge - \iota_{v_B}) \bar{\Psi}_0 \, , \quad \Psi_{--}=\eta_- \otimes \eta_-^{\dagger}  = ( v_B \wedge \iota_{v_B}-\iota_{v_B}v_B \wedge) \bar\Psi_0 , \nonumber \\[2mm]
\Psi_{++}=\eta_+ \otimes \eta_+^{\dagger}  &= (\iota_{v_F}v_F \wedge -  v_F \wedge \iota_{v_F})\bar\Psi_0+(Y^1)^2 \Psi_0 + i Y^1 \left( (v_F \wedge+\iota_{v_F}) \Psi_0 - (v_F \wedge-\iota_{v_F}) \bar{\Psi}_0 \right) . \nonumber
\end{align}
The bispinors on the eight-dimensional internal space can be easily build upon these ones. Using a subscript to indicate the form-degree we are picking from the bi-spinor, we have that
\begin{align}
&K= -2 Y^1\text{Im}\Big[ \sin \left(\beta _1-\beta _3\right) \Psi_{+0}+ (\sin \left(\beta _1-\beta _2\right)+\sin \left(\beta _2-\beta _3\right)) \Psi_{-0}+\sin \left(\beta _1-\beta _2\right) \Psi_{+-}\Big]_1 \nn \\[2mm]
&-\text{Re}\Big[2\left(\sin \beta _1+\sin \beta _3 \right) \left(\cos \beta _1-\cos\beta _3 \right)(Y^1)^2 \Psi_0+ \sin \left(\beta _1-\beta _3\right) Y^1 \Psi_{+0}- \sin \beta_1 \Psi_{++} - \sin \beta_2 \Phi_{--}\Big]_0 \wedge(e^k d r) \nn \\[2mm]
&\Psi_3= \text{Im} \Big[-2\left(\sin \beta _1-\sin \beta _3\right) \left(\cos \beta _1-\cos\beta _3\right) (Y^1)^2 \Psi_0 + 2 (\sin 2 \beta _1-\sin \left(\beta _1+\beta _3\right)) Y^1 \Psi_{+0} \nn \\[2mm]
&+ 2 (\sin \left(\beta _1+\beta _2\right)-\sin \left(\beta _2+\beta _3\right)) Y^1 \Psi_{-0}-2\sin \left(\beta _1+\beta _2\right) \Psi_{+-}-\sin 2 \beta_1 \Psi_{++}-\sin 2 \beta_2 \Psi_{--} \Big]_3 \nn \\[2mm]
&+2 \text{Re}\Big[\sin \left(\beta _1-\beta _3\right) Y^1\Psi_{+0}+2 (\sin \left(\beta _1-\beta _2\right)+\sin \left(\beta _2-\beta _3\right)) Y^1 \Psi_{-0}+\sin \left(\beta _1-\beta _2\right) \Psi_{+-}\Big]_2\wedge (e^k dr)\nn \\[2mm]
&\Psi_4 = \text{Re} \Big[2(1- \cos \left(\beta _1-\beta _3\right)) (Y^1)^2 \Psi_0 + 2 \left(\cos \left(\beta _1-\beta _3\right)-1\right) Y^1 \Psi_{+0}  \\[2mm]
&+ 2 (\cos \left(\beta _2-\beta _3\right)-\cos \left(\beta _1-\beta _2\right))  Y^1 \Psi_{-0}+ 2 \cos \left(\beta _1-\beta _2\right) \Psi_{+-} +  \Psi_{++} +  \Psi_{--} \Big]_4 \nn \\[2mm]
&+\text{Im} \Big[(\cos \left(2 \beta _1\right)+\cos \left(2 \beta _3\right)-2 \cos \left(\beta _1+\beta _3\right)) (Y^1)^2 \Psi_0 + 2 \left(\cos \left(\beta _1+\beta _3\right)-\cos \left(2 \beta _1\right)\right) Y^1 \Psi_{+0}  \nn\\[2mm]
&+ 2 (\cos \left(\beta _1+\beta _2\right)-\cos \left(\beta _2-\beta _3\right))  Y^1 \Psi_{-0}+ 2 \cos \left(\beta _1+\beta _2\right) \Psi_{+-} + \cos \left(2 \beta _1\right) \Psi_{++} + \cos \left(2 \beta _2\right) \Psi_{--} \Big]_3\wedge (e^k dr) \nn
\end{align}
The next step is to insert the spinor bilinears just computed into the supersymmetry conditions. For example, upon plugging $K$ and $f$ into \eqref{eq:MtheoryBPS2}, it is quick to establish that
\begin{align}
\sin(\beta_1-\beta_3)(m e^C+e^{A} \sin(\beta_1+\beta_3))&=\sin(\beta_2-\beta_3)(m e^B-e^{A} \sin(\beta_2+\beta_3))\nn\\[2mm]
&=\sin(\beta_1-\beta_2)(m e^B+e^{A} \sin(\beta_1+\beta_2))=0,\label{eq:finalbranchingcondtion}
\end{align}
which indicates a branching of possible solutions. However, once \eqref{eq:MtheoryBPS3} is considered only two options end up being viable. The first is $\sin(\beta_1-\beta_3)=\sin(\beta_2-\beta_3)=\sin(\beta_1-\beta_2)=0$ from which one again recovers AdS$_4\times$S$^7$. The more interesting case follows from solving \eqref{eq:finalbranchingcondtion} as
\beq
\beta_3=-\beta_1-2\beta_2,~~~~ me^{B}= -e^{A}\sin(\beta_1+ \beta_2),~~~~m e^{C}= e^{A} \sin2\beta_2
\eeq
without loss of generality. Using just \eqref{eq:MtheoryBPS1} and \eqref{eq:MtheoryBPS3} it is possible to extract fixed expressions for the fluxs $(F_1,F_4)$ along with the following  conditions
\begin{align}
&3\sin2\beta_2+\sin 2\beta_1=0,\\[2mm]
&2 \sin 2 \beta_2 \sin ^3\left(\beta _1+\beta _2\right) e^{A-k} \partial_{r}\beta_1+m \sin 2 \beta_1 \sin \left(\beta _1+3 \beta _2\right) \cos 2 \beta_2=0,\nn\\[2mm]
&12 \sin 2 \beta_2 \sin ^3\left(\beta _1+\beta _2\right) e^{-k} \partial_r e^A+m \left(\sin \left(\beta _1-5 \beta _2\right)-3 \sin \left(\beta _1+3 \beta _2\right)-2 \sin \left(3 \beta _1+5 \beta _2\right)\right)=0,\nn
\end{align}
one finds that these actually imply the rest of \eqref{eq:MtheoryBPS1}-\eqref{eq:MtheoryBPS6}. Despite appearances, these conditions are relatively easy to solve as one can choose $e^{k}$ (see below) such that $\partial_{r}\beta_1 =\frac{1}{2}$ - then this and the remaining differential condition defining $e^{A}$ can be integrated exactly.\\
~~\\
In summary the solution is given by 
\begin{align}
ds^2&= L^2\Delta_1^{\frac{1}{3}}\Delta_2\bigg[\frac{1}{\sin^2 r}ds^2(\text{AdS}_3)+\frac{1}{12 m^2}\bigg(\frac{(2+\cos r(\cos r-\sqrt{\Delta_3}))}{2\sin^2 r}ds^2(\text{S}^4)+ \frac{1}{3}(L_2^i+{\cal A}^i)^2 \bigg)\bigg]+ e^{2k}dr^2,\nn\\[2mm]
e^k&=-\frac{L\Delta_1^{\frac{1}{6}}\Delta_2^{\frac{1}{2}}}{m\sin r}\frac{\frac{2\sin^2r}{\sqrt{\Delta_3}}+\cos^2 r \sqrt{\Delta_3}- \cos r(2+\cos^2 r)}{(1+2 \cos^2r)\sqrt{\Delta_3}-\cos r (7+2 \cos^2 r)}\nn,\\[2mm]
\Delta_1&=\cos r+\sqrt{8+ \cos^2 r},~~~~ \Delta_2=3 \cos r+\sqrt{8+ \cos^2 r},~~~\Delta_3= 8+ \cos^2 r\nn\\[2mm]
G&= -\frac{12m L^2\Delta_1^{\frac{1}{3}}\Delta_2e^{k}}{\sin r\big(2+\cos r(\cos r-\sqrt{\Delta_3})\big)}\text{vol}(\text{AdS}_3)\wedge dr+ dC_3,\label{eq:squshed3}.\\[2mm]
C_3&=  -\frac{L^3\Delta_1^{\frac{1}{2}} \Delta_2^{\frac{3}{2}}}{36m^3}\bigg(\frac{27 \cos r-\sqrt{\Delta_3}(5+4 \cos^2r)}{3\sin^3 r} \Lambda^0_3+ \frac{4 \sqrt{\Delta _3}+\cos r \left(4 \sqrt{\Delta _3} \cos r-23\right)-\cos (3 r)}{\sin r \left(\cos r \left(\sqrt{\Delta _3}-\cos r\right)+4\right)}\tilde{\Lambda}^0_3\bigg)\nn
\end{align}
Clearly the solution is rather complicated and due to the appearance of $\sin^{-2}r$ it may at first sight appear singular - however a careful analysis reveals this not to be the case. Close to $r=\pi$ the fibered 3-sphere vanishes regularly with the rest of the warpings becoming constant - the electric flux meanwhile tends to zero and the magnetic flux reduces to a single constant component along $\Lambda^0_4$. As $r\to 0$ the S$^4\times $S$^3$ fibration becomes a round 7-sphere of constant radius, and the warp factor of AdS$_3$ and the electric flux blow up  such that $(\text{AdS}_3,r)$  again tends to AdS$_4$, this time of radius $2^\frac{1}{6}\sqrt{3}m$ (when $L=1$) - the electric flux becoming proportional to it's volume form. We have thus constructed a third candidate for a holographic dual to a defect in Chern-Simons matter theory, this time preserving the algebra $\mathfrak{osp}(4^*|4)$.
 
\section*{Acknowledgments}
We would like to thank  Nicolo Petri for clarifying correspondences and Alessandro Tomasiello for in depth discussions and for sharing a draft of his book with us.  NTM is supported by the Spanish government grant PGC2018-096894-B100. The work of GLM is supported by the Swedish Research Council grant number 2015-05333 and partially supported by the ERC Grants 772408 ``String landscape''. AL is supported by UKRI Science and Technology Facilities Council (STFC) Consolidated Grants ST/P00055X/1 and ST/T000813/1.

\appendix

\section{AdS$_3$ in M-theory with  ${\cal N}=1$ supersymmetry}\label{sec:Neq1AdS3mtheory}
In this appendix we derive necessary and sufficient geometric conditions for a solution with AdS$_3$ factor in M-theory to preserve $\mathcal{N}=1$ supersymmetry. Such a classification was already performed in \cite{Martelli:2003ki}, we will present an equivalent set of conditions, which are easier to work with for the cases we deal with in the main text.\\
~\\  
We begin by decomposing the bosonic fields of eleven-dimensional supergravity as 
\beq
ds^2 = e^{2A} ds^2(\text{AdS}_3)+ ds^2(\text{M}_8)\,,~~~ G= e^{3A}\text{vol}(\text{AdS}_3)\wedge F_1 +F_4\,,
\eeq
and take the Majorana Killing spinor to decompose in terms of these factors as
\beq
\epsilon =\zeta\otimes \chi
\eeq
where $\zeta$ are spinors on AdS$_3$ that obey
\beq
\nabla^{(3)}_{\mu}\zeta = \frac{m}{2} \gamma_{\mu}^{(3)}\zeta
\eeq
and $\chi$ are non-chiral spinors on M$_8$, both are Majorana.\\
~~\\
One can show that supersymmetry implies
\beq
\chi^{\dag} \chi= 2 e^{A},~~~ \chi^{\dag}\hat\gamma \chi= 2 e^{A} f,~~~ 0<f<1\,,
\eeq
where the factor of 2 is choice we make without loss of generality.\\
~~\\
Following \cite{Gauntlett:2002fz,Gauntlett:2003wb} necessary conditions for supersymmetry are defined in terms of the following 11d bi-linears 
\beq
K^{(11)}_M = \overline{\epsilon}\Gamma_M\epsilon,~~~ \Omega^{(11)}_{M N}=\overline{\epsilon}\Gamma_{MN}\epsilon,~~~~ \Sigma^{(11)}_{MNOPQ}= \overline{\epsilon}\Gamma_{MNOPQ}\epsilon\,,
\eeq
for $\overline{\epsilon}= (\Gamma_0 \epsilon)^{\dag}$. These should obey
\begin{align}
\label{eq:11dsusygen1}
\nabla_{M}K^{(11)}_{N}&=\frac{1}{3!}(\Omega^{(11)})^{PQ}G_{PQMN}+\frac{1}{6!}(\Sigma^{(11)})^{PQRST}(\star_{11} G)_{PQRSTMN},\\[2mm]
\label{eq:11dsusygen2}
\nabla_{M}(\Omega^{(11)})_{NL}&=\frac{1}{3 \cdot 4!}(g^{(11)})_{M[N}(\Sigma^{(11)})_{L]}^{~~PQRS}G_{PQRS}+\frac{1}{3 \cdot 3!}(\Sigma^{(11)})_{NL}^{~~PQR}F_{NPQR},\\[2mm]
-&\frac{1}{3 \cdot 3!}(\Sigma^{(11)})_{M[N}~~^{RST}G_{L]RST}+\frac{1}{3}(K^{(11)})^{R}G_{RMNL},\nn\\[2mm]
\nabla_{M}(\Sigma^{(11)})_{NOPQR}&=\frac{1}{3!}(K^{(11)})^S\star_{11}G_{SMNOPQR}-\frac{10}{3}G_{M[NOP}(\Omega^{(11)})_{QR]}-\frac{5}{6}G_{[NOPQ}(\Omega^{(11)})_{R]M}\nn\\[2mm]
\label{eq:11dsusygen3}
-&\frac{10}{3}(g^{(11)})_{M[N}(\Omega^{(11)})_{O}^{~S} G_{|S|PQR]}+\frac{5}{6}G_{M[N|ST|}(\star_{11} \Sigma^{(11)})^{ST}_{~~OPQR}\\[2mm]
+&\frac{5}{6}G_{[NO|ST|}(\star_{11}\Sigma^{(11)})^{ST}_{~~PQR]M}-\frac{5}{9}(g^{(11)})_{M[N}G_{O]STU}(\star^{(11)}\Sigma^{(11)})^{STU}_{~~~PQR}\nn
\end{align}
for ${\mathcal N}=1$ to hold in general.  Note a consequence of these conditions is that $(K^{(11)})^M\partial x^{M}$ defines a Killing vector of the metric and flux that can be time-like or null.  The conditions \eqref{eq:11dsusygen1}-\eqref{eq:11dsusygen3} imply the following differential form constraints
\begin{align}\label{eq11dsusy1}
d\Omega^{(11)}&= \iota_{K^{(11)}}G,\\[2mm]
d\Sigma^{(11)}&= \iota_{K^{(11)}}\star_{11}G-\Omega^{(11)}\wedge G,\\[2mm]
\star_{11}dK^{(11)}&= \frac{2}{3}\Omega^{(11)}\wedge \star_{11}G-\frac{1}{3}\Sigma^{(11)}\wedge G, \label{eq11dsusy3}
\end{align}
which for the time-like case are necessary and sufficient for supersymmetry. When the Killing vector is null, as it will be for AdS$_3$, \eqref{eq:11dsusygen1}-\eqref{eq:11dsusygen3} may contain additional conditions not implied by \eqref{eq11dsusy1}-\eqref{eq11dsusy3}. In general this significantly complicates the matters \cite{Gauntlett:2003wb},  however for AdS$_3$ we find that most of  \eqref{eq:11dsusygen1}-\eqref{eq:11dsusygen3} are implied by \eqref{eq11dsusy1}-\eqref{eq11dsusy3}. We shall begin by extracting 8d conditions for ${\mathcal N}=1$  from \eqref{eq11dsusy1}-\eqref{eq11dsusy3}; to this end, we decompose the 11d gamma matrices as
\beq
\Gamma_{\mu}= e^{A} \gamma^{(3)}_{\mu} \otimes \hat\gamma,~~~ \Gamma_{\mu} =\mathbb{I}\otimes \gamma_a
\eeq
and take the intertwiner defining Majorana conjugation as $\epsilon^c= B^{11}\epsilon^*$ to be $B^{11}=\mathbb{I}\otimes B$ such that $\gamma_a^*= B^{-1}\gamma_a B$, $B B^*=\mathbb{I}$. We parameterise AdS$_3$ as
\beq
ds^2(\text{AdS}_3)= e^{2m r} (-dt^2+ dx^2)+ dr^2
\eeq
which means the (Mink$_2$ respecting part of) the AdS$_3$ spinor is
\beq
\zeta= e^{\frac{m}{2} r}\left(\begin{array}{c}1\\0\end{array}\right),
\eeq
in the obvious frame, from which one can easily derive  bi-linears on AdS$_3$ - we skip the details. On M$_8$ we define the bi-linears
\beq
\label{eq:N=1formsMtheory}
2 e^{A} f = \chi^{\dag}\hat\gamma \chi,~~~ 2 e^{A} K_a= \chi^{\dag} \gamma_a \chi,~~~ 2 e^{A} (\Psi_3)_{abc}=\chi^{\dag}\gamma_{abc}\hat\gamma\chi,~~~ 2 e^{A} (\Psi_4)_{abcd}= \chi^{\dag}\gamma_{abcd}\chi.
\eeq
Obviously these can be further decomposed in terms of an angle, a G$_2$ structure 3-form and a component of the vielbein on M$_8$ as in \cite{Martelli:2003ki} - but this refinement does not make the conditions easier to work with for our purposes, so we will not peruse it here. One can then show that the 11d bi-linears decompose in terms of the 8d ones and data on AdS$_3$ as
\begin{align}
K^{(11)}&= 2 e^{A+m r} (e^t- e^x),\\[2mm]
\Omega^{(11)}&= 2 e^{A+m r} (e^t- e^x) \wedge (K+ f e^r),\\[2mm]
\Sigma^{(11)}&= 2 e^{A+m r} (e^t- e^x) \wedge (-e^r\wedge \Psi_3+ \Psi_4),
\end{align}
where $e^{t,x,r}$ are the parts of the 11d vielbein along the AdS$_3$ directions, not the unwarped AdS$_3$ ones. 
Plugging these into \eqref{eq11dsusy1}-\eqref{eq11dsusy3} we find $\mathcal{N}=1$ supersymmetry for AdS$_3$ requires
\begin{align}
& d(e^{2A} K)=0\label{eq:susy11done},\\[2mm]
&d(e^{3A} f)- e^{3A} F_1-2 m  e^{2A} K=0,\\[2mm]
& d(e^{3A} \Psi_3)- e^{3A}(-\star_8 F_4+ f F_4)+2 m e^{2A} \Psi_4=0,\\[2mm]
&d (e^{2A} \Psi_4)- e^{2A} K \wedge F_4=0,\\[2mm]
&6 \star_8 dA-2 f \star_8 F_1+ \Psi_3\wedge F_4=0,\label{eq:notimpliedcondtion}\\[2mm]
&6m \text{Vol}(\text{M}_8)+ e^{A}(2 K\wedge \star_8 F_1+ \Psi_4\wedge F_4)=0\label{eq:impliedcondtion},
\end{align}
which are necessary but not sufficient. To get a sufficient set of conditions, one needs to first extract further 8d conditions from \eqref{eq:11dsusygen1}-\eqref{eq:11dsusygen3} and then isolate a system of equations that imply the rest - the first step is straightforward, the second is a very long and involved computation. As this has essentially already been performed twice in \cite{Martelli:2003ki} and \cite{Tsimpis:2005kj} respectively, here we will just present the results and show they imply those of \cite{Martelli:2003ki}, which are known to be sufficient, correcting a typo in one of their equations in the process. In doing so the following identities are useful
\begin{align}
&K\wedge \Psi_3\wedge (f X_4+\star_8 X_4)=(1-f^2)\Psi_4\wedge X_4,~~~(-\star_8 X_4+f X_4)\wedge \Psi_4=\Psi_3\wedge K\wedge F_4,\label{eq:formconds}\\[2mm]
&K\wedge \star_8 K=\frac{1}{7}K\wedge \Psi_3\wedge\Psi_4=(1-f^2)\text{vol}(\text{M}_8),~~~\Psi_4\wedge\Psi_4= 14 f \text{vol}(\text{M}_8),~~~ 7\star_8K=\Psi_3\wedge \Psi_4,\nn
\end{align}
where $X_4$ is any  4-form on M$_8$. As we eluded to earlier, \eqref{eq:susy11done}-\eqref{eq:impliedcondtion} actually imply most of \eqref{eq:11dsusygen1}-\eqref{eq:11dsusygen3}, what is not implied can be extracted from \eqref{eq:11dsusygen2}, namely
\beq
6 e^{-A} m \star_8 K - 6 f \star_8 dA +2 \star_8 F_1+ \Psi_3 \wedge \star_8 F_4=0
\eeq
which together with \eqref{eq:notimpliedcondtion} and \eqref{eq:formconds} actually implies \eqref{eq:impliedcondtion}. Thus necessary and sufficient conditions for supersymmetry are
\begin{align}
& d(e^{2A} K)=0\label{eq:NS1},\\[2mm]
&d(e^{3A} f)- e^{3A} F_1-2 m  e^{2A} K=0,\label{eq:NS2}\\[2mm]
& d(e^{3A} \Psi_3)- e^{3A}(-\star_8 F_4+ f F_4)+2 m e^{2A} \Psi_4=0,\label{eq:NS3},\\[2mm]
&d (e^{2A} \Psi_4)- e^{2A} K \wedge F_4=0,\label{eq:NS4}\\[2mm]
&6 \star_8 dA-2 f \star_8 F_1+ \Psi_3\wedge F_4=0,\label{eq:NS5}\\[2mm]
&6 e^{-A} m \star_8 K - 6 f \star_8 dA+2 \star_8 F_1+\Psi_3 \wedge \star_8 F_4=0\label{eq:NS6}.
\end{align}
The first 3 of these look like 3 of the conditions in \cite{Martelli:2003ki}  (see (3.11), (3.15), (3.16) therein)  the final 3 appear quite different (they are more similar to the AdS$_2$ conditions  in \cite{Hong:2019wyi} derived in a similar fashion) and it is these that are much easier to work with for our purposes. One can show that these supersymmetry conditions imply the remaining equations of motion when  the magnetic parts of
\beq
dG=0,~~~~ d\star_{11}G+\frac{1}{2}G\wedge G=0
\eeq 
are imposed by hand - the electric components along with the rest of the EOM being implied by supersymmetry and these. Note that in the presence of localised sources the right hand side of these expressions should be appropriately modified.\\
~~\\
 The map relating our bi-linears to \cite{Martelli:2003ki} is
\begin{align}
e^{2A}&=e^{2\Delta},~~~~ds^2(\text{M}_8)= e^{2\Delta}ds^2(\tilde{\text{M}}_8) ,~~~ F_1= \tilde{f},~~~F_4 =e^{3\Delta} F,~~~ m=-\frac{1}{2} \tilde{m}\nn \\[2mm]
f&= \sin\zeta,~~~K= e^{\Delta}\cos\zeta \tilde{K},~~~ \Psi_3= - e^{3\Delta}\cos\zeta\phi,~~~ \Psi_4= e^{4\Delta} Y \label{eq:MSmap}
\end{align}
where we add tildes to objects defined in \cite{Martelli:2003ki} when we use the same symbol in our classification. Using this it is not hard to see that \eqref{eq:NS1}-\eqref{eq:NS3} can be directly mapped to (3.11), (3.15), (3.16)  of \cite{Martelli:2003ki}. For the final 3 conditions of  \cite{Martelli:2003ki} we need to derive some descendent conditions. First one can show that
\beq
(1-f^2)\Psi_4 = f K\wedge \Psi_3- \iota_K \star_8 \Psi_3,
\eeq
so given \eqref{eq:NS1} and \eqref{eq:NS4} we have
\beq
K\wedge d\left(\frac{e^{2A}}{1-f^2}\iota_K \star_8 \Psi_3\right)=0,
\eeq
which can be mapped to (\!\!\cite{Martelli:2003ki}, Eq.3.12). By combining $e^{2A}$\eqref{eq:NS3}$\wedge \Psi_4+$ $e^{3A}$\eqref{eq:NS4}$\wedge \Psi_3$, and using the identities in \eqref{eq:formconds} one finds
\beq
d(e^{5A}\Psi_3\wedge \Psi_4)+2 e^{4A}m \Psi_4\wedge \Psi_4=0,
\eeq
which can be mapped to (\!\!\cite{Martelli:2003ki}, Eq.3.13). Finally by combining $e^{3A}$\eqref{eq:NS6}$+\star_8$\eqref{eq:NS2}$-e^{3A}f$\eqref{eq:NS5}-$\Psi_3\wedge$\eqref{eq:NS3} we find
\beq
d\Psi_3\wedge \Psi_3+2 (1-f^2)\star_8 F_1-4 \star_8 df+16m e^{-A}\star_8 K=0.
\eeq
Using \eqref{eq:MSmap} on this equation we find it becomes
\beq
d\phi\wedge \phi \cos\zeta= 32 \tilde{m}\text{vol}_7+4 \hat{\star}_8 d\zeta-2 \cos\zeta \star_8 \tilde{f}
\eeq
where we define $\text{vol}_7=\hat{\star}_8\tilde{K}$. This corrects a numerical factor in the first term on the RHS of (\!\!\cite{Martelli:2003ki}, Eq.3.14), as well as the signs of the second 2 terms. We have checked that all M-theory solutions in this paper solve the supersymmetry constraints as we present them and fail to solve (\!\!\cite{Martelli:2003ki}, Eq.3.14) with the numerical factors given there - importantly this includes the embeddings of AdS$_3$ into AdS$_4\times$S$^7$ and AdS$_7\times$S$^4$ which are well known to be supersymmetic independent of our analysis here. 

\section{Bi-spinors on spheres}
\label{sec:BispSphere}
In this appendix we present bi-spinors on various spheres that appear in the computations of the main text. In order to do this it is helpful to consider the even and odd dimensional spheres separately, as the former can actually be presented in terms of the latter.
\subsection{Odd Spheres}
It is well known that odd dimensional spheres support a  Sasaki--Einstein structure. Specifically on an S$^{2n+1}$ one can define a 1-form $V_n$, 2-form $J_n$ and holomorphic $n$-form $\Omega_{(n)}$ such that
\beq\label{eq:SEcond}
 dV_n= 2 J_n,~~~~ d\Omega_{(n)}= i(n+1) V_n \wedge \Omega_{(n)},
\eeq
where this also holds for S$^1$, albeit with $dV_0=J_0=0$ which reflects the fact that there is no Kahler--Einstein base over which to define a cone in this case. For $n>0$ one can always parametise S$^{2n+1}$ as a U(1) bundle over $\mathbb{CP}^{n}$ such that the metric becomes
\beq
ds^2(\text{S}^{2n+1}) = (d\psi_n+ \eta_n)^2+ ds^2(\mathbb{CP}^{n}),~~~ d\eta_n= 2 J_n 
\eeq 
where $\partial_{\psi_n}$ is a U(1) isometry of period $2\pi$, $J_n$ is a Kahler form on $\mathbb{CP}^{n}$, $\eta_n$ has support on the base only  and the metric on $\mathbb{CP}^{n}$ can be defined recursively as
\beq\label{eq:cpnrecermet}
ds^2(\mathbb{CP}^{n})= d\theta_n^2+\sin^2\theta_n ds^2(\mathbb{CP}^{n-1})+\sin^2\theta_n\cos^2\theta_n(d\psi_{n-1}+ \eta_{n-1})^2,
\eeq
where $\eta_n= \sin^2\theta_n(d\psi_{n-1}+ \eta_{n-1})$.  For the lowest order case, $n=1$ we have
\beq
ds^2(\mathbb{CP}^{1})=\frac{1}{4}ds^2(\text{S}^2)= \frac{1}{4} (d\theta_1^2+ \sin\theta_1^2d\phi_1^2),~~~~ \eta_1= \cos\theta_1 d\phi_1.
\eeq 
Using these definitions it is simple to establish that  Ricci$(\mathbb{CP}^{n})=2(n+1) g(\mathbb{CP}^{n})$ and so Ricci$(\text{S}^{2n+1})=2n g(\text{S}^{2n+1})$, as it should for a unit radius sphere. In general the forms $(V,J,\Omega_n)$ are in fact not uniquely defined: First off clearly if $\Omega_n$ satisfies \eqref{eq:SEcond}, then it also does when multiplied by a constant phase. Second, modulo such constant phases  there are $2^{n+1}$ distinct forms $(V,J,\Omega_n)$ on S$^{2n+1}$ that solve \eqref{eq:SEcond}. To see this more clearly consider the case of S$^3$; here the metric can be arranged in two equivalent Hopf fibrations U(1) $\rightarrow$ S$^3\hookrightarrow~\mathbb{CP}^1$, namely
\beq\label{eq:S3rearangement}
\frac{1}{4} (d\theta_1^2+ \sin^2\theta_1 d\phi_1^2)+ (d\psi_1+ \cos\theta_1 d\phi_1)^2= \frac{1}{4}d\theta_1^2+ \sin^2\theta_1 d\psi_1^2+ (\frac{1}{2}d\phi_1+ \cos\theta_1 d\psi_1)^2,
\eeq
and as the metric is insensitive to the sign of $(V_n,J_n)$, this is suggestive of 4 distinct Sasaki Einstein structures on the 3-sphere, specifically
\beq
\begin{array}{c|c|c}
V_1&J_1& \Omega_{(1)}\\
\hline
\hline
\pm (d\psi_1+\frac{1}{2}\cos\theta_1d\phi_1)& \mp \frac{1}{4}\sin\theta_1d\theta_1\wedge d\phi_1& \frac{1}{2}e^{\pm2 i \psi_1}(\pm i d\theta_1+ \sin\theta_1 d\phi_1)\\
\hline
\pm (\frac{1}{2}d\phi_1+\cos\theta_1d\psi_1)&\mp \frac{1}{2}\sin\theta_1d\theta_1\wedge d\psi_1& -e^{\pm  i \phi_1}(\pm i \frac{1}{2}d\theta_1+ \sin\theta_1 d\psi_1)
\end{array}
\eeq
which all solve \eqref{eq:SEcond} for the same parametrisation of the metric. More generally the recursion relation of \eqref{eq:cpnrecermet} essentially ensures that as one moves from considering $n\to n+1$ the number of metric rearrangements of the type \eqref{eq:S3rearangement} one can perform doubles, giving the $2^{n+1}$ distinct Sasaki--Einstein structures on S$^{2n+1}$. We shall not present these explicitly as for our purposes it is more important that these $2^{n+1}$ structures are in one to one correspondence with the number of supercharges that the $(2n+1)$-sphere preserves.

\subsubsection{Killing spinors on S$^{2n+1}$} 
On the unit norm $(2n+1)$-sphere one can define two types of Killing spinor $\xi_{\pm}$ that solve the Killing spinor equations 
\beq\label{eq:oddKSE}
\nabla_{a}\xi_{\pm}= \pm \frac{i}{2} \gamma_a\xi_{\pm},
\eeq 
where $\gamma_a$ are gamma matrices on S$^{2n+1}$ with each of $\xi_{+}$ and $\xi_{-}$ preserving $2^{n}$ supercharges for $n>0$, or 2 when $n=0$, we will not consider this case further here as it does not feature in the main text. For the 3-sphere, when $n=1$, we can set our flat space gamma matrices equal to the Pauli matrices as $\gamma_a =\sigma_a$, then in Hopf fibration frame $e^a= (\frac{1}{2}d\theta_1,\frac{1}{2}\sin\theta_1 d\phi_1,d\psi_1+ \frac{1}{2} \cos\theta_1 d\phi_1)^a$ one finds that
\beq
\xi^{\text{S}^3}_+= e^{\frac{i}{2}\theta_1 \sigma_1}e^{\frac{i}{2}\phi_1 \sigma_3}\xi^0_+,~~~~\xi^{\text{S}^3}_-= e^{-\frac{i}{2}\psi_1 \sigma_3}\xi^0_-
\eeq 
where $\xi^0_{\pm}$ are arbitrary constant $d=2$ spinors that span the $4$ supercharges on the 3-sphere. For higher dimensional spheres one can exploit the recursion relation for the metric of $\mathbb{CP}^{n}$ \eqref{eq:cpnrecermet}. Decomposing $a=(\hat{a},\psi_{n-1},\beta_n,\psi_n)$, where $\hat{a}$ span $\mathbb{CP}^{n-1}$ one may define a frame on S$^{2n+1}$ as
\beq
e^{\hat{a}}= \sin\theta_n e^{\hat{a}}_{\mathbb{CP}^{n-1}},~~~e^{\psi_{n-1}}=\sin\theta_n\cos\theta_n(d\psi_{n-1}+ \eta_{n-1}),~~~e^{\theta_{n}}=d\theta_n,~~~e^{\psi_{n}}=(d\psi_n+ \eta_n),
\eeq
and a corresponding basis of gamma matrices 
\begin{align}
\gamma_{\hat{a}}&= \sin\theta_n\gamma^{\mathbb{CP}^{n-1}}_{\hat{a}}\otimes \sigma_1,~~~\gamma_{\psi_{n-1}}= \sin\theta_n\cos\theta_n\hat\gamma_{\mathbb{CP}^{n-1}}\otimes \sigma_1,\nn\\[2mm]
\gamma_{\theta_{n}}&=\mathbb{I}_{2(n-1)}\otimes \sigma_2,~~~\gamma_{\psi_{n}}=\mathbb{I}_{2(n-1)}\otimes \sigma_3,
\end{align}
where $\hat\gamma_{\mathbb{CP}^{n-1}}= \mathbb{I}_{2(n-2)}\otimes \sigma_3$ is the chirality matrix on $\mathbb{CP}^{n-1}$. In terms of this frame the Killing spinors on S$^{2n+1}$ are then given by the recursive expression
\begin{align}
\xi^{\text{S}^{2n+1}}_{\pm}&= e^{\frac{\psi_n}{2}(\slashed{J}_n\pm i \gamma_{\psi_{n}})}\Big(P^{\mathbb{CP}^{n-1}}_{\pm}\otimes e^{\pm i\beta_n\sigma_2}+P^{\mathbb{CP}^{n-1}}_{\mp}\otimes\mathbb{I}_2\Big)Q^{2n+1},\nn\\[2mm]
Q^{2n+1}&= \xi^{\text{S}^{2n-1}}_{+}\otimes q^0_-+\xi^{\text{S}^{2n-1}}_{-}\otimes q^0_+,~~~P^{\mathbb{CP}^{n-1}}_{\pm}=\frac{1}{2}(\mathbb{I}_{2(n-1)}\pm \hat\gamma_{\mathbb{CP}^{n-1}}),~~~ \sigma_3q^0_{\pm}=\pm q^0_{\pm},
\end{align}
where $q^0= q^0_++q^0_-$ is an arbitrary $d=2$ constant spinor - hence each of $\xi^{\text{S}^{2n+1}}_{\pm}$ preserve $2^n$ supercharges, for a total of $2^{n+1}$ on  $\text{S}^{2n+1}$.

\subsubsection{Bi-spinors on S$^{2n+1}$}
We established in the previous section that the Killing spinors on S$^{2n+1}$ $\xi_{\pm}$ each preserve $2^n$ supercharges, as such one can in general decompose them in an orthonormal basis as 
\beq
\xi_{\pm}= \sum_{a=1}^{2^n} c_{a,\pm}\xi_{a,\pm},~~~~ \xi^{\dag}_{a,\pm}\xi_{b,\pm}= \delta_{ab}
\eeq 
with each of $\xi_{a,\pm}$ preserving a single supercharge and $c_{a,\pm}$ arbitrary constants. Either by solving \eqref{eq:oddKSE} explicitly as in the previous section, or by making use of Fierz identities, one can then show that each supercharge defines bi spinors of the form
\begin{align}
\Phi_{a,\pm}=\xi_{a,\pm}\otimes \xi^{\dag}_{a,\pm} &= \frac{1}{2^n}(1\pm V_{a,\pm})\wedge e^{-i J_{a,\pm}},~~~~
\tilde{\Phi}_{a,\pm}=\xi_{a,\pm}\otimes \xi^{\dag c}_{a,\pm} =\frac{1}{2^n}(1\pm V_{a,\pm})\wedge \Omega_{a,\pm},\\[2mm]
\xi_{a,\pm}^c\otimes \xi^{c\dag}_{a,\pm}&=(-1)^{{n \, \text{deg}}}(\Phi_{a,\pm})^*,\qquad \qquad\qquad \, \, \, \, \xi_{a,\pm}^c\otimes \xi^{\dag}_{a,\pm}=(-1)^{n \, {\text{deg}}+\frac{n}{2}(n+1)}(\tilde{\Phi}_{a,\pm})^*\nn\label{eq:genbiodd}
\end{align}
 where we have defined the function deg which takes the form degree, assumed without loss of generality $|\xi_{a,\pm}|=1$ and defined Majorana conjugation as $\xi^{ c}= B_{2n+1} \xi^*$ with intertwiner such that $B_{2n+1} B_{2n+1}^*= (-1)^{\frac{n}{2}(n+1)}$, $B_{2n+1}\gamma_a B_{2n+1}^{-1}=(-1)^n \gamma^{*}_a$.   We additionally have that
\beq
\pm\frac{1}{n!}V_{a,\pm}\wedge (J_{a,\pm})^n =  \text{vol}(\text{S}^{2n+1}),
\eeq 
with $\text{vol}(\text{S}^{2n+1})$ the only SO(2n+2) invariant form on S$^{2n+1}$.
As the notation suggests these bi-spinors give rise to the $2^{n+1}$ independent Sasaki-Einstein structures mentioned earlier - we have dropped the $n$ subscript on the forms to ease notation. In the main text we are ultimately interested in solving differential bi spinor relations in 7 or 8 dimensions that will contain $(\Phi_{a,\pm}, \tilde{\Phi}_{a,\pm})$ as part of wedge products with other bi spinors. For a given $n$, all $(V_{a,\pm},J_{a,\pm}, \Omega_{a,\pm})$ obey the same differential conditions \eqref{eq:SEcond}, but  $(\Phi_{a,+},\tilde{\Phi}_{a,+})$ and $(\Phi_{a,-},\tilde{\Phi}_{a,-})$ do contain some sign differences. So, it should be clear that if our higher dimensional conditions are solved by  $(\Phi_{a,+},\tilde{\Phi}_{a,+})$, when $a=1$, then they are also solved for any $a$, but it need not necessarily follow from this that  $(\Phi_{a,-},\tilde{\Phi}_{a,-})$ also solve these higher dimensional conditions. As such, from here on we will drop the $a$ index in $\xi_{a,\pm}$ but keep the $\pm$, with the  understanding that $\xi_{\pm}$ is representative of any of the $\pm$ supercharges supported by S$^{2n+1}$, we thus define the bi-spinors  
\beq\label{eq:PhiPsi}
\Phi_{\pm}=\frac{1}{2^n}\xi_{\pm} \otimes \xi^{\dag}_{\pm} = (1 \pm V_{\pm})\wedge e^{-i J_{\pm}},~~~~\tilde{\Phi}_{\pm}=\frac{1}{2^n}\xi_{\pm} \otimes \xi^{\dag c}_{\pm} =(1 \pm V_{\pm})\wedge \Omega_{\pm,n}
\eeq
which should be understood as representing any of the $2^{n}$ supercharges contained in $\xi_{\pm}$. A word of caution is in order for the case of the 7-sphere where it is possible to take a Majorana basis for the killing spinor. When one does this, one ends up with $2^{2n}$, G$_2$ structures, but these can also be derived simply in terms of the above bi-linears.

\subsection{Even Spheres}\label{eq:evenspherebilinears}
Killing spinors on S$^{2(n+1)}$ $(n=0,1...)$ also obey a condition of the form \eqref{eq:oddKSE}, however  one can now define a chirality matrix in $2(n+1)$ dimensions $\hat\Gamma$ relating $\xi^{2(n+1)}_{\pm}$ as $\xi^{2(n+1)}_-=\hat\Gamma \xi^{2(n+1)}_+$, so that $\xi^{2(n+1)}_{\pm}$ are no longer independent. As such we shall take  Killing spinors on even dimensional spheres to obeys
\beq\label{eq:evenKSE}
\nabla_A \xi= \frac{i}{2}\Gamma_A \xi,
\eeq 
for $\Gamma_A$ a $2(n+1)$ dimensional gamma matrix - which has $2^{(n+1)}$ independent solutions each parameterising a distinct supercharge preserved on S$^{2(n+1)}$ - notice this is the same number that S$^{2n+1}$ preserves, this is no accident as each supercharge on an even-dimensional sphere can be mapped to a supercharge on an odd sphere of one less dimension by expressing S$^{2(n+1)}$ as a foliation of S$^{2n+1}$ over an interval. Explicitly if one takes the even sphere metric of unit radius to be
\beq
\label{eq:metric_even}
ds^2(\text{S}^{2(n+1)})= d\alpha^2+ \sin^2\alpha ds^2(\text{S}^{2n+1}),
\eeq 
which suggests splitting the gamma-matrix index as $A=(a,\alpha)$, then one define a basis for the Clifford algebra in $2(n+1)$ dimensions as  
\begin{align}\label{eq:splitgamma}
\Gamma_1&=\sin\alpha\sigma_1,~~~~\Gamma_{\alpha}= \sigma_2,~~~~ \text{for}~ n=0,\\[2mm]
\Gamma^{2n+1}_{a}&= \sin\alpha\gamma_a\otimes \sigma_1,~~~~ \Gamma_{\alpha}= \mathbb{I}\otimes \sigma_2,~~~\text{for}~n>0
\end{align}
with $\gamma_a$ gamma matrices on the unit $2n+1$-sphere and  $\sigma_i$ the Pauli matrices, we take the chirality matrix to be $\hat\Gamma= \mathbb{I}\otimes \sigma_3$.
With respect to this basis it is simple to show that \eqref{eq:evenKSE} is solved when $\xi$ is any linear combination of
\beq\label{eq:evenspinsol}
\varphi_{\pm}=\xi_{\pm}\otimes e^{\frac{i}{2}\alpha\sigma_2 }\theta_{\mp},
\eeq
with $\theta_{\pm}$ constant 2d spinors obeying $\sigma_3\theta_{\pm}=\pm \theta_{\pm}$  whenever $\xi_{\pm}$ solves \eqref{eq:oddKSE} on S$^{2n+1}$. Note that \eqref{eq:evenspinsol} also holds for $n=0$, i.e. for the 2-sphere - one merely needs to take the S$^1$ ``spinors'' to be the scalars $\xi_{\pm}= e^{\pm\frac{i}{2}\phi}$, where $\phi\sim \phi+2\pi$ spans S$^1$. As $\xi^{2n+1}_{\pm}$ each contain $2^{n}$ supercharges, we then have a total of $2^{n+1}$ on S$^{2(n+1)}$. To define the Majorana conjugate on S$^{2(n+1)}$ we must choose an intertwiner, there are actually two consistent with the intertwiner on S$^{2n+1}$
 that we parametrise as
\beq\label{eq:intertwiners}
B^{(1)}_{2(n+1)}= B_{2n+1}\otimes \sigma_1^{n+1},~~~~B^{(2)}_{2(n+1)}=(-1)^{n+1} i \hat\Gamma B^{(1)}_{2(n+1)}
\eeq
(the $B_{2n+1}$ factor should be dropped on S$^2$) which are unitary matrix such that  
\beq
B^{(k)}_{2(n+1)}B^{(k)*}_{2(n+1)}=(-1)^{\frac{n}{2}(n+1)+(k-1)(n+1)},~~~  B^{(k)}_{2(n+1)}\Gamma_A (B^{(k)}_{2(n+1)})^{-1}=(-1)^{n+k-1} \Gamma_A^*
\eeq
where the index $k=1,2$. Of course in principle it should make no physical difference which choice of intertwiner one takes in \eqref{eq:intertwiners}, however one choice may be better suited to embed a particular sphere into a higher dimensional solution. In terms of these definition we can then compute the bi-spinors on S$^{2(n+1)}$, fixing
\beq
\theta_+ = \left(\begin{array}{c} i\\ 0\end{array}\right),~~~~\theta_- = \left(\begin{array}{c} 0\\ 1\end{array}\right)
\eeq
without loss of generality, we find
\begin{align}\label{eq:evenbi}
\Phi_{\pm}&=\varphi_{\pm}\otimes \varphi_{\pm}^{\dag} = \frac{1}{2^{n+1}}\Big(1+ \sin\alpha_{\pm} V_{\pm}\wedge (\sin\alpha_{\pm}+ i d \sin\alpha_{\pm})\Big)\wedge e^{-i \sin^2\alpha_{\pm} J_{\pm}}\nn,\\[2mm]
\Phi_{\hat\Gamma\pm}&=\hat\Gamma\varphi_{\pm}\otimes \varphi_{\pm}^{\dag} = -\frac{1}{2^{n+1}}\Big(\cos\alpha_{\pm}+i d\cos\alpha_{\pm}\wedge(1+V_{\pm})\Big)\wedge e^{-i \sin^2\alpha_{\pm} J_{\pm}},\nn\\[2mm]
\tilde{\Phi}^{(1)}_{\pm}&=\varphi_{\pm}\otimes \varphi_{\pm}^{c_1\dag} =\frac{1}{2^{n+1}}\Big(\sin\alpha_{\pm}(1+V_{\pm})-i d\sin\alpha_{\pm}\Big)\wedge(\sin^n\alpha_{\pm}\Omega_{\pm}),\nn\\[2mm]
\tilde{\Phi}^{(1)}_{\hat\Gamma\pm}&=\hat\Gamma\varphi_{\pm}\otimes \varphi_{\pm}^{c_1\dag} =\frac{1}{2^{n+1}}i\Big(d\alpha_{\pm}+\sin\alpha_{\pm} V_{\pm}\wedge(i \cos\alpha_{\pm}+ d\cos\alpha_{\pm})\Big)\wedge(\sin^n\alpha_{\pm}\Omega_{\pm}),\nn\\[2mm]
\tilde{\Phi}^{(2)}_{\pm}&=\varphi_{\pm}\otimes \varphi_{\pm}^{c_2\dag} =\frac{1}{2^{n+1}}\Big(d\alpha_{\pm}+\sin\alpha_{\pm} V_{\pm}\wedge(i \cos\alpha_{\pm}- d\cos\alpha_{\pm}\Big)\wedge(\sin^n\alpha_{\pm}\Omega_{\pm}),\nn\\[2mm]
\tilde{\Phi}^{(2)}_{\hat\Gamma\pm}&=\hat\Gamma\varphi_{\pm}\otimes \varphi_{\pm}^{c_2\dag} =-\frac{1}{2^{n+1}}i \Big(\sin\alpha_{\pm}(-1+V_{\pm})-i d\sin\alpha_{\pm}\Big)\wedge(\sin^n\alpha_{\pm}\Omega_{\pm}),
\end{align}
where we define $\varphi^{c_k}= B_{2(n+1)}^{(k)}\varphi^*$ and
\beq
\alpha_{+}= \alpha,~~~~ \alpha_-=\pi-\alpha,
\eeq
which has absorbed all the sign changes that appear between the bi-linears following from $\xi_{+}$ and $\xi_{-}$  \eqref{eq:PhiPsi}.
The other possible bispinors can be computed in terms of these ones using these identities
\begin{align}
&\varphi_{\pm}^{c_k}\otimes \varphi_{\pm}^{c_k \dag} = (-)^{(n+k-1)\text{deg}} (\Phi_{\pm})^*, \qquad \qquad \qquad \, \, \, \hat\Gamma\varphi_{\pm}^{c_k}\otimes \varphi_{\pm}^{c_k \dag} = (-)^{(n+k-1)\text{deg}} (\Phi_{\hat\Gamma\pm})^*, \\
&\varphi_{\pm}^{c_k}\otimes \varphi_{\pm}^{\dag} = (-)^{(\frac{n}{2}+k-1)(n+1)+(n+k-1)\text{deg}} (\tilde{\Phi}^{(k)}_{\pm} )^*, \quad \hat\Gamma\varphi_{\pm}^{c_k}\otimes \varphi_{\pm}^{\dag} = (-)^{(\frac{n}{2}+k-1)(n+1)+(n+k-1)\text{deg}} (\tilde{\Phi}^{(k)}_{\hat\Gamma\pm} )^*, \nn
\end{align}
and also by recalling that, for generic spinors $\varphi_1,\varphi_2$ we have
\begin{equation}
\hat\Gamma\varphi_{1}\otimes \varphi_{2}^{\dag}=(-)^{\text{deg}}\varphi_{1}\otimes (\hat\Gamma\varphi_{2})^{\dag}, \qquad B^{(k)}_{2(n+1)} \hat{\Gamma} (B^{(k)}_{2(n+1)})^{-1} = (-)^{n+1} \hat\Gamma^* ,
\end{equation}
one may infer the remaining bilinears.\\
~\\
Finally we note that the volume form of S$^{2(n+1)}$ is given by 
\beq
\label{eq:volume_even}
\frac{1}{n!}d\alpha_{\pm}\wedge(\sin\alpha_{\pm} V_{\pm})\wedge (\sin^2\alpha_{\pm} J_{\pm})^n = \sin^{2n+1}\alpha \,  d\alpha\wedge\text{vol}(\text{S}^{2n+1})=\text{vol}(\text{S}^{2(n+1)}).
\eeq
Again we are ultimately interested in solving higher dimensional bi spinor relations that will contain the S$^{2(n+1)}$ bi-spinors. As for a given $n$, $(\alpha_{\pm},V_{\pm},J_{\pm},\Omega_{\pm})$ all act in the same way under exterior differentiation, and this time give rise to $\text{vol}($S$^{2(n+1)})$ (which can appear in the fluxes of a supergravity solution) in the same fashion, it actually makes no difference if we choose $+$ or $-$ in \eqref{eq:evenbi} so we drop this index in the main text for even spheres. 

\section{Lifting isometries}\label{sec:lifting}
In this section we will summarize some results about fiber-bundles as presented in \cite{Alessandrobook}. We will assume that the fibers of our Riemannian submersion are totally geodesics, so that using Hermann theorem we get that they are isometric and the bundle is a $G$-bundle where $G$ is the isometry group of the fiber. Let's parameterize such a Lie algebra by the following set of Killing vectors on the fiber
\begin{equation}
K_F^a = K_F^{a i} \partial_{y_i} , \qquad [K_F^a,K_F^b] = f^{ab}_{\quad c} K_F^c
\end{equation}
where $\{y_i\}$ are coordinates on the fiber. At the end of this section we will see some explicit examples.

In order to define a covariant exterior derivative $D$ on the whole fiber-bundle we have to introduce the connection $A^i = A_a K_F^{ai}$, which is nothing but a one-form on the base which takes value in the Lie algebra of $G$. Specifically if $x^m$ are the coordinates on the base then $A_{a}=A_{am}(x) d x^m$. $D$ acts on the fiber coordinates as
\begin{equation}
\label{eq:fiber_coords}
D y^i = d y^i + A_a K_F^{a i} \, 
\end{equation}
while it acts on the base coordinate as the usual external derivative.
Using these definitions we can write the fiber-bundle metric as
\begin{equation}
d s^2 = (g_B)(x)_{mn} d x^m d x^n   + (g_F)(y)_{ij} D y^i D y^j \, ,
\end{equation}
where $g_B$ is the metric on the base while $g_F$ the one on the fiber; recall that $K_F^a$ are Killing vectors of $ (g_F)_{ij} d y^i d y^j$, but not necessarily of $d s^2$.

Let us now denote $K_B$ to be a Killing vector of the base metric $(g_B)_{mn}(x) d x^m d x^n$. We would like to understand under which conditions such an isometry can be uplifted to a symmetry of the whole space. We will be agnostic about the form of the lift, and we will simply assume that it is $K = K_B-\lambda^i \partial_{y^i}$ where $\lambda^i = \lambda^i(x,y)$.
What we have to calculate is therefore the condition under which
\begin{equation}
\mathcal{L}_{K} d s^2 = 0.
\end{equation}
Since the base metric is already isometric we have
\begin{equation}
\begin{split}
\mathcal{L}_{K} d s^2 
&= 
 -\lambda^k \partial_{y^k} ((g_F)_{ij})D y^i D y^j+ 2 (g_F)_{ij}  D y^j \mathcal{L}_{K}(D y^i) .
\end{split}
\end{equation}
Let us now expand the final term of this expression; using $\mathcal{L}_V = \iota_V d + d \iota_V$, we find
\begin{equation}
\begin{split}
\mathcal{L}_{K}(D y^i) =& \mathcal{L}_{K_B-\lambda^k \partial_{y^k}} ( d y^i + A_a K_F^{a i}) = -d \lambda^i +  \mathcal{L}_{K_B} A_a K_F^{ai}-A_a \lambda^k \partial_{y^k}K_F^{ai} \\
=& \mathcal{L}_{K_B} A_a K_F^{ai} - \partial_{x^m} \lambda^i d x^m - A_a (\lambda^k \partial_{y^k}K_F^{ai} -K_F^{ak} \partial_{y^k} \lambda^i)- \partial_{y^k} \lambda^i D y^k .
\end{split}
\end{equation}
Using this result we get 
\begin{equation}
\begin{split}
\mathcal{L}_{K} d s^2 = &
-\mathcal{L}_{\lambda^k \partial_{y^k}} ((g_F)_{ij})D y^i D y^j\\
&+ 2 (g_F)_{ij}  D y^j (\mathcal{L}_{K_B} A_a K_F^{ai} - \partial_{x^m} \lambda^i d x^m - A_a (\lambda^k \partial_{y^k}K_F^{ai} -K_F^{ak} \partial_{y^k} \lambda^i)) ;
\end{split}
\end{equation}
since the first line is independent of the last, in order to make it vanishes we have to impose that $\lambda^k \partial_{y^k}$ is a Killing vector for the fiber metric, which means that $\lambda^k = \lambda_a(x) K_F^{ak}$. 
The remaining part of the metric Lie derivative reads
\begin{equation}
\begin{split}
\mathcal{L}_{K} d s^2 =& 2 (g_F)_{ij}  D y^j (\mathcal{L}_{K_B} A_a K_F^{ai} - d (\lambda_a) K_F^{ai}  - A_a \lambda_b (K_F^{bk} \partial_{y^k}K_F^{ai} -K_F^{ak} \partial_{y^k} K_F^{bi})) \\
=& 2 (g_F)_{ij}  D y^j (\mathcal{L}_{K_B} A_a - d \lambda_a  - f^{bc}{}_a \lambda_b  A_c ) K_F^{ai} \, .
\end{split}
\end{equation}
So what we get is that $K_B$ can be uplifted to an isometry of the fiber-bundle if and only if $\mathcal{L}_{K_B}$ acts on $A_a$ as a gauge transformation:
\begin{equation}
\label{eq:isometries_cond}
\mathcal{L}_{K_B} A_a = d \lambda_a  + f^{bc}{}_a \lambda_b  A_c \, .
\end{equation} 
%
 
It is not hard to show that this argument can in-fact be generalised to metrics of the form
\beq
ds^2=   e^{2C_1(x,y)}(g_B)(x)_{mn} d x^m d x^n   + e^{2C_2(x,y)}(g_F)(y)_{ij} D y^i D y^j,~~~ Dy_i = dy_i + A_a K_F^{a i}
\eeq
where in addition to \eqref{eq:isometries_cond}, one should also impose $\mathcal{L}_K C_{1,2}=0$, so that these functions are invariants of the lifted isometry.\\
~\\
In the following we will consider two particularly interesting cases. 
 
 \subsection{Abelian case}
 In this case we have that the condition \eqref{eq:isometries_cond} can be rephrase as the statement that the field $F = d A$ is a singlet under the action of the base isometries, indeed
 \begin{equation}
 \label{eq:AbelianCase}
 0=d \mathcal{L}_{K_B} A = \mathcal{L}_{K_B} F .
 \end{equation}
 
\subsection{Fibration of coset manifolds and Lie groups}\label{sec:fibcolie}
 
Let's start our discussion by considering a group manifold $G$. On a group manifold we have that the isometry group $G_L \times G_R$ is generated by the Killing vectors $K^a_{L,R}$ defined as the dual of the right- and left-invariant one form\footnote{Notice that in this notation $(K^a_F)_{L}$ generates left translations but it is right-invariant, and vice-versa.}:
\begin{equation}
\omega^L_a = -\text{Tr}(t_a g^{-1} d g ) , \qquad \omega^R_a = -\text{Tr}(t_ad g g^{-1}), \qquad K^a_{L,R} \cdot \omega^{R,L}_b = \delta^a_b, \qquad \omega^R_a = D_a{}^b \omega^L_b
\end{equation}
where $t_a \in \text{Lie}(G)$\footnote{We are using the convention $\delta_{ab} =  -\text{Tr}(t_at_b)$.} and $D_{ab}(g)$ is the adjoint action of $G$:  $D_{ab}(g) = -\text{Tr}(g^{-1}t_a g  t_b )$ .

Now let's consider a coset defined by the right quotient M$=G/H$, again we can define left-invariant forms $\omega^L_\alpha$, where $\alpha$ is an index in the lie algebra of $G/H$, using the restriction of the left action of $G$ to $G/H$. The isometry group now reduces to $G$ and it is generated by the vectors $K^a$
\begin{equation}
K^a = D^{a \alpha} (\omega^L)_\alpha^\mu \partial_{\mu} \, , \qquad K^a \cdot \omega_b = \delta^a_b
\end{equation} 
where the forms $\omega_a$ dual to the Killing vectors are the Maurer-Cartan forms for the coset. Notice that even if $K^a$ are defined starting from a the right-invariant Killing vectors inherited from the Lie group, the presence of the adjoint action makes them respect the same algebra of the left ones, i.e. with an opposite sign in front of the structure constant. 

 Let's consider now two identical coset manifold M$_x$ and M$_y$ of the same group manifold $G$, so that they preserve (at least) an isometry group equal to $G$. Now suppose we want to fiber M$_y$ over M$_x$ so that part the isometries of the base are preserved. In order to do so let's consider, as in \eqref{eq:fiber_coords}, the following coordinates on the fiber
 \begin{equation}
 D y^i = d y^i + A_a K^{a i}_y \, ,
 \end{equation}
and let's take the connection $A_a$ to be proportional to the Maurer-Cartan forms over $M_x$, i.e. $A_a= c \omega_a$, with $c$ constant\footnote{Really we mean constant with respect to the fiber bundle coordinates. If the bundle is embedded into some large space then allowing $c$ to depend on the  remaining coordinates changes nothing of what follows.}, so that
\begin{equation}
\label{eq:coset_lie_der}
\mathcal{L}_{K^a_x} A_b = -f_b{}^{ac} A_c .
\end{equation}
Now, if we want to uplift $\mathcal{L}_{K^a_x}$ we have to define  Killing vectors $K^a$ such that $\mathcal{L}_{K^a}=\mathcal{L}_{K^a_x - \lambda_d K^d_y}$ where $\lambda_d$ has to satisfy \eqref{eq:isometries_cond}. If we assume that $\lambda_d$ is constant then this is equivalent to
\begin{equation}
\mathcal{L}_{K^a_x} A_b = f^{dc}{}_b \lambda_d A_c.
\end{equation}
Comparing this last equation with \eqref{eq:coset_lie_der} we get that we have a solution for every $K^a_x$ if $\lambda_d =-\delta^a_d$ provided that the Lie group $G$ is semisimple (and therefore $f^{ab}{}_c$ are completely antisymmetric in a certain basis). Thus we get that the isometry, spanned by $K^a=K^a_x+K^a_y$ is nothing but the diagonal action of $G$ acting on both M$_x$ and M$_y$. Notice that we can perform an almost identical computation by considering the base manifold as the Lie group $G$ and $A_a$ the left- or right-invariant forms on it.

A particularly interesting case is when M$=$S$^k$ since every sphere can be obtained as a quotient (and we have also the particular case S$^3\!=$SU(2)). In this case it is convenient to express the algebra index $^a$ as a couple of antisimmetrised indices $^{[mn]}$ on the embedding coordinates of the spheres in $R^{k+1}$. Using this reparametrisation we have
\begin{equation}
K^{[mn] i} = g^{ij} x^{[m} \partial_j x^{n]} , \qquad A_{[mn]} =  x_{[m} d x_{n]} .
\end{equation}
In the particular case $k=2$ one usually rewrite this couple of indices back to a single one using the three-dimensional Levi-Civita symbol.

\section{On the Sp(2)$\times$Sp(1) preserving squashed S$^7$}
\label{Appendix C}

In this appendix we discuss the fibration of S$^3$ over S$^4$ that preserves an Sp(2)$\times$Sp(1) isometry. This manifold can be viewed as a squashed S$^7$, and was introduced in \cite{Awada:1982pk} in the context of AdS$_4$ vacua of M-theory with  weak G$_2$-structure internal space. The metric can be expressed in terms of two sets of SU(2) left invariant forms $(L^i_1,L^i_2)$ $i=1,2,3$ which obey
\beq
dL^i_{1,2}= \frac{1}{2}\epsilon_{ijk}L^j_{1,2}\wedge L^k_{1,2}
\eeq
as
\beq\label{eq:squashedS7met}
ds^2= \frac{1}{4}\bigg[d\alpha^2+ \frac{1}{4}\sin^2 \alpha(L_1^i)^2+ \lambda^2\big(L_2^i- \cos^2 \left(\frac{\alpha}{2}\right) L_1^i\big)^2\bigg]
\eeq   
where $\lambda$ is a constant squashing parameter\footnote{When embedded into a higher dimensional space $\lambda$ can become a function of the co-dimensions of the squashed S$^7$.} with the unit radius round 7-sphere recovered when $\lambda=1$ \footnote{The fibration becomes topologically trivial in this limit, which is made more obvious if on rearranges \eqref{eq:squashedS7met} with $\lambda=1$ as
\beq
 \frac{d\alpha^2}{4}+\frac{1}{4}\cos^2\left(\frac{\alpha}{2}\right) ( L_1^i)^2+\frac{1}{4}\sin^2\left(\frac{\alpha}{2}\right) (L_2^i-L_1^i)^2.
\eeq
There exists an SU(2) transformation, mapping $L^i_2\to L_2^i-L_1^i$, so the fibration is clearly trivial.}.  The Killing vectors corresponding to the Sp(2)$\times$Sp(1) isometry on the S$^4\times$S$^3$ fibration can be constructed in the fashion of the previous section, in terms of the isometries of the base S$^4$ and fiber S$^3$, i.e. Sp(2) $\equiv$ SO(5) = SO(5)/SO(4)$\times$SO(4) and SO(4)=SU(2)$_L\times$SU(2)$_R$ respectively.  Choosing a specific parameteriation for the invariant forms 
\beq
L^1_a+i  L^2_a= e^{i \psi_a}(i d\theta_a+ \sin\theta_a d\phi_a),~~~~ L^3_a= d\psi_a+ \cos\theta_a d\phi_a,~~~~a=1,2,
\eeq
which is compatible with the following complex emedding coordinates for the 7-sphere
\begin{align}\label{eq:sp4sp2emeddingcomplex}
Z^1&= e^{\frac{i}{2}(\phi_2-\psi_1)}\sin\left(\frac{\alpha}{2}\right)\sin\left(\frac{\theta_2}{2}\right),~~Z^2= i e^{-\frac{i}{2}(\phi_2+\psi_1)}\sin\left(\frac{\alpha}{2}\right)\cos\left(\frac{\theta_2}{2}\right),\nn\\[2mm]
Z^3&= e^{\frac{i}{2}(\phi_1+\phi_2- \psi_1-\psi_2)}\cos\left(\frac{\alpha}{2}\right)(e^{i\psi_1}\cos\left(\frac{\theta_1}{2}\right)\sin\left(\frac{\theta_2}{2}\right)-e^{i\psi_2}\cos\left(\frac{\theta_2}{2}\right)\sin\left(\frac{\theta_1}{2}\right)),\nn\\[2mm]
Z^4&= i e^{-\frac{i}{2}(\phi_1-\phi_2+ \psi_1+\psi_2)}\cos\left(\frac{\alpha}{2}\right)(e^{i\psi_1}\sin\left(\frac{\theta_1}{2}\right)\sin\left(\frac{\theta_2}{2}\right)+e^{i\psi_2}\cos\left(\frac{\theta_1}{2}\right)\cos\left(\frac{\theta_2}{2}\right))
\end{align}
the Killing vectors of the SO(4) isometries appearing in the base and fibre are
\begin{align}
K^{1L}_a+ i K^{2L}_a&= e^{i\phi_a} \left(i \partial_{\theta_a}+\frac{1}{\sin\theta_a}\partial_{\psi_a}-\frac{\cos\theta_a}{\sin\theta_a} \partial_{\phi_a} \right),~~~K^{3L}_a= -\partial_{\phi_a},\label{eq:isosqushed1}\\[2mm]
K^{1R}_a+ i K^{2R}_a&= e^{i\psi_a}\left(i \partial_{\theta_a}+\frac{1}{\sin\theta_a}\partial_{\phi_a}-\frac{\cos\theta_a}{\sin\theta_a} \partial_{\psi_a} \right),~~~K^{3R}_a= \partial_{\psi_a}\label{eq:isosqushed2}
\end{align} 
while for the SO(5)/SO(4) Killing vectors on the base  we have
\beq
K^A_{SO(5)/SO(4)}= - (\mu_A\partial_{\alpha}+\cot\alpha \partial_{y_i}\mu_A g_3^{ij} \partial_{y_j}), ~~~A=1,...,4 \label{eq:isosqushed3}
\eeq
where $\mu_A$ are embedding coordinates for the 3-sphere spanned by $K^i_1$, $g_3^{ij}$ is its metric and $y_i= (\theta_1,\phi_1,\psi_1)_i$, specifically we have taken
\beq
\mu_A= \bigg(\sin\left(\frac{\theta_1}{2}\right)\cos\left(\frac{\phi_-}{2}\right),~\sin\left(\frac{\theta_1}{2}\right)\sin\left(\frac{\phi_-}{2}\right),~\cos\left(\frac{\theta_1}{2}\right)\cos\left(\frac{\phi_+}{2}\right),~-\cos\left(\frac{\theta_1}{2}\right)\sin\left(\frac{\phi_+}{2}\right)\bigg)_A,\nn
\eeq
where $\phi_{\pm}=\phi_1\pm\psi_1$.\\
~~\\
In the full space \eqref{eq:squashedS7met} some of the isometries \eqref{eq:isosqushed1}-\eqref{eq:isosqushed3} are broken, some are unaffected and some get lifted. Clearly as we fiber in terms of the left invariant forms, the Killing vectors \eqref{eq:isosqushed1} are still preserved by the fiber bundle. As 6 of the 7 directions of  \eqref{eq:squashedS7met} form an S$^3\times$S$^3$ fibration of the type described in appendix \ref{sec:fibcolie}, the base SO(3)$_R$ is lifted as
\beq
K^i_D=K^{iR}_1+ K^{iR}_2.
\eeq
which is a diagonal, the anti diagonal $K^{iR}_1- K^{iR}_2$ is broken. The base SO(5)/SO(4) on the other hand is lifted in a new fashion as
\beq\label{eq:colift}
\hat K^A_{SO(5)/SO(4)}=K^A_{SO(5)/SO(4)}+ \cot\left(\frac{\alpha}{2}\right)y_B(\kappa_{A})^B_i K^{i R}_2
\eeq
where
\beq
\kappa_1=\left(\begin{array}{ccc}0&0&0\\0&0&-1\\0&-1&0\\-1&0&0\end{array}\right),~~~\kappa_2=\left(\begin{array}{ccc}0&0&1\\0&0&0\\-1&0&0\\0&1&0\end{array}\right),~~~\kappa_3=\left(\begin{array}{ccc}0&1&0\\1&0&0\\0&0&0\\0&0&-1\end{array}\right),~~~\kappa_4=\left(\begin{array}{ccc}1&0&0\\0&-1&0\\0&0&1\\0&0&0\end{array}\right)\nn.
\eeq
It is  not hard to confirm that \eqref{eq:colift} does indeed obey \eqref{eq:isometries_cond} as required. We note 
 as a possible point of interest that the matrices $\kappa_A$ give rise to a 3d basis of  $\mathfrak{so}$(3) as
\beq
\kappa^T_A\kappa_B= \left(\begin{array}{cccc}\mathbb{I}_3&\Lambda_3&-\Lambda_2&\Lambda_1\\
-\Lambda_3&\mathbb{I}_3&\Lambda_3&\Lambda_2\\
\Lambda_2&-\Lambda_1&\mathbb{I}_3&\Lambda_3\\
-\Lambda_1&-\Lambda_2&-\Lambda_3&\mathbb{I}_3\end{array}\right)_{AB},~~~ [\Lambda_i,\Lambda_j]= -\epsilon_{ijk}\Lambda_k.
\eeq
The Killing vectors spanning Sp(2)$\times$Sp(1) on the squashed 7-sphere are then
\beq
\text{Sp(2)}:~ (K^{iL}_1,K^i_D,\hat K^A_{SO(5)/SO(4)}),~~~~ \text{Sp(1)}:~ K^{iL}_2.
\eeq

\subsection{Killing spinors on the round S$^7$}\label{eq:sp2sp1killingspinor}
In the obvious frame that \eqref{eq:squashedS7met} suggests (i.e. $e^{1}= \frac{1}{2}d\alpha,~ e^{2,3,4}=\frac{1}{4}\sin\alpha L^{1,2,3},...)$, the corresponding round  7-sphere (i.e. \eqref{eq:squashedS7met} in the $\lambda=1$ limit) Killing spinors are given by
\begin{align}\label{eq:squahedS7spinor}
\xi_\pm&={\cal M}_{\pm}\xi^0_{\pm},\\[2mm]
{\cal M}_{\pm}&=e^{\frac{\alpha}{4}(\pm i\gamma_1+Y)}e^{\mp i\frac{\psi_1}{2}\gamma_7P_{\mp}}e^{\mp i\frac{\theta_1}{2}\gamma_6 P_{\mp}}e^{\mp i\frac{\phi_1}{2}\gamma_7 P_{\mp}}e^{\frac{\psi_2}{4}(\pm i\gamma_7+X)}e^{\frac{\theta_2}{2}(\gamma_{13}P_+\pm \gamma_6 P_{\pm})}e^{\frac{\phi_2}{2}(\gamma_{14}P_+\pm i \gamma_7P_{\pm})}\nn
\end{align}
where $\gamma_a$ are a basis of flat 7d gamma matrices, specifically 
\beq
\gamma_1=\sigma_1\otimes\mathbb{I}_2\otimes \mathbb{I}_2,~~~
\gamma_{2,3,4}=\sigma_2\otimes\sigma_{1,2,3}\otimes \mathbb{I}_2~~~\gamma_{5,6,7}=\sigma_3\otimes\mathbb{I}_2\otimes \sigma_{1,2,3} \, ,
\eeq
$\xi^0_{\pm}$ are unconstrained constant spinors and
\beq
P_{\pm}=\frac{1}{2}(\mathbb{I}_4\pm \gamma_{1234}),~~~X=\gamma_{14}-\gamma_{23}-\gamma_{56},~~~~Y=-\gamma_{25}-\gamma_{36}-\gamma_{47} .
\eeq
 We remind the reader that $\xi_{\pm}$ solve the Killing spinor equations $\nabla_{a}\xi_{\pm}= \pm \frac{i}{2}\gamma_a \xi_{\pm}$. Note that this is not the spinor preserved on the internal space of the AdS$_4$ squashed 7-sphere solution in \cite{Awada:1982pk}, this preserves just $\mathcal{N}=1$ supersymmetry and solves the equation $\nabla_{a}\xi= s \frac{i}{2}\gamma_a \xi$ for $\lambda^{-2}=5$, $s^2\lambda^{-2}= 9$ only, it is in fact constant in the above frame.\\
~\\
We want to establish the transformation properties of the Killing spinors of this parameterisation of the 7-sphere under the Killing vectors which are preserved by the squashing. It is not hard to establish that for any parametrisation of the 7-sphere, the 8 independent $\pm$ Killing spinors each transform in the \textbf{8} of SO(8), we thus define
\beq
\xi^I_{\pm}= {\cal M_{\pm}} \hat\eta^I,~~~ \hat\eta^I=\sum_J\delta^{IJ}
\eeq 
Interestingly the two octuplets $\xi^I_{\pm}$ behave quite differently under the squashing. For $\xi^I_{+}$ we have \textbf{8}$\to$\textbf{3}$\oplus$\textbf{5}. To see this it useful to define the following triplet and quintuplet
\begin{align}
\xi^i_3= \frac{1}{\sqrt{2}}\left(\begin{array}{c}i(-\xi^5_++\xi^8_+)\\
\xi^5_++\xi^8_+\\
i(\xi^6_++\xi^7_+)\\
\end{array}\right)^i,~~~ \xi^{\alpha}_5=\frac{1}{\sqrt{2}} \left(\begin{array}{c}-i(\xi^1_++\xi^4_+)\\
\xi^1_+-\xi^4_+\\
-i(\xi^2_+-\xi^3_+)\\
\xi^2_++\xi^3_+\\
\xi^6_+-\xi^7_+
\end{array}\right)^{\alpha},
\end{align} 
which one can check are Majorana. One can establish what symmetries these objects are charged under by computing the spinoral Lie derivative with respect to the appropriate Killing vectors: For a spinor $\eta$ this is defined as 
\beq
{\cal L}_{K}\eta= K^a\nabla_{a}\eta+ \frac{1}{4}\nabla_a K_b\gamma^{ab}\eta,
\eeq
where $K$ is a Killing vector. We find the following transformation properties for the Killing vectors preserved by the squashing
\begin{align}
\mathcal{L}_{K^{iL}_2}\xi^j_3&= -\epsilon_{ijk}\xi^k_3,~~~\mathcal{L}_{K^{iL}_1}\xi^j_3=\mathcal{L}_{K^{i}_D}\xi^j_3=\mathcal{L}_{K^{A}_{SO(5)/SO(4)}}\xi^j_3=0,~~~~\mathcal{L}_{K^{iL}_2}\xi^{\alpha}_5=0\\[2mm]
\mathcal{L}_{K^{iL}_1}\xi^{\alpha}_5&=\left(\begin{array}{c|c}T^{iL}_1 &0_4\\
\hline
 0^T_4&0\end{array}\right)^{\alpha}_{~\beta}\xi_5^{\beta},~~\mathcal{L}_{K^{i}_D}\xi^{\alpha}_5=\left(\begin{array}{c|c}T^{i}_D &0_4\\
\hline
 0^T_4&0\end{array}\right)^{\alpha}_{~\beta}\xi_5^{\beta},~~\mathcal{L}_{K^{A}_{SO(5)/SO(4)}}\xi^{\alpha}_5=\left(\begin{array}{c|c}0_{4\times 4} &-c_A\\
\hline
 c_A^T&0\end{array}\right)^{\alpha}_{~\beta}\xi_5^{\beta},\nn
\end{align}
where the A$^{\text{th}}$ entry of $c_A$ is 1 with the rest zero and where 
\beq
T^{iL}_1=\frac{i}{2}(\sigma_1\otimes \sigma_2,~ \sigma_2\otimes\mathbb{I}_2,~\sigma_3\otimes \sigma_2)^i,~~~T^{i}_D=-\frac{i}{2}(\sigma_2\otimes \sigma_1,~\sigma_2\otimes \sigma_3,~\mathbb{I}_2\otimes \sigma_2)^i
\eeq
such that $(T^{iL}_1, T^{i}_D)$ span $\mathfrak{so}$(4) (which $\xi^5_5$ is a singlet under) and together with the matrices involving $c_A$  give a fundamental representation of $\mathfrak{so}$(5). This confirms that $\xi^j_3$ and $\xi^{\alpha}_5$ indeed transform in the fundamentals of SO(3) and SO(5) respectively, and are singlets under the opposing groups. We thus conclude that the $\xi^I_+$ Killing spinors are relevant for solutions preserving ${\cal N}=(3,0)$ and $(5,0)$, but not $(8,0)$.\\
~\\
The $\xi^I_-$ spinors behave quite differently, they do not decompose into separate multiplets instead they transform in the (\textbf{4},\textbf{2}) of $\mathfrak{sp}$(2)$\oplus \mathfrak{sp}$(1). This fact is made considerably more obvious by defining
\beq
\xi^{I}_{(4,2)}= \left(\begin{array}{c}\xi^1_-\\ \xi^2_- \\ \xi^3_-\\ \xi^4_-\\
-\xi^5_-\\-\xi^7_-\\ -\xi^6_-\\ -\xi^8_-\end{array}\right)^I
\eeq 
which transforms under the Killing vectors the squashing preserves as
\begin{align}
\mathcal{L}_{K^{iL}_2}\xi^{I}_{(4,2)}&= \big[\mathbb{I}_4\otimes (\frac{i}{2}\sigma_i)\big]^I_{~J}\xi^{J}_{(4,2)},~~~\mathcal{L}_{K^{iL}_1}\xi^{I}_{(4,2)}= \big[(0_{2\times 2}\oplus \frac{i}{2}\sigma_i)\otimes\mathbb{I}_2\big]^I_{~J}\xi^{J}_{(4,2)},\nn\\[2mm]
\mathcal{L}_{K^{i}_D}\xi^{I}_{(4,2)}&= \big[( \frac{i}{2}\sigma_i\oplus 0_{2\times 2})^*\otimes\mathbb{I}_2\big]^I_{~J}\xi^{J}_{(4,2)},~~~\mathcal{L}_{K^{A}_{SO(5)/SO(4)}}\xi^{I}_{(4,2)}= \big[\Sigma_{SO(5)/SO(4)}^A\otimes\mathbb{I}_2\big]^I_{~J}\xi^{J}_{(4,2)}
\end{align}
for
\beq
\Sigma_{SO(5)/SO(4)}^A=\frac{i}{2}(-\sigma_2\otimes \sigma_2,~\sigma_2\otimes \sigma_1,~-\sigma_1\otimes \mathbb{I}_2,~\sigma_2\otimes \sigma_3)^A.
\eeq
Clearly $\frac{i}{2}\sigma_i$ spans $\mathfrak{sp}(1)=\mathfrak{su}(2)$, while $((\frac{i}{2}\sigma_i)^*\oplus 0_{2\times 2},~0_{2\times 2}\oplus \frac{i}{2}\sigma_i,~\Sigma_{SO(5)/SO(4)}^A)$  give 10 independent anti-Hermitian 4$\times$4 matrices formed of anti-Hermitian 2$\times$2 blocks, so span $\mathfrak{sp}(2)$ - as every component of $\xi^{I}_{(4,2)}$ transforms  under both these groups we conclude that it  does indeed transform in the  (\textbf{4},\textbf{2}) of $\mathfrak{sp}$(2)$\oplus \mathfrak{sp}$(1). As such we see that only the $\xi^I_-$ Killing spinors are suitable for realising $\mathcal{N}=(8,0)$ on the squashed 7-sphere.

 \subsection{Generalized Killing spinors on the squashed S$^7$}\label{sec:andreaextras}
 The discussion in the previous section perfectly matches the group theoretical argument in equation \eqref{eq:branchSpin8-Sp2}, which predicted that there is a Killing spinor multiplet on the S$^7$ that transforms in the (\textbf{4},\textbf{2}), while the other two multiplets are given by acting on this one with the squashed S$^7$ invariant forms $\Lambda^0_4,\tilde{\Lambda}_0^4$, as already discussed in section \ref{sec:sp2sp1cases}.
 
 Notice however that such multiplets cannot be made of Killing spinors on the squashed S$^7$ since such condition would force the manifold to be Einstein, which is not in general the case. 
 What we have to do is therefore to generalize the Killing spinor equation in such a way that it preserves the multiplet, i.e. if $\xi_-$ solves the equation also $\mathcal{L}_{K}\xi_-$ is a solution, where $K$ is a generic Killing vector of Sp(2)$\times$Sp(1). Requiring this property forces us to use just invariant forms on the squashed S$^7$ to deform the equation, which in general reads:
 \begin{equation}
 \label{eq:genKillingSpinor}
 \begin{split}
  \nabla_a \xi_- =& - \frac{i}{2} \Big[ \gamma_a + \left(a_1 \Lambda^0_4- \frac{a_2}{2} \tilde{\Lambda}^0_4 \right)\gamma_a + \gamma_a  \left(a_3 \Lambda^0_4- \frac{a_4}{2} \tilde{\Lambda}^0_4 \right) \\
  +& \frac{1}{2} \left(a_5 \Lambda^0_4 \gamma_a\tilde{\Lambda}_0^4+a_6 \tilde{\Lambda}^0_4 \gamma_a \Lambda^0_4 + a_7 \tilde{\Lambda}^0_4 \gamma_a \tilde{\Lambda}^0_4+ a_8 \Lambda^0_4 \gamma_a \Lambda^0_4 \right) \Big] \xi_- ,
 \end{split}
 \end{equation}
 since we don't have to consider higher powers of the invariant forms thanks to the relations
 \begin{equation}
 (\Lambda^0_4)^2 = 1 , \quad \Lambda^0_4 \tilde{\Lambda}^0_4 = \tilde{\Lambda}^0_4 \Lambda^0_4 = \tilde{\Lambda}^0_4 , \quad (\tilde{\Lambda}_0^4)^2 = 6 + 6\Lambda^0_4 - 4 \tilde{\Lambda}^0_4 .
 \end{equation}
 
Let's now prove that $(\mathcal{L}_{K}\xi_-)$ is also a solution if we restrict ourselves to consider \eqref{eq:genKillingSpinor}. Following \cite{Alessandrobook} for the usual Killing spinor case, we have:
 \begin{equation}
\begin{split}
\nabla_a (\mathcal{L}_{K}\xi_-)=& \mathcal{L}_K \nabla_a \xi_- + \partial_a K^b \nabla_b \xi_- \\
=& \mathcal{L}_K \Big[ - \frac{i}{2} \left( \gamma_a + \left(a_1 \Lambda^0_4- \frac{a_2}{2} \tilde{\Lambda}^0_4 \right)\gamma_a + \dots\right) \xi_- \Big] \\
+& \partial_a K^b \Big[ - \frac{i}{2} \left( \gamma_b + \left(a_1 \Lambda^0_4- \frac{a_2}{2} \tilde{\Lambda}^0_4 \right)\gamma_b + \dots\right) \xi_- \Big] \, .
\end{split}
 \end{equation}
 Now, thanks to the fact that the action of $\mathcal{L}_K$ on the invariant forms is zero, we have that $\mathcal{L}_K$ just acts on the free-index gamma matrix $\gamma_a$. The commutator between $\mathcal{L}_K$ and  $\gamma_a$ can be written has following:
 \begin{equation}
[\mathcal{L}_K,\gamma_a]=(\nabla_a K_b - \nabla_{[a} K_{b]}-\partial_a K_b ) \gamma^b = (\nabla_{(a} K_{b)}-\partial_a K_b ) \gamma^b=\partial_a K^b  \gamma_b
 \end{equation}
 so that we are left with
 \begin{equation}
 \nabla_a (\mathcal{L}_{K}\xi_-) = - \frac{i}{2} \left( \gamma_a + \left(a_1 \Lambda^0_4- \frac{a_2}{2} \tilde{\Lambda}^0_4 \right)\gamma_a + \dots\right) \mathcal{L}_{K}\xi_-
 \end{equation}
 which is what we wanted to prove.
 
In order to fix the coefficients $a_i$ in \eqref{eq:genKillingSpinor} we can use the following equation:
\begin{equation}
\label{eq:riccicond}
\gamma^b[\nabla_a, \nabla_b] = \frac{1}{2} \gamma^b R_{ab} , \qquad R_{ab} = 2 (5+2 \lambda^2) g_{ab}-2 (-1 + 5 \lambda^2) g_{ab}|_{\lambda=1}
\end{equation}
where $g_{ab}$ is the metric on the squashed S$^7$ as defined in \eqref{eq:squashedS7met} while $g_{ab}|_{\lambda=1}$ is the metric on the round S$^7$. It is particularly easy to notice from the explicit expression of the Ricci that the manifold is Einstein for $\lambda^2 = 1 , 1/5$.
Plugging \eqref{eq:genKillingSpinor} inside \eqref{eq:riccicond} we find a consistency condition which is satisfied for
\begin{equation}
\label{eq:genKilling_ai}
a_1=a_2=a_3=a_4= \frac{a_7}{2} = \frac{\lambda-1}{2 \lambda} , \quad a_5 = a_6 = \lambda +\frac{1}{2 \lambda }-\frac{3}{2}, \quad a_8 = 0 .
\end{equation}
An explicit computation shows that the multiplets generated by $\{\xi_-, \Lambda^0_4\xi_-,\tilde{\Lambda}^0_4\xi_-\}$ all satisfy \eqref{eq:genKillingSpinor} with the choice of parameters \eqref{eq:genKilling_ai}\footnote{Notice that the condition \eqref{eq:riccicond} does not depend on our multiplet $\xi_-$ and indeed we can find solution to this equation with a different choice of parameters respect to \eqref{eq:genKilling_ai}. However the (\textbf{4},\textbf{2}) multiplets will not solve these generalized Killing spinor equations.}.
 



\bibliographystyle{at}
\bibliography{ref}

\end{document}